\documentclass[%
reprint,
nofootinbib,
 amsmath,amssymb,
 aps,
prd,
superscriptaddress
]{revtex4-2}
\usepackage{tikz}
\usepackage{mathrsfs}
\usepackage{graphicx}% Include figure files
\usepackage{dcolumn}% Align table columns on decimal point
\usepackage{bm}% bold math
\usepackage{bbm}
\usepackage[normalem]{ulem}
\usepackage[activate={true,nocompatibility},final,tracking=true,kerning=true,spacing=true
]{microtype}
\microtypecontext{spacing=nonfrench}
\usetikzlibrary{external}
\usepackage{xcolor}
\usepackage{braket}
\usepackage[unicode, colorlinks, allcolors=blue!70!black, linktocpage, pdfusetitle]{hyperref}
\usepackage[all]{hypcap}
\definecolor{indigo(dye)}{rgb}{0.0, 0.25, 0.42}
\usepackage{enumitem}
\usepackage{amsthm}

\newcommand*{\Tr}{\operatorname{Tr}}
\theoremstyle{definition}
\newtheorem{definition}{Definition}
\newtheorem*{definition*}{Definition}

\newif\ifslow
\slowtrue%comment out if want to not have all the tikz load

\newcommand{\op}[1]{\boldsymbol{#1}}

\tikzset{>=latex}% for LaTeX arrow head
\colorlet{myred}{red!80!black}
\colorlet{myblue}{blue!80!black}
\colorlet{mygreen}{green!80!black}
\colorlet{mydarkred}{red!50!black}
\colorlet{mydarkblue}{blue!50!black}
\colorlet{mylightblue}{mydarkblue!3}
\colorlet{mypurple}{blue!40!red!80!black}
\colorlet{mydarkpurple}{blue!40!red!50!black}
\colorlet{mylightpurple}{mydarkpurple!40!red!3}
\colorlet{myorange}{orange!40!yellow!95!black}
\tikzstyle{cone}=[mydarkblue,line width=0.2,top color=blue!60!black!30,
                  bottom color=blue!60!black!50!red!30,shading angle=60,fill opacity=0.9]
\tikzstyle{cone back}=[mydarkblue,line width=0.1,dash pattern=on 1pt off 1pt]
\tikzstyle{photon}=[-{Latex[length=4,width=3]},myorange,line width=0.4,decorate,
                    decoration={snake,amplitude=0.9,segment length=4,post length=3.8}]
\tikzstyle{world line}=[myblue!30,line width=0.4]
\tikzstyle{world line t}=[mypurple!30,line width=0.4]
\tikzstyle{particle}=[mygreen,line width=0.5]
\tikzstyle{singularity}=[myred,line width=0.6,decorate,
                         decoration={zigzag,amplitude=2,segment length=6.17}]

\tikzset{declare function={%
  penrose(\x,\c)  = {\fpeval{2/pi*atan( (sqrt((1+tan(\x)^2)^2+4*\c*\c*tan(\x)^2)-1-tan(\x)^2) /(2*\c*tan(\x)^2) )}};%
  penroseu(\x,\t) = {\fpeval{atan(\x+\t)/pi+atan(\x-\t)/pi}};%
  penrosev(\x,\t) = {\fpeval{atan(\x+\t)/pi-atan(\x-\t)/pi}};%
  kruskal(\x,\c)  = {\fpeval{asin( \c*sin(2*\x) )*2/pi}};% Penrose coordinates for Kruskal
}}

\usepackage{xfp}% higher precision (16 digits?)
\usepackage[outline]{contour}% glow around text
\usetikzlibrary{decorations.markings,decorations.pathmorphing}
\usetikzlibrary{angles,quotes}% for pic (angle labels)
\usetikzlibrary{arrows.meta}% for arrow size
\contourlength{1.4pt}

\newcommand{\scri}{\ms I}

\newcommand{\be}{\begin{equation}}
\newcommand{\ee}{\end{equation}}

\newcommand{\ms}{\mathscr}

\renewcommand{\=}{\hateq}

\newcommand{\union}{\cup}

\newcommand{\Lie}{\pounds}
\newcommand{\hatLie}{\Lie\kern-0.25em\hat{\vphantom{\Lie{}}}\kern0.25em}

\newcommand{\hateq}{\mathrel{\mathop {\widehat=} }}
\let\oldint\int
\renewcommand{\int}{\oldint\limits}

\begin{document}
\title{Horizons and Soft Quantum Information}

\author{Daine L. Danielson}
\email{daine@mit.edu}
\affiliation{Center for Theoretical Physics -- a Leinweber Institute, Massachusetts Institute of Technology, Cambridge, Massachusetts 02139, USA}
\affiliation{Black Hole Initiative and Department of Physics, Harvard University, Cambridge, Massachusetts 02138, USA}
\affiliation{Leinweber Institute for Theoretical Physics, Enrico Fermi Institute, and Department of Physics, The University of Chicago, Chicago, Illinois 60637, USA}
\author{Gautam Satishchandran}%
 \email{gautam93@mit.edu}
 \affiliation{Princeton Gravity Initiative, Princeton University, Princeton, New Jersey 08544, USA}

\date{\today}

\begin{abstract}
    \noindent It was recently shown that black holes decohere any quantum superpositions in their vicinity. This decoherence is mediated by soft radiation through the horizon, and can be understood as the result of the fact that quantum states in the exterior source distinguishable states of long-range fields in the interior. To study this phenomenon and others, we extend Tomita-Takesaki theory to accommodate states of soft radiation such as arise in the electromagnetic and gravitational memory effects, and provide a general framework for computing the distinguishability of general coherent states. Applying these tools, we use the methods of unambiguous state discrimination and approximate quantum error correction to prove some new relations regarding the distinguishability of quantum states, and the quantum information content of soft radiation, and thereby show that a black hole (or, under the conditions stated below, any causal horizon) decoheres its environment as though its interior were full of optimal observers.
\end{abstract}

\maketitle

\section{Introduction}
The fact that black holes have a finite temperature \cite{hawking1975particle} and, thereby, a nonvanishing entropy \cite{Beckenstein_1973} has provided some of the deepest clues into the nature of these objects in a quantum theory of gravity. In the case of black holes, this entropy has been argued to arise from ``internal degrees of freedom'' (described by microstates) and, for certain quantum gravitational realizations of black holes, the counting of microstates can directly reproduce their entropies \cite{Strominger_1996}. Despite these successes, the origin and nature of these  microstates remain poorly understood. Indeed, classically, a black hole has no internal degrees of freedom. What is their semiclassical imprint? What is their effect on observables in the exterior? These questions lie at the heart of many paradoxes and recent investigations regarding the semiclassical nature of black holes \cite{Witten, Kudler-Flam_2023}. In these investigations, quantum information theory has proven to be an indispensable tool in providing a bridge between the quantum nature of black holes and their classical spacetime geometry \cite{Harlow_2014, Engelhardt_2014, Bousso_2015, Faulkner_2013}.

The fact that, in the semiclassical limit, {\em any} horizon must behave as though it contains internal degrees of freedom directly follows the basic principles of quantum mechanics and causality \cite{Satishchandran:2025cfk, Danielson_thesis}. To illustrate this we consider the following gedanken experiment involving two experimenters: Alice and Bob. For definiteness, we consider the case where they perform their experiments in a black hole spacetime.
Alice controls a quantum superposition which she attempts to keep coherent in the exterior of a black hole. 
Despite her best efforts, she cannot possibly control the long-range gravitational field of her superposition. The resulting superposed gravitational field penetrates the horizon of the black hole where an observer, Bob, could choose to measure it. If Bob can gain any information about Alice's superposition, then her superposition must be decohered by a complementary amount. However, by causality,  Bob's actions cannot have any effect whatsoever on the state of her superposition. Furthermore, while the interior of a black hole is not an optimal place to perform such a sensitive experiment, we could imagine equipping Bob with better apparatuses and more assistants.  From this gedanken experiment we conclude that Alice's superposition must become entangled with---and subsequently decohered by---the black hole {\em as if} the black hole contains ``maximally entangling'' degrees of freedom. 
We note that the crucial ingredient in this argument was the existence of an event horizon which causally separated Alice and Bob. Thus, these conclusions apply whenever such a horizon bounds the causal past of Alice's worldline---e.g., Rindler horizons,  cosmological horizons.

Motivated by this thought experiment we, together with R.\ M. Wald, have shown in a series of papers that, indeed, any (Killing) horizon will decohere a quantum superposition in its vicinity \cite{DSW_2022,DSW_2023,DSW_2024}. To explicitly show this, we considered the case where Alice creates a quantum spatial superposition of a massive (or charged) body---e.g., a nanoparticle---which, to a good approximation, produces a superposition of coherent states of the gravitational field. In the case of a black hole it was shown that, semiclassically, the black hole acquires low-frequency, fluctuating gravitational (and electromagnetic) multipole moments \cite{DSW_2024} which stimulate the emission of entangling, soft radiation into the black hole \cite{DSW_2022}. Furthermore, these multipole fluctuations closely mimic those of a material body---though it would appear that, in the gravitational case, extraordinary properties of matter would be required to reproduce the effect \cite{Biggs_2024}. 
In total these {results} suggest that the decoherence caused by black holes is, in some sense, ``extreme'' given its size and temperature. However, these analyses only partly validate the conclusions of the above gedankenexperiment. How does this decoherence compare to the decoherence inflicted by an ``optimal'' Bob in the interior of the black hole? If a black hole behaves like a material body when viewed from the outside, then it should be possible to show that any family of observers---or ``degrees of freedom''---in the interior can gain (at most) as much ``which-path'' information about Alice's particle as is allowed by the degree of decoherence of Alice's superposition in the exterior. 

As a warm-up, in Sec. \ref{sec:simpleIDT}, we first address {this question} by revisiting the specific gedanken experiment considered in \cite{DSW_2022,DSW_2023,DSW_2024} where Alice creates a quantum spatial superposition. From Bob's point of view,
he is attempting to perform a ``one-shot'' experiment to distinguish the two possible states of Alice's particle via their long-range field. In quantum information theory, there are many different measures of the distinguishability of states (e.g., the Holevo and {Uhlmann} fidelities, the relative entropy, the trace distance), each with its own operational meaning. The relevant notion of distinguishability in this case is the {Uhlmann} fidelity. For Alice's particle at a finite time, the {Uhlmann} fidelity computes the ``optimal purification'' of her particle so as to minimize her decoherence \cite{Uhlmann_1976}. In this sense, it is precisely a measure of ``how decohered'' her particle is up until that time. In the interior, the {Uhlmann} fidelity also bounds the probability that Bob, by performing a single optimal measurement, can become absolutely certain of the state of Alice's particle. 
We thereby obtain the following statement: the decoherence of Alice's superposed particle at any finite time is always greater than or equal to the decoherence due {to} an optimal one-shot, unambiguous state discrimination test in the interior up until that time. At late times this inequality is saturated, as if the black hole (or interior observers / degrees of freedom) is performing an optimal measurement of Alice's superposed long-range fields. Indeed, this relationship between the information accessible to Bob and the decoherence of Alice by soft quanta has been recently used as a ``teleportation protocol''  {to} send information into a black hole with negligible energy cost \cite{KP_2025}.  

The gedanken experiment suggests, however, that such a relationship, between Alice's experiment in the exterior and the decoherence due to optimal measurements made in the interior, should hold for {\em any} experiment performed in the vicinity of a black hole. In Sec.~\ref{sec:IDT} we consider this more general relationship. In quantum information theory, a general experiment performed by Alice is a ``quantum channel.'' We introduce definitions for the intrinsic decoherence of a quantum channel, and the information of its complementary channel.  The ``channel decoherence'' quantifies the  failure to be able to recover her superposition by performing measurements strictly in the black hole exterior. The complementary ``channel information'' quantifies the information accessible to Bob, and bounds the probability of unambiguously distinguishing the output of his channel from a totally uncorrelated state. Section~\ref{sec:IDT} gives precise definitions for the channel decoherence and information. As is often the case in quantum information theory \cite{1996quant.ph.11010F, Fuchs_1995,Beny:2010byk}, there is a trade-off between the information carried by a channel into the environment and the disturbance of the original system by that same environment. In Sec.~\ref{sec:IDT}, however, we show that the channel decoherence and channel information enjoy an \textit{exact} ``information-decoherence trade-off.'' Namely,  we show the two quantities are precisely equal to one another.

Applied to black holes, the channel decoherence and channel information provide an \textit{intrinsic} definition of the exterior decoherence, and interior information, of the black hole in the semiclassical regime, irrespective of any experimental protocol. We recall that, in ordinary experience, decoherence due to entanglement generated with any material body is, to some extent, reversible in principle. In Sec. \ref{sec:secondLawforHorizons} we show that this is not the case for any semiclassical black hole (with any deviations from being suppressed in $G_\mathrm{N}$). Rather, a black hole exhibits what might be called a ``second law'' for exterior decoherence. Additionally, we
comment on how these ideas, being fundamentally quantum-information-theoretic in nature, might plausibly be extended beyond the semiclassical limit. In total, these results lend support to the perspective that a black hole {affects} its environment as though it were composed of maximally entangling degrees of freedom.

What of the more familiar notions of distinguishability, such as the Holevo fidelity, relative entropy, and the R\'enyi entropy? While
less operationally relevant
to the decoherence effects considered in Sec.~\ref{sec:IDT}, these notions of distinguishability have played a prominent role in analyzing the quantum mechanical properties of horizons ranging from the generalized entropy to the black hole information paradox \cite{Lewkowycz:2013nqa,Almheiri:2019hni,Harlow_2014,Casini:2008cr}. How does the growth of these other notions of distinguishability compare to that of the Uhlmann fidelity on the horizon? In Sec.~\ref{sec:horizonAlgebra}, we address this question in the context of the gedanken experiment considered in \cite{DSW_2022, DSW_2023, DSW_2024, DKSW_2025} where the relevant states are coherent states on the horizon.

We answer these questions in Secs.~\ref{sec:horizonAlgebra} and \ref{sec:ttTheory}. To explain the answer we first recall that the growth of distinguishability on the horizon is due to the emission of soft radiation. These soft radiative effects are mathematically equivalent to the soft radiative effects found in flat spacetime \cite{PSW_2022, Prabhu_2024, Prabhu_2024b, Strominger_2, Ashtekar_2018}. These effects result in well-known infrared divergences which have plagued scattering calculations\footnote{In flat spacetime, the IR divergences occur in four dimensions with a massless field. On the horizon, they occur in any spacetime dimension with a massive or massless field   \cite{DSW_2023}.} since the earliest days of quantum field theory \cite{Fierz_1936}. Indeed, both effects are controlled by a ``memory tensor'' which on the horizon is given by 
\begin{equation}
\Delta_{AB} =  h_{AB}\vert_{V=\infty} - h_{AB}\vert_{V=-\infty},
\end{equation}
where $h_{AB}(V,x^{A})$ are the radiative degrees of freedom on the horizon, $V$ is the affine time, and $x^{A}$ are angular coordinates. The memory captures the failure of the radiation to decay at asymptotically early or late times. States with nonvanishing memory contain an {\em infinite} number of soft quanta and lie in Hilbert space representations unitarily inequivalent to the ``standard'' Fock space of radiation. Analogously, the IR divergences encountered in scattering theory arise from the failure to include states with memory in the theory.

In Sec.~\ref{sec:horizonAlgebra}, we consider the practical calculation of distinguishability measures of radiation states on the horizon and the extent to which these measures capture effects due to soft radiation. The emission of an infinite number of soft quanta and the corresponding IR divergences occur only in the strict infinite-time limit if $h_{AB}$ fails to decay. At any finite time, only a finite number of soft quanta can be produced. Therefore, one might expect that the IR issues highlighted above play no role in calculating the distinguishability of radiation states at any finite time. However, as we now explain, incorporating states with memory enables one to calculate finite time distinguishability measures in a simple and efficient manner.

We note that the practical computation of any of the above distinguishability measures for any two states requires the technology of ``Tomita-Takesaki theory'' and is an extremely difficult and unsolved problem. However we can, for our purposes, restrict to coherent states of radiation since those were the radiation states most relevant to our original gedanken experiment\footnote{Coherent states of soft radiation also describe the soft emission of scattering phenomena.} \cite{Satishchandran:2025cfk, Danielson_thesis, DSW_2021, DSW_2023, DSW_2024, DKSW_2025}. If $h_{AB}$ is a classical radiation field (with no memory) on the horizon, we denote its corresponding coherent states as $\op{U}(h)\ket{\Omega}$ where $\op{U}(h)$ is the unitary operator generating the coherent state and $\ket{\Omega}$ is the vacuum state on the horizon. It was pointed out by Casini \textit{et al.} that if the operator $\op{U}(h)$ factorizes across $\mathscr{C}$, then one can straightforwardly apply the technology of standard Tomita-Takesaki theory. One can obtain an explicit computation of the Tomita operator and the relative modular operator in this case and one can straightforwardly compute any of the above distinguishability measures.

However, $\op{U}(h)$ factorizes if and only if $h_{AB}$ vanishes at the cut. If $h_{AB}$ is nonvanishing, then there does not appear to be a simple description of the Tomita operator or modular operator within standard Tomita-Takesaki theory. {However, $h_{AB}$ can be shifted by a smooth $V$-independent trace-free tensor, whose electric-parity part is a supertranslation,}
{\begin{equation}
h_{AB}(V,x^{A}) \to h_{AB}(V,x^{A})+k_{AB}(x^{A}).
\end{equation}}
{By choosing $k_{AB}$ appropriately,} we show that the unitary can be essentially factorized about the cut. While this removes the above obstruction, the {shifted} radiation state fails to decay and the unitary operator factors now have nonzero horizon memory. Therefore, to obtain a simple description of distinguishability measures at a finite time, we must extend Tomita-Takesaki theory to include states with memory. We obtain such an infrared extension of Tomita-Takesaki and obtain explicit formulas for the Tomita operator and the relative modular operator between any two coherent states for any cut $\mathscr{C}$. Using this construction we explicitly compute the fidelity, relative entropy, and R\'enyi entropies of general coherent states on the horizon, with or without memory. We use this to analyze the quantum information content of soft radiation. We find that this quantum information content is well-captured by the fidelity, as might be expected from the original thought experiment. In that context, the fidelity is sensitive to the absolute number of soft gravitons emitted into the black hole (which may be arbitrarily large), as opposed to their energy (which may be arbitrarily small). The relative entropy is, by contrast, only very weakly sensitive to the number of soft photons, and is much more sensitive to the \textit{energy} of the radiation emitted into the horizon.

Throughout this work, we develop several new techniques that may find wider utility. For example, the equivalence of channel decoherence and complementary channel information shown in Sec.~\ref{sec:IDT} is a general result for quantum channels, which are the most general models of physical processes acting on quantum states. For a particular class of channels, in which the system becomes entangled with two states of another system, Appendix \ref{sec:approximateCapacity} also proves a bound between the channel decoherence and the fidelity between those two quantum states. To analyze other distinguishability measures, Secs.~\ref{sec:horizonAlgebra} and \ref{sec:ttTheory} extend the Tomita-Takesaki theory of coherent states so that it includes soft radiation, thus putting questions pertaining to the quantum information content of soft modes on rigorous mathematical footing. We believe these methods may find utility in contexts beyond those considered here.

The reader who is merely interested in the statement of the new results in Tomita-Takesaki theory may safely skip ahead to Secs.~\ref{sec:horizonAlgebra} and~\ref{sec:ttTheory}. There, as in much of the present article, we simplify to the case of electromagnetic radiation, and the gravitational case follows by close analogy. Section~\ref{sec:horizonAlgebra} extends the horizon algebra to include soft modes, and Sec.~\ref{sec:ttTheory} gives the formula for the relative Tomita operator in general coherent states.
We use this to bound the more general and abstract information-decoherence relations of prior sections in terms of explicit fidelities. Section~\ref{sec:discussion} extends the discussion of the present work to the context of scattering theory, where directly {analogous} considerations apply. Finally, Sec.~\ref{sec:causal} applies the techniques of Sec.~\ref{sec:IDT} to show how a generic causal horizon can give rise to fundamental decoherence in its vicinity.

Throughout this paper we work in units where $\hbar = c = G_\mathrm{N} = \epsilon_0 = 1$. Lowercase Roman indices denote indices in spacetime, while uppercase Roman indices denote indices tangent to the spatial cross sections of the horizon. Greek indices label the numerical components of a tensor in a basis. We denote quantized observables in a $*$-algebra by the boldface version of the corresponding classical observable. For example, the classical electric field $E$ becomes $\op{E}$ in the quantum theory. We denote von Neumann {algebras} in fraktur as $\mathfrak{A}$, and restrictions of algebras to the past of a horizon cut $\mathscr{C}$ by $\mathfrak{A}_c$.

\section{Information-Decoherence Trade-off}
\label{sec:IDT}
In this section we show that the decoherence inflicted by a horizon is a consequence of an ``information-disturbance trade-off'' between the coherence of any  quantum mechanical system in the exterior and the information about that body accessible to any observer in the interior. In Sec.~\ref{sec:simpleIDT} we explain how one can already see a version of this relationship in context of the specific gedanken experiment involving the coherence of a quantum spatial superposition considered in \cite{DSW_2022,DSW_2023,DSW_2024,DKSW_2025}. To prove the more general relationship suggested in the Introduction, in Sec.~\ref{sec:simpleIDT} we generalize the notions of decoherence and information accessible in the interior to any experiment and prove their equality. In the case of horizons, this proves the statement that the decoherence inflicted by a black hole---or any horizon satisfying the assumptions stated below---is related by the exact information-decoherence trade-off to the information available in its interior. For horizons in particular we show that the decoherence of the exterior grows monotonically with time, as would also be expected from a body comprising maximally entangling degrees of freedom.
\subsection{Analysis of the original gedanken experiment}\label{sec:simpleIDT}
In this section we revisit the original gedanken experiment considered in \cite{DSW_2022,DSW_2023,DSW_2024,DKSW_2025} involving a quantum spatial superposition in the vicinity of a horizon. From this example, we will identify (1) the appropriate notion of decoherence of such a superposition at any finite time and (2) a precise operational meaning of this decoherence in terms of an ``unambiguous state discrimination test'' performed in the interior. While our considerations will be valid for any Killing horizon \cite{DSW_2023}---e.g., a stationary black hole horizon or de Sitter horizon---we will, for simplicity, restrict attention to Alice performing her experiment in the vicinity of a Rindler horizon in Minkowski spacetime. 

An experimenter, Bob, assembles his lab equipment on the opposite side of the Rindler horizon of a uniformly accelerating experimenter, Alice. Now suppose that Alice prepares a charged nanoparticle in a definite $x$-spin state. She then uses a Stern-Gerlach apparatus to adiabatically manipulate the particle. If the Stern-Gerlach were oriented along the $x$-direction, then it would simply displace her particle. If, however, the apparatus is oriented in the $z$-direction, she can use it to prepare a quantum spatial superposition of her particle's wave function. 
After doing this, her superposed particle sources a superposition of coherent states of the electromagnetic field (nearly identical considerations apply for gravity, with mass replacing charge, and quadrupole radiation replacing dipole radiation). The global Hilbert space is now \begin{equation}
    \mathcal{H} = \mathcal{H}_\text{Alice}\otimes\mathcal{H}_{\mathscr{H}^+}.
\end{equation}
We denote the coherent states $|\Psi_1\rangle$ and $|\Psi_2\rangle$ in the horizon Hilbert space $\mathcal{H}_{\mathscr{H}^+}$,  while denoting the $z$-spin states (and corresponding spatial wave functions $A_i$) of Alice by $|A_1,\downarrow\rangle$ and $|A_2,\uparrow\rangle$ in $\mathcal{H}_\text{Alice}$.
In our thought experiment, the resulting entangled state at late times will be
\begin{equation}
\label{eq:superpose}
    \frac{1}{\sqrt{2}}\left(|A_1,\downarrow\rangle\otimes|\Psi_1\rangle_{{\mathscr{H}^+}} + |A_2,\uparrow\rangle\otimes|\Psi_2\rangle_{{\mathscr{H}^+}}\right),
\end{equation}
where ${\mathscr{H}^+}$ denotes the Rindler horizon and labels the corresponding, entangled states of radiation. Meanwhile, suppose that Bob is interested in distinguishing between these two field configurations by performing a measurement on his side of the Rindler horizon. 

Previous analyses have focused on the decoherence of Alice's superposition, as quantified by 
\begin{equation}
    1-|\langle \Psi_1|\Psi_2\rangle_{\mathscr{H}^+}|,
\end{equation}
using the inner product of the final states of entangling radiation that her experiment sources on the horizon. In other words, the final decoherence is controlled by the off-diagonal elements of Alice's final reduced density matrix. 

In addition to the final decoherence of her particle, it is natural to ask for the decoherence of her superposition as a function of time. Such a refined definition of decoherence {would} be useful for monitoring the dynamical buildup of Alice's decoherence outside a horizon, but this is complicated by the fact that the ``off-diagonal elements of Alice's reduced density matrix'' are not well-defined while Alice's superposition is still held open. Formally, such a quantity would involve ``tracing out'' and taking the inner product between two different Coulomb fields, which are not independent degrees of freedom from Alice's system. It could be argued that the two fields should be considered to be orthogonal, but the resulting apparent, total decoherence of Alice's superposition is ``false decoherence'' \cite{Unruh_2000} that in no way obstructs the recovery of Alice's initial state. 

\begin{figure}
    \centering
\includegraphics[width=0.9\columnwidth]{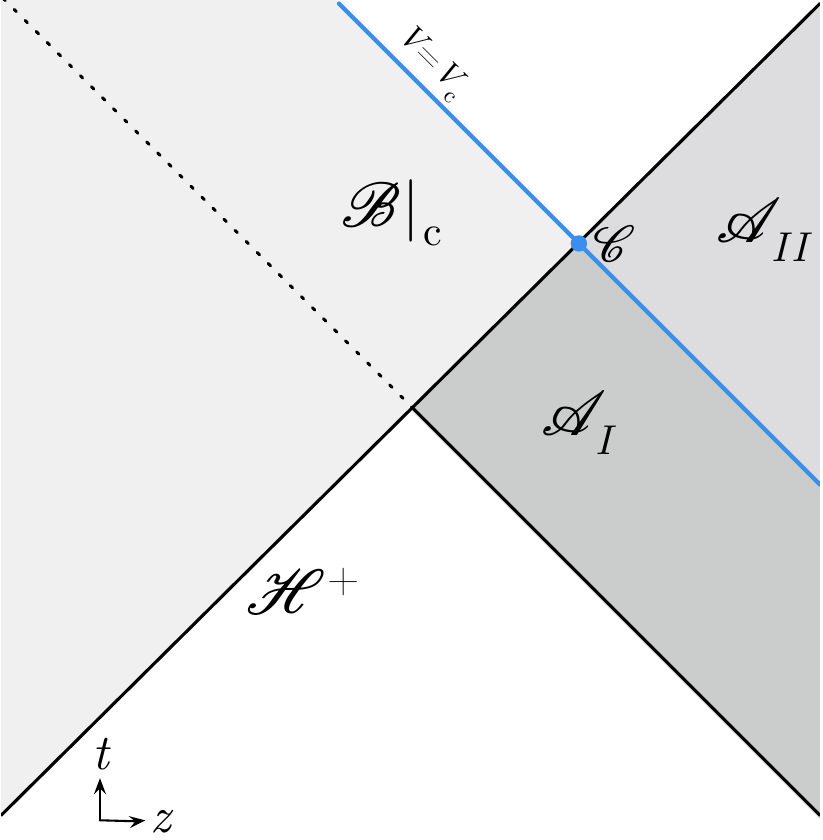}
    \caption{Bob can perform any possible measurement in region $\mathscr{B}_c$, defined as the region on the $-z$ side of the horizon that lies to the past of the future light cone of the horizon cut $\mathscr{C}$}.
    \label{fig:heuristic_fig}
\end{figure}

A natural candidate at finite times, which avoids the problem of false decoherence, is to compute the off-diagonals of Alice's reduced density matrix if, after the chosen time, she were to perform the remainder of her experiment so as to maximize the recovery of the coherence of her original state.
More specifically one may ask, what is the minimum decoherence that the horizon will inflict on Alice, if, in region $\mathscr{A}_{II}$ of Fig.~\ref{fig:heuristic_fig}, she is allowed to manipulate her particle so that it sources radiation in region $\mathscr{A}_{II}$ that minimizes her final decoherence? This protocol gives one operational notion of the ``minimum decoherence'' (or ``maximum recoverable coherence''), and forms the basis of the analysis carried out in \cite{DKSW_2025}. This seems like a reasonable candidate, but it is by no means clear that this is the distinguished, operationally relevant notion of decoherence at a finite time.

Even the idea of using the off-diagonals of Alice's density matrix to diagnose coherence can be questioned, because while the quantities clearly control the coherence of her superposition, their precise operational meaning is not clearly the right one. In our case, however, we have already argued that the decoherence should have an operational meaning in terms of distinguishability by Bob. To single out the operationally pertinent signature of decoherence, therefore, we will not look to Alice; we will look to Bob.

While singling out an operationally distinguished measure of decoherence appears challenging, what is clear from the thought experiment is that there should be a complementary notion of the which-path distinguishability information available to the ``observer,'' Bob, in the interior. We will see that this notion of distinguishability is operationally transparent even at finite times, and it gives a natural complementary notion of decoherence for Alice that will, in fact, correspond directly to the notion of \cite{DKSW_2025}. To see this, we will put ourselves in Bob's position.

Suppose Bob plans to complete his measurement in a particular finite time interval, so that his measurement concludes before he enters the future light cone of a given horizon cut $\mathscr{C}$ as illustrated in Fig.~\ref{fig:heuristic_fig}.
If we allow Bob (and any number of assistants) to perform any conceivable measurement or joint measurement in his spacetime region $\mathscr{B}|_c$, what is the upper bound on the probability for him to successfully determine the position of Alice's particle?

The precise answer to this question depends on the precise specification of the task assigned to Bob. Suppose Bob is allowed to perform his measurement anywhere outside the future light cone of a horizon cut $\mathscr{C}$. The resulting states of the quantum field, restricted to Bob's region, will be mixed states $\Psi_1|_c$ and ${\Psi_2|_c}$. Now suppose that, after performing some measurement, Bob is allowed to guess whichever state would be more probable to give the outcome he has observed: we say ``guess,'' because even if his measurement yields an outcome that leaves some ambiguity as to which state he has measured, Bob might be allowed to guess the most probable state, given that outcome. The success probability of this ``state discrimination'' task will be bounded in terms of the trace distance between the two states \cite{helstrom1969quantum}. But Bob's ability to \textit{guess} the right state has just as much to do with classical correlation as it does with entanglement and decoherence, so it cannot be the right operational task.

If, on the other hand, we disallow guessing, we isolate a different quantity. If we set the rules such that Bob may only declare success if he is \textit{certain} of the state, then the so-called ``Uhlmann fidelity'' is the relevant quantity. He could obtain certainty of the state in such a ``one-shot'' test if he were to obtain a measurement outcome that is \textit{only} possible in one of the two states he must distinguish between. When this additional ``no guessing'' rule is imposed, Bob's task is known as an unambiguous state discrimination task, and the Uhlmann fidelity $f_\mathrm{U}(\Psi_1|_c, \Psi_2|_c)$ bounds his success probability to unambiguously distinguish between the states $\Psi_1|_c$ and $\Psi_2|_c$ on the horizon, restricted to the past of $\mathscr{C}$, as illustrated in Fig.~\ref{fig:heuristic_fig}.

Given two mixed states $\rho$ and $\sigma$, the Uhlmann fidelity is the supremum of the inner products between purifications of the states. We denote these purifications as $\ket{\rho}_{\pi}$ and $\ket{\sigma}_{\pi}$ with respect to the Hilbert space representation $\pi$. In other words, the Uhlmann fidelity is defined as \cite{Uhlmann_1976}
\begin{equation}\label{eq:uhlmannDefinition1}
    f_\mathrm{U}(\rho,\sigma) := \sup_{\pi}|\langle \rho|\sigma\rangle_\pi|^2.
\end{equation}
Here the supremum runs over all Hilbert space representations $\pi$ of a given $*$-algebra such that $|\rho\rangle$ and $|\sigma\rangle$ are vectors in the corresponding Hilbert space $\mathcal{H}$ and are purifications of the states $\rho$ and $\sigma$, respectively.

The probability that Bob will successfully distinguish the state in this ``unambiguous'' sense is then bounded by
\begin{equation} \label{eq:uhlmannIntermediateTimes}
P_\mathrm{Bob}^\mathrm{certain}(\mathscr{B}|_c) \le 1-\sqrt{f_\mathrm{U}}(\Psi_1|_c, \Psi_2|_c).
\end{equation}
As we shall see, when we relax the time constraints on Bob, this bound in fact becomes an equality.

It is this unambiguous notion of state discrimination, which measures Bob's ability to achieve \textit{certainty}, that is most closely tied to the decoherence of Alice's particle. This is easily seen from the fact that the Uhlmann fidelity has another, complementary interpretation from the point of view of Alice in the exterior: it gives the infimum of Alice's unavoidable decoherence discussed above. Now, this measure of decoherence stands on firm operational footing, in terms of the knowledge that can be obtained by an observer coupling to Alice's environment.

The Uhlmann fidelity gives the infimum of the decoherence $D_\mathrm{Alice}|_c$ that the horizon will inflict on Alice, if, in region $\mathscr{A}_{II}$ of Fig.~\ref{fig:heuristic_fig}, she is allowed to manipulate her particle so that it sources radiation in region $\mathscr{A}_{II}$. The resulting final decoherence of Alice is
\begin{equation}\label{eq:dAlice}
    D_\mathrm{Alice}|_c := 1-\sqrt{f_\mathrm{U}}(\Psi_1|_c,\Psi_2|_c).
\end{equation}
From Alice's point of view, the ``optimal trajectory of her particle'' can be seen as sourcing precisely the ``optimal purification'' of Eq. (\ref{eq:uhlmannDefinition1}) on the horizon \cite{DKSW_2025}.
Thus, we learn that in general, the probability that Bob succeeds is bounded by the supremum of Alice's recoverable coherence,
\begin{equation}\label{eq:twoStataeMixedStateBound}
P_\mathrm{Bob.}^\text{certain}(\mathscr{B}|_c) \le D_\text{Alice}|_c.
\end{equation}
In fact, when we place no time restrictions on Bob and allow him access to the entire region on his side of the horizon,\footnote{In the case of a black hole in an asymptotically flat spacetime, the saturation is never achieved, because of the entanglement in the vacuum between the horizon and null infinity.} it is guaranteed that Bob can saturate the inequality by some (possibly very complicated) measurement \cite{Eldar_2003}, and the probability that an optimal measurement by Bob\footnote{We here neglect any backreaction of Bob's experiment on the environment. It is, for example, possible that a practical realization of the {optimal} protocol might produce a dramatic change to the spacetime in Bob's region such that a singularity forms. This could prevent Bob from achieving the optimal protocol in practice. Nevertheless, the physics in Alice's region will proceed \textit{as if} Bob were performing an optimally discriminating measurement.} will unambiguously reveal the state of Alice's system is precisely equal to her decoherence:
\begin{equation}\label{eq:twoStatePureStateEquivalence}
P_\mathrm{Bob.}^\text{certain}(\mathscr{B}) = 1-\sqrt{f_\mathrm{U}}(\Psi_1,\Psi_2) = D_\text{Alice}.
\end{equation} At least in this scenario, one may view the horizon as carrying out an unambiguous state discrimination test on the particular quantum states prepared by Alice in its exterior. 
We have assumed that in Eq. (\ref{eq:superpose}), Bob must distinguish between two equiprobable states. In more general situations, the states may have unequal prior probabilities $p_+, p_{-}$ with $p_+ > p_-$. If $f_\mathrm{U}\ge p_-/p_+$, the probability that Bob will succeed is bounded by the fidelity itself, $P^\text{certain}_\text{Bob} \le p_+(1-f_\mathrm{U})$, rather than the square root \cite{rudolph_unambiguous_2003}, and the interpretation in terms of unambiguous state discrimination again holds.

This result, and the results of \cite{DSW_2022,DSW_2023,DSW_2024,DKSW_2025} are, however, incomplete: they describe the decoherence due to horizons as inflicted on a particular class of experiments of the form of Eq. (\ref{eq:superpose}). This is far from the general result claimed in the Introduction. The original gedanken experiment suggests that there should be a general equivalence between the decoherence in the vicinity of a horizon, and the distinguishability of the resulting states behind the horizon. In fact, this equivalence suggests that the causal structure of the horizon should be intimately connected with this operational fact and, taken together, causality and complementarity dictate that this principle should hold regardless of the particulars of Alice's experiment. Equations (\ref{eq:twoStataeMixedStateBound}) and (\ref{eq:twoStatePureStateEquivalence}) only pertain to when Alice creates a superposition of two states, as in Eq. (\ref{eq:superpose}), so this is far too restrictive a notion of decoherence to capture the original idea of the gedanken experiment. To capture the general relationship, we will need a more general characterization of decoherence, and of the information available to Bob. Nevertheless, we will remain rooted in the operational interpretation of unambiguous state discrimination, and the generalizations we define will recover the familiar definition of the final decoherence of Alice's particle when restricted to the experiment of Eq. (\ref{eq:superpose}). In doing this we will arrive at a definition of the decoherence intrinsic to the horizon itself.

\subsection{Equivalence of information and decoherence for general experiments}
\label{sec:Equivalence}

In this subsection we will no longer assume anything about the particular form of Alice's experiment or Bob's. We will formulate an equivalence between the decoherence of Alice and the information available to Bob or arbitrary cuts of the horizon, and for arbitrary quantum experiments by Alice and Bob. These more general notions of decoherence and information will recover the original notions when applied to the specific experiment of Alice and Bob, and will carry suitably generalized operational definitions in terms of unambiguous state discrimination. No clarity will be lost if the reader continues to assume that the quantum state is of the form given Eq. (\ref{eq:superpose}), but in fact the following discussion will hold for any state Alice can create on her side of the horizon and for any experiment Bob can do on his.

A suitable notion of decoherence should satisfy some essential properties. It should be the following:
\begin{enumerate}[label=D.\roman*]
    \item Intrinsic to the horizon in that it is independent of the exterior state.
    \item Localized to the past of a given horizon cut.
    \item Equivalent to the ordinary notion of decoherence in the appropriate limit.
\end{enumerate}
The complementary notion of information should be the following:
\begin{enumerate}[label=I.\roman*]
    \item Independent of the choice of experiment used to measure the information behind the horizon.
    \item Localized to experiments performable behind the horizon and to the past of the causal future of a given horizon cut.
    \item Equivalent to an ordinary notion of distinguishability behind the horizon in the appropriate limit.
\end{enumerate}

After developing these general definitions, we will apply them to formulate channel-level bounds on unambiguous state discrimination tasks involving general quantum states in the exterior.

The key to achieving properties D.i and I.i is the notion of ``quantum channel'' $\mathcal{N}$, which is the most general description of any physical experiment performed by either Alice {or} Bob. A channel is a {linear, completely positive, trace-preserving} map, but is otherwise unconstrained. Every channel can be realized by coupling some ``environment'' (in our case, the horizon) to the system and carrying out the subsequent open evolution  \cite{stinespring1955positive}. In our case, 
\begin{equation}
\mathcal{N}^{\mathscr{H}^+}_\text{Alice}:\mathcal{S}(\mathcal{H}_\text{Alice})\to\mathcal{S}(\mathcal{H}_\text{Alice}),
\end{equation} where $\mathcal{S}(\mathcal{H}_\text{Alice})$ is the state space (density matrices) on Alice's Hilbert space. $\mathcal{N}^{\mathscr{H}^+}_\text{Alice}$ will be responsible for the decoherence of Alice's state outside the horizon. For every channel $\mathcal{N}$ there is a ``complementary'' channel $\mathcal{N}^c$ (unique up to isometric equivalence) corresponding to the change in this environment as the joint evolution applies $\mathcal{N}$ to the original system. In our case, for example, a complementary channel \begin{equation}(\mathcal{N}^{\mathscr{H}^+}_\text{Alice})^c:\mathcal{S}(\mathcal{H}_\text{Alice})\to\mathcal{S}(\mathcal{H}_{\mathscr{H}^+})
\end{equation}
would map Alice's initial state onto the resulting state of the \textit{horizon} (equivalently, the state in Bob's region). This complementary channel transmits the entangling photons/gravitons produced by Alice's experiment into Bob's region of the spacetime, thereby providing the complementary which-path information that Bob must measure. The relevant formal definitions, as well as the substance of the proofs developed in this section, are presented in Appendix \ref{app:IDT}. The goal of this section is to present the results in an intuitive and physically motivated context.

\subsubsection{Definition of channel decoherence}
An important measure of quantum coherence is the recoverability of a quantum state after that state has passed through a channel $\mathcal N$. Decoherence manifests as an inability to recover the full initial state, and we will provide a definition that formalizes this notion.

If a channel perfectly preserves the coherence of states, then there exists a recovery channel $\mathcal{R}$ (sometimes called a ``decoding channel''), independent of Alice's initial state such that
\begin{align}
    \mathcal{R} \circ \mathcal{N} = \text{Id} .
    \label{eq:noDecoherence}
\end{align}
This means that the experiment is perfectly reversible and is the condition of \textit{exact} quantum error correction. A trivial example of a quantum channel that is perfectly reversible is a unitary channel, such that the decoding channel is simply the adjoint unitary channel. Of course, practical experiments are not perfectly reversible, so we will use the language of \text{approximate} quantum error correction to define a general measure of channel decoherence that ranges between full coherence and total decoherence.

To do this, we need a more refined characterization of the similarity of quantum channels than strict equality. We seek a notion of ``similarity'' between the recovered channel $\mathcal{R}\circ\mathcal{N}$ and the identity channel $\operatorname{Id}$ that runs between 1 (equality/recoverability) and 0 ({total} distinguishability/unrecoverability). The ``channel fidelity'' gives the appropriate notion of similarity of quantum channels \cite{ksw_channelFidelity_2007}. Here we build up its definition piece-by-piece. Along the way we explain each piece of the definition in the special case where the channels model the Stern-Gerlach experiment of Sec.~\ref{sec:simpleIDT}.

Given any two quantum channels $\mathcal{N}$ and $\mathcal{M}$ and a chosen input state $\omega$, the similarity of their outputs is captured by the Uhlmann fidelity as
\begin{equation}
f_\text{U}(\mathcal{N}(\omega),\mathcal{M}(\omega)).
\end{equation}
We now seek a notion of the similarity between the channels themselves, independent of any particular initial state. Not all states are equally good when it comes to discriminating one channel from another. Consider, for example, a situation in which $\mathcal{N}$ corresponds to Alice running a spin through her Stern-Gerlach apparatus in flat spacetime, while we take $\mathcal{M}$ to be the identity channel. Then, if Alice prepares her spin in a definite $z$-spin up state, the output of $\mathcal{N}$ and the output of $\operatorname{Id}$ will be identical (we assume she produces only negligible radiation). Should we then conclude that the Stern-Gerlach channel is identical to the identity channel? Of course, the answer is that we have merely chosen a poor state for discriminating between these two channels. A natural quantity, therefore, for the {intrinsic distinguishability} of channels, is to consider the fidelity when the \textit{most discriminating} state is considered. In the case where $\mathcal{M}$ is the identity channel and $\mathcal{N}$ models Alice's $z$-aligned Stern-Gerlach experiment, the most discriminating state would be a state of definite $x$-spin of Alice's particle. In general, the formula becomes \cite{Raginsky_2001} 
\begin{equation}
\inf_{\omega}f_{\mathrm{U}}(\mathcal{N}(\omega),\mathcal{M}(\omega)).
\end{equation}
This, however, has an undesirable property, peculiar to quantum entanglement: this quantity is unstable under the enlargement of the Hilbert space, even when the channels act trivially on the additional Hilbert space factors. That is to say, for some auxiliary Hilbert space dimension $k$,
\begin{equation}
    \inf_{\omega}f_{\mathrm{U}}(\operatorname{Id}_k\otimes\mathcal{N}(\omega),\operatorname{Id}_k\otimes\mathcal{M}(\omega)) \le \inf_{\omega}f_{\mathrm{U}}(\mathcal{N}(\omega),\mathcal{M}(\omega)).
\end{equation}
This might appear to be a rather formal property to concern ourselves with, but it has practical implications for Alice. Suppose, for example, that Alice does a more complicated experiment, in which she prepares her spin in a bell pair with another qubit in her laboratory. As before, suppose that $\mathcal{M}$ is the identity channel, so we are comparing the effect of $\mathcal{N}$ to the identity, where $\mathcal{N}$ acts trivially on the other qubit. If $\mathcal{N}$ destroys the entanglement of Alice's bell pair, we should consider this channel to be \textit{farther} from the identity channel than one that preserves the entanglement. A better measure of the similarity of channels, therefore, would take this into account. A simple remedy presents itself, which is to consider the auxiliary Hilbert space (the other qubit in Alice's lab) that is most sensitive to such effects. By extremizing over this auxiliary Hilbert space dimension,\footnote{The Hilbert space dimension will never need to be larger than the Hilbert space dimension of the domain of the original channels, e.g., Alice's original spin Hilbert space \cite{Puzzuoli_2016}.} we arrive at the definition of the channel fidelity \cite{ksw_channelFidelity_2007},
\begin{align}
    F(\mathcal N,\mathcal M):=\inf_k\inf_\omega f_\text{U}\left((\mathrm{Id}_k\otimes\mathcal N)(\omega),(\mathrm{Id}_k\otimes\mathcal M)(\omega)\right),
    \label{eq:entFidelity}
\end{align}
which quantifies the similarity of two quantum channels.

Using this definition, the exact error correction condition of Eq. (\ref{eq:noDecoherence}) can be relaxed into an approximate error correction condition. Given a recovery channel $\mathcal{R}$, this approximate recoverability condition is
\begin{equation}
    F(\mathcal{R}\circ\mathcal{N},\operatorname{Id}) \simeq 1.
\end{equation}
Conversely, however, if $F(\mathcal{R}\circ\mathcal{N},\operatorname{Id}) \ll 1$, we cannot conclude that $\mathcal{N}$ is a decohering channel. It may well be that $\mathcal{R}$ is simply a bad choice of recovery channel. The intrinsic decoherence of a channel should depend only on whether it is recoverable \textit{in principle}, so we should take the supremum over recovery channels:
\begin{equation}
    \sup_\mathcal{R}F(\mathcal{R}\circ\mathcal{N},\operatorname{Id})
\end{equation} characterizes the supremum of the recoverable coherence of a quantum channel $\mathcal{N}$. This immediately implies a definition of the \textit{irrecoverable decoherence} of a channel:
\definition[Channel Decoherence]
    {
        
        \begin{equation}\label{eq:channelDecoherence}
            D(\mathcal{N}) :=
            1-\frac{1}{1-1/d}\left[\sup_{\mathcal{R}}F(\mathcal{R}\circ\mathcal{N},\operatorname{Id})-1/d\right]_+.
        \end{equation}
        
    }For $d>1$, the dependence on the Hilbert space dimension $d$ of the domain of $\mathcal{N}$ is chosen so that \begin{equation}
        0\le D(\mathcal{N}) \le 1,    \end{equation} where $D(\mathcal{N})=0$ indicates no decoherence and $D(\mathcal{N})=1$ indicates saturation of the decoherence measure.\footnote{For an infinite-dimensional input Hilbert space, $1/d$ is understood as zero.}
Here $[x]_+:=\max\{x,0\}$.
\footnote{Alternatively, one may omit the positive-part truncation and define
$\widetilde D(\mathcal N):=1-\big(\sup_{\mathcal R}F(\mathcal R\circ\mathcal N,\operatorname{Id})-1/d\big)/(1-1/d)$, together with the analogous modification to the forthcoming definition of $\widetilde{\mathcal I}$. Whereas the truncated definition treats complete dephasing as maximal decoherence, the untruncated definition continues to distinguish channels that also destroy the surviving classical information: $\widetilde D=1$ at complete dephasing, while $\widetilde D>1$ is possible for more destructive channels (with complete erasure giving $\widetilde D=1+1/d$ in finite dimensions). For the pure, equiprobable two-state case of Appendix~\ref{sec:consistency}, however, the cutoff is inactive, so $\widetilde D=D=1-|\langle\Psi_1|\Psi_2\rangle|=P_\mathrm{Bob}^\mathrm{certain}$; only the general probability-valued interpretation is lost once $\widetilde D>1$. The exact information-decoherence relation $\widetilde D(\mathcal N)=\widetilde{\mathcal I}(\mathcal N^c)$ remains intact.}
Intuitively, the channel decoherence (\ref{eq:channelDecoherence}) can be understood as the normalized loss of the optimal worst-case fidelity between an original state $\omega$ and what remains after acting with $\mathcal{N}$ and optimizing over recovery channels $\mathcal{R}$. As promised, it is a measure of irrecoverable decoherence inflicted by $\mathcal{N}$. 
\footnote{
In the context of the complexity theory of quantum fields, Hollands and Ranallo \cite{Hollands_2023} have introduced a closely related quantity as a measure of the \textit{complexity} of a quantum channel. The exploration of the possible relationships between channel decoherence and channel complexity would seem to be worthy of further investigation, especially in light of the discussion of Sec.~\ref{sec:linearGrowth}.
}

\subsubsection{Definition of channel loss and channel information}

Complementarity dictates that when Bob can distinguish between the states on his side of the horizon, i.e., he has ``which-state information,'' then Alice's superposition must be decohered. To formalize this notion, we must formalize the notion of the information available to Bob. As a first step, we formalize the \textit{loss} of information by a channel transmitting information from Alice to Bob.

As explained in Sec.~\ref{sec:simpleIDT}, performing a Stern-Gerlach experiment in the vicinity of a horizon can provide a significant amount of which-path information to Bob. As shown in \cite{DSW_2023}, however, the strength of the  decoherence diminishes if Alice produces a smaller superposition---i.e., she reduces the spatial separation of the branches of her particle's spatial wave function. Then the resulting long-range fields in Bob's region are more difficult to distinguish and the amount of which-path information available to Bob is reduced. 

The channel responsible for conveying this information from Alice to Bob is the complementary channel $(\mathcal{N_\mathrm{Alice}^{\mathscr{H}^+}})^c:{\mathcal S(\mathcal{H}_\mathrm{Alice})\to\mathcal S(\mathcal{H}_{\mathscr{H}^+})}$. If Alice produces an extremely small superposition, that channel will be very lossy: Bob will get very little information about Alice's superposition, because the output state on the horizon will depend only very weakly on Alice's superposition.

Generalizing from this example, we say that a channel $\mathcal{N}_\text{lossy}$ is lossy if it always outputs nearly the same, fixed state $\chi_0$, in which case\begin{equation}
\inf_{\omega}f_\mathrm{U} ( \mathcal{N}_\text{lossy}(\omega), \chi_0) 
\simeq 1.
\end{equation}
It is possible, however, to choose $\chi_0$ so that it is distinguishable from all possible outputs of $\mathcal{N}_\text{lossy}$ (perhaps, for example, it lies outside the range of $\mathcal{N}_\text{lossy}$). Then a small fidelity would say nothing about the lossiness of $\mathcal{N}_\text{lossy}$ and rather more about the choice of $\chi_0$. When defining the ``channel loss,'' therefore, we should consider the fixed state $\chi$ that is most similar to the output of the channel, while still carrying no information about the actual, particular state that was input to $\mathcal{N}$:

\begin{align}\label{eq:channelLoss}
    \mathscr{L}(\mathcal{N}):=\sup_{\chi}F(\mathcal{N},\mathcal{C}_{\chi}),
    \qquad \mathcal{C}_{\chi}(\omega):=\Tr(\omega)\chi .
\end{align}

Intuitively, the channel loss $\mathscr {L}(\mathcal{N})$ can be understood as a measure of whether, after acting $\mathcal{N}$ on $\omega$, the result is similar to some state that carries no information about $\omega$ whatsoever.

If a channel is nearly lossless, it preserves a great deal of information about the input state. We define the information of a quantum channel $\mathcal N$ as follows:
\definition[Channel Information]{\label{def:chInfo}
    
    \begin{equation}
        \mathcal I(\mathcal N) :=
        1- \frac{1}{1-1/d}\left[\mathscr L(\mathcal N)-1/d\right]_+,
    \end{equation}
    
}
Definition~\ref{def:chInfo} measures the information \textit{preserved} by $\mathcal{N}$. 
As with the channel decoherence, 
\begin{equation}
        0\le \mathcal{I}(\mathcal{N}) \le 1
    \end{equation} where $\mathcal{I}(\mathcal{N})=1$ indicates saturation of the information measure and $\mathcal{I}(\mathcal{N})=0$ indicates a maximally lossy channel.
    We have already seen that, when Bob is presented with two equally probable states to discriminate between, the unambiguous state discrimination probability is controlled by the square root of the fidelity by Eq.(\ref{eq:uhlmannIntermediateTimes}). As discussed in Sec.~\ref{sec:simpleIDT}, an {unequal-prior} bound on unambiguous state discrimination can be given in terms of the fidelity itself. Therefore, Eq. (\ref{eq:channelLoss}) can be understood as giving an operational bound on Bob's ability to perform unambiguous state discrimination.

\subsubsection{ Information-decoherence trade-off}
\label{sec:exactChannelIDT}
As anticipated, a tight relation holds between the amount of decoherence $D(\mathcal{N})$ inflicted by a quantum channel and the information shunted by the \textit{complementary} channel $\mathcal{N}^c$ into the environment, $\mathcal{I}(\mathcal{N}^c)$. In fact, in Appendix~\ref{app:IDTproof} we leverage an information-disturbance trade-off relation proved by B\'eny and Oreshkov \cite{Beny:2010byk} to prove these two operationally distinct quantities are precisely equal. In what might be called an information-decoherence trade-off, we prove the following equality:
\begin{equation}\label{eq:exactChannelIDT}
    \mathcal{I}(\mathcal{N}^c) = D(\mathcal{N}).
\end{equation}

This equality holds for arbitrary finite-dimensional quantum channels, and for infinite-dimensional channels whenever the preceding fidelities and complementary-channel theorem apply,\footnote{Although B\'eny and Oreshkov work in finite dimensions, an infinite-dimensional extension follows from adapting their Stinespring-Uhlmann argument to normal channels, replacing $\min$ and $\max$ by $\inf$ and $\sup$, and using Sion's theorem \cite{Sion_1958} to justify their minimax interchange. For regional algebras, fidelity must be formulated algebraically, and a ``complementary channel'' produces a state on the commutant algebra of the Stinespring dilation of the output domain.} but we will concern ourselves with its application to the original gedanken experiment of Sec.~\ref{sec:simpleIDT}. Recall that Alice used the spin degree of freedom of her particle to control her superposition, using a {Stern-Gerlach} apparatus, so that the final coherence of the paths of her particle is determined by its coherence in the spin basis. We rescale the affine generators of the horizon so that the cut $\mathscr{C}$ is labeled by a single affine time coordinate $V_c$. This is, in turn, related to the Killing time $t$ in Alice's laboratory  by $V_c = e^{\kappa t}$, where $\kappa$ is the surface gravity of the horizon. The channel $\mathcal{N}^{\mathscr{H}^+}_\text{Alice}(t)$ evolves the initial state of Alice's spin
into the state of the spin as it evolves past region $\mathscr{A}_{I}$, accounting for any decoherence caused by the horizon. The complementary channel ${(\mathcal N_\text{Alice}^{\mathscr{H}^+})^c(t)}$ maps that same initial spin onto a mixture of two distinct states of coherent radiation on the horizon, which are responsible for Alice's decoherence.

How does the channel decoherence inflicted by the horizon relate to the earlier notion of decoherence $D_\mathrm{Alice}|_c$? At finite times, the channel decoherence carries a distinct, but closely related, operational meaning. Whereas $D_\mathrm{Alice}|_c$ computed the fidelity between optimal purifications of Alice's state given her final {state} at $V=V_c$, the channel decoherence $D(\mathcal{N}^{\mathscr{H}^+}_\text{Alice}(t))$ takes an infimum over all states and is therefore an intrinsic decoherence due to the horizon itself, as required by D.i.

At late times, the two definitions become precisely equal. Appendix \ref{sec:consistency} computes the exact value of $D(\mathcal{N}_\text{Alice}^\mathscr{H^+})$ in the infinite-time limit, as $V_c\to\infty$ (correspondingly,  no time constraints are placed on Bob).
In that case, optimal recovery merely needs to rotate away the  phases of the off-diagonal components of Alice's reduced density matrix, and we find
\begin{equation} \label{eq:lateTimeDecoherenceConsistency}D(\mathcal{N}_\text{Alice}^{{\mathscr{H}^+}}) = D_\mathrm{Alice}\quad \text{for }t\to \infty.
\end{equation}
This, of course, also implies that 
\begin{equation}
\mathcal{I}((\mathcal{N}_\text{Alice}^{{\mathscr{H}^+}})^c) = P_\mathrm{Bob}^\mathrm{certain}\quad\text{for }t\to\infty,
\end{equation}
so we have satisfied our requirements I.iii and D.iii. Looking back on Sec.~\ref{sec:simpleIDT} with this more general perspective, we see that the complementary roles of the decoherence and distinguishability measures in (\ref{eq:twoStataeMixedStateBound}) and (\ref{eq:twoStatePureStateEquivalence}) are consequences of the exact information-decoherence trade-off of Eq. (\ref{eq:exactChannelIDT}).
Operationally, and now in full generality, the complementary channel furnishes a channel-level bound on unambiguous state discrimination tasks involving quantum states in the exterior.

At finite times, the optimal recovery channel is more difficult to compute, but is undoubtedly still closely related to the optimal protocol established in \cite{DKSW_2025}, and therefore the Uhlmann fidelity between the two  states sourced by Alice's superposition. Despite the difficulty involved in evaluating the channel decoherence at finite times, Appendix~{\ref{sec:approximateCapacity}} derives a bound between them:
\begin{equation}\begin{split}\label{eq:decoherenceBound}
D(\mathcal{N}^{\mathscr{H}^+}_\text{Alice}(t)) \ge {D_\mathrm{Alice}|_c}.
\end{split}\end{equation} {This bound is saturated at late times.}

The channel decoherence is an intrinsic definition of the decoherence due to a quantum channel, satisfying all of our requirements D.i--D.iii. It is, by definition, a measure of the unrecoverable decoherence inflicted by a quantum channel. Furthermore, it is precisely equal to the channel information of the complementary channel, which itself satisfies all of our requirements I.i--I.iii and provides an operational bound on one-shot unambiguous state discrimination.

Despite its operational clarity and suitability in satisfying an exact information-decoherence trade-off (IDT), however, the channel decoherence and channel information are difficult to compute because of the extremization required. {Computing the channel decoherence at finite times is likewise technically challenging; Eq. (\ref{eq:decoherenceBound}) gives the bound above.} The simple Uhlmann-fidelity decoherence $D_\text{Alice}|_c$ is, by contrast, much easier to compute in general than the rather difficult $D(\mathcal{N})$. Since we are interested in practical computations, later sections will be concerned with developing a framework for computing practical distinguishability measures, which can be used to bound the exact decoherence $D(\mathcal{N})$ and information $\mathcal{I}(\mathcal{N}^c)$. These distinguishability measures, such as the relative entropy and R\'enyi entropies, will obey similar bounds to the channel decoherence and channel information, albeit sometimes with a less clear or relevant operational meaning and without the exact IDT of {Eq.~(\ref{eq:exactChannelIDT})}. The computation of such distinguishability measures is, of course, also of independent interest to a broader audience of those interested in the quantum information theory of fields.

\subsection{Monotone decoherence: A hallmark of horizons}
\label{sec:secondLawforHorizons}
The original gedanken experiment suggests that a horizon, as viewed from the outside, may be viewed as though it is made of highly entangling degrees of freedom. More abstractly, the apparent equivalence between ``spacetime horizons'' and ``maximally entangling degrees of freedom'' raises the possibility that the presence of a horizon may be characterized in purely operational and quantum information-theoretic terms. Having developed a more general set of tools, we are now equipped to formulate such a characterization.

We have thus far focused on the effects of the horizon on a given subsystem of the exterior, namely Alice's superposition. For a description of the effects of a horizon that is completely intrinsic to it, however, it is useful to note that evolution in the presence of the horizon can be understood as a channel acting on \textit{any} subsystem of the exterior (or the entire exterior), including Alice and her experiment.

Let $t$ label the Killing time coordinate corresponding to time in Alice's laboratory. As the horizon cut $V_c = e^{\kappa t}$ advances in time along the horizon, Bob is able to measure a {strictly larger} region of spacetime on his side of the horizon. Let us now take 
\begin{equation}
\mathcal{N}^{\mathscr{H}^+}_\text{ext.}(t):{\mathcal S(\mathcal U_\text{ext.}(0))}{\to \mathcal S(\mathfrak A_\text{ext.}(V_c))},
\end{equation} {where $\mathcal S(\mathcal U_\text{ext.}(0))$ denotes the state space of the full exterior (in which case $\mathcal{U}_\text{ext.}=\mathfrak{A}_\text{ext.}$) at $t=0$ or of any fixed code subspace thereof.} The channel maps the initial state of the exterior to the state at a later time under the influence of the black hole---namely, the state exterior to a horizon cut $V=V_c$. We label the channels by $t$ because, as $V_c$ advances, the flow gives a one-parameter family of channels 
\begin{equation}\label{eq:decoherenceIncreases}
(\mathcal{N}^{\mathscr{H}^+}_\text{ext.})^c(t):{\mathcal S(\mathcal U_\text{ext.}(0))}{\to\mathcal S(\mathfrak A_{\mathscr{H}^+|_c}(V_c))}.
\end{equation}
As Alice's Killing time $t$ advances outside the horizon, $(\mathcal{N}^{\mathscr{H}^+}_\text{ext.})^c(t)$ targets states on one-parameter family of horizon algebra inclusions. This corresponds to a restriction of the algebra of observables in the exterior, where the flow corresponds to Killing time. Heuristically, this corresponds to the fact that, as time progresses, an exterior observer must ``trace out'' more degrees of freedom as they fall into the horizon. The composition of an algebra restriction with a recovery channel is itself a valid recovery channel, so the supremum over recovery channels in Definition (\ref{eq:channelDecoherence}) guarantees that the decoherence of the exterior is nondecreasing\footnote{The decoherence may not be $C^1$, but because it is nondecreasing, it is almost-everywhere differentiable \cite{Lebesgue_1904}.}:
\begin{equation}\label{eq:optimalDecoherence}
    \frac{d}{dt} D(\mathcal{N}^{\mathscr{H}^+}_\text{ext.}(t)) \ge 0.
\end{equation}
This is a succinct statement of the extraordinary efficiency of horizons, including black hole and de Sitter horizons, at decohering their environments. Semiclassically---i.e., the limit as $G_{\textrm{N}}\to0$---this is an exact equation. In existing models of quantum gravitational effects, however \cite{ Almheiri_2019, Penington_2019}, it seems likely that the right-hand side of this equation will receive corrections suppressed in powers of $G_{\textrm{N}}$. The degree to which this equation holds at finite $G_{\textrm{N}}$, then, may prove useful in signaling the emergence of an effective horizon.

It is notable that Eq. (\ref{eq:optimalDecoherence}) provides a {consequence} of a semiclassical horizon entirely in terms of quantum information-theoretic quantities. It is natural to wonder whether these conditions might be useful in giving a purely operational definition of an emergent ``horizon'' in the collective dynamics of even nongravitational bodies.

Whenever a quantum superposition is produced outside an ordinary material body, the long-range gravitational and electromagnetic fields sourced by the superposition become entangled to some degree with the internal degrees of freedom of the material body. However, most material bodies are incredibly inefficient in absorbing long-wavelength gravitational radiation. This renders it generally possible to recover the coherence of the superposition, by adiabatically recombining the branches of the superposition. Biggs and Maldacena \cite{Biggs_2024} have, for example, investigated the extraordinary material properties that would be required for a material body to mimic the decoherence outside a black hole. The extraordinary efficiency of a black hole in decohering its environment is captured by Eq. (\ref{eq:optimalDecoherence}). The information-decoherence trade-off of Eq. (\ref{eq:exactChannelIDT}) reveals that this decoherence can also be understood as a statement of the fact that the distinguishability of the complementary, entangled states on the opposite side of the horizon is also nondecreasing:
\begin{equation}\label{eq:informationIncreases}\frac{d}{dt}\mathcal{I}((\mathcal{N}_\text{ext.}^{{\mathscr{H}^+}})^c(t))\ge 0.\end{equation}

The channel information bounds a one-shot state discrimination probability for an optimal observer in Bob's region. Due to the inefficiency of most material bodies in absorbing long-wavelength gravitational radiation, the degrees of freedom of most material bodies will go into superposition while Alice's superposition is held open, but will adiabatically recombine and, for the most part, follow the Newtonian field and disentangle themselves from Alice's spin as she adiabatically closes her superposition. In that case one expects the state-discrimination information available to Bob at intermediate times to be much larger than at late times. Equation (\ref{eq:informationIncreases}), along with (\ref{eq:decoherenceIncreases}) and (\ref{eq:exactChannelIDT}), can be understood to show that a black hole decoheres its environment as though it were made of \textit{maximally entangling} degrees of freedom.

\subsection{Linear growth of distinguishability long after equilibration}
\label{sec:linearGrowth}
The preceding discussion holds at a high level of generality, for arbitrary states outside black holes. By specializing to a particular case, we gain more explicit quantitative insight into the rate at which black holes decohere their environments---and their interiors harvest information. In the specific case of Alice's superposition experiment outside a black hole, we may focus on $D(\mathcal{N}^{\mathscr{H^+}}_\mathrm{Alice})$ (rather than the entire exterior). Then, under the protocols of the gedanken experiment in which Alice attempts to \textit{minimize} her decoherence, we recover $D(\mathcal{N}^{\mathscr{H^+}}_\mathrm{Alice})=D_\mathrm{Alice}$ {at late times}, and $-\ln(1-D_\mathrm{Alice})$ grows linearly with the time $T$ over which she maintains a superposition \cite{DSW_2022, DKSW_2025}.

Significant recent interest has focused on the interpretation of various quantities (e.g., ``complexity,'' ``volume,'' and ``action'') relevant to black hole physics that exhibit linear growth long after the equilibration time of the black hole \cite{Stanford_2014, Brown_2016, Brown_2018, Belin_2022, Belin_2023}. We point out that, in the context of the initial   gedanken experiment involving Alice's superposition, Alice decoheres due to a linearly growing number of soft photons radiated through the horizon \cite{DKSW_2025}. As a result of the preceding discussion, the distinguishability measure \begin{equation}
-\ln\left(1-\mathcal{I}\left((\mathcal{N}_\text{Alice}^{\mathscr{H}^+})^c\right)\right)\propto\kappa T
\end{equation}
therefore exhibits secular, linear growth, where $T$ denotes proper time over which Alice maintains her superposition in the black hole exterior and $\kappa$ is the surface gravity. Here, $(\mathcal{N}_\text{Alice}^{\mathscr{H}^+})^c$ denotes the channel that carries out the joint evolution of Alice and the entire horizon before restricting to the interior. When Alice creates and recombines her superposition sufficiently slowly, an arbitrarily small amount of hard radiation falls into the horizon. The black hole rapidly equilibrates, as the field on the horizon settles down to a new stationary state, but with a displaced value of the field on the horizon. The {logarithmic} distinguishability of interior states, however, continues to grow without bound at least until timescales comparable to the evaporation timescale, due to a linearly growing number of soft gravitons/photons produced due to the superposition in the exterior. From the exterior perspective, it is these soft, entangling gravitons/photons which are responsible for the corresponding decoherence.

From an interior point of view, it is quite natural that the distinguishability of states should increase long after the horizon equilibrates, because the size of the interior region (both in terms of volume and proper time) available for Bob to perform his measurement grows in direct analogy to Fig. \ref{fig:heuristic_fig}. 

We believe the fact that the decoherence of exterior states, and hence the distinguishability of interior states [by Eq.~(\ref{eq:exactChannelIDT})], increases long after a black hole settles down may be of relevance to the understanding of interior dynamics of black holes in quantum gravity, which continues to be an area of active inquiry \cite{Stanford_2014, Brown_2016, Brown_2018, Belin_2022, Belin_2023, H_2019}. Naturally, it is possible---likely, even---that such qualitative statements may receive corrections when working nonperturbatively in $G$, but the semiclassical description applied here should nevertheless describe the most salient, emergent features available to an {observer} with bounded resources. It is anticipated that the nonperturbative description of a black hole in quantum gravity may, as viewed from the ``exterior,'' even more closely mirror the properties of a material body---albeit one with possibly extraordinary material properties \cite{Biggs_2024}. In that case, the ``degrees of freedom'' of the black hole are expected to become identified with material degrees of freedom. The analogous ``material description'' for, e.g., a Rindler horizon or de Sitter horizon at finite $G_\textrm{N}$, however, would seem to be even more difficult to construct.

\section{IR-Complete Horizon Algebras}\label{sec:horizonAlgebra}
While the previous section focused primarily on the operational interpretation of distinguishability measures and the exact trade-off between information and decoherence, this section and the following are concerned with the practical computation of distinguishability measures of the horizon radiation. For concreteness, we will restrict attention to the gedanken experiment considered in Sec.~\ref{sec:IDT}, in which Alice creates a superposition of coherent states on the horizon. 

In the phenomenon of decoherence due to horizons, it is the accumulation of soft radiation on the black hole horizon that is responsible for the loss of coherence of external quantum matter \cite{DSW_2021,DSW_2022}. Indeed, in the limit in which Alice never recombines her superposition, the radiation field does not decay at asymptotically late times. Consequently, an infrared-divergent number of soft gravitons or photons fall through on the horizon, resulting in complete decoherence. This infrared divergence is symptomatic of the fact that a proper treatment of soft quanta requires an enlargement of the theory on the horizon. In our gedanken experiment, however, Alice recombines her superposition at a finite time, and only a finite expected number of soft quanta are produced. Remarkably, the proper treatment of soft quanta will prove useful even in finite time computations that would otherwise be difficult, for reasons we will now explain.

 \begin{figure}[]
         \centering
         \includegraphics[width=\columnwidth]{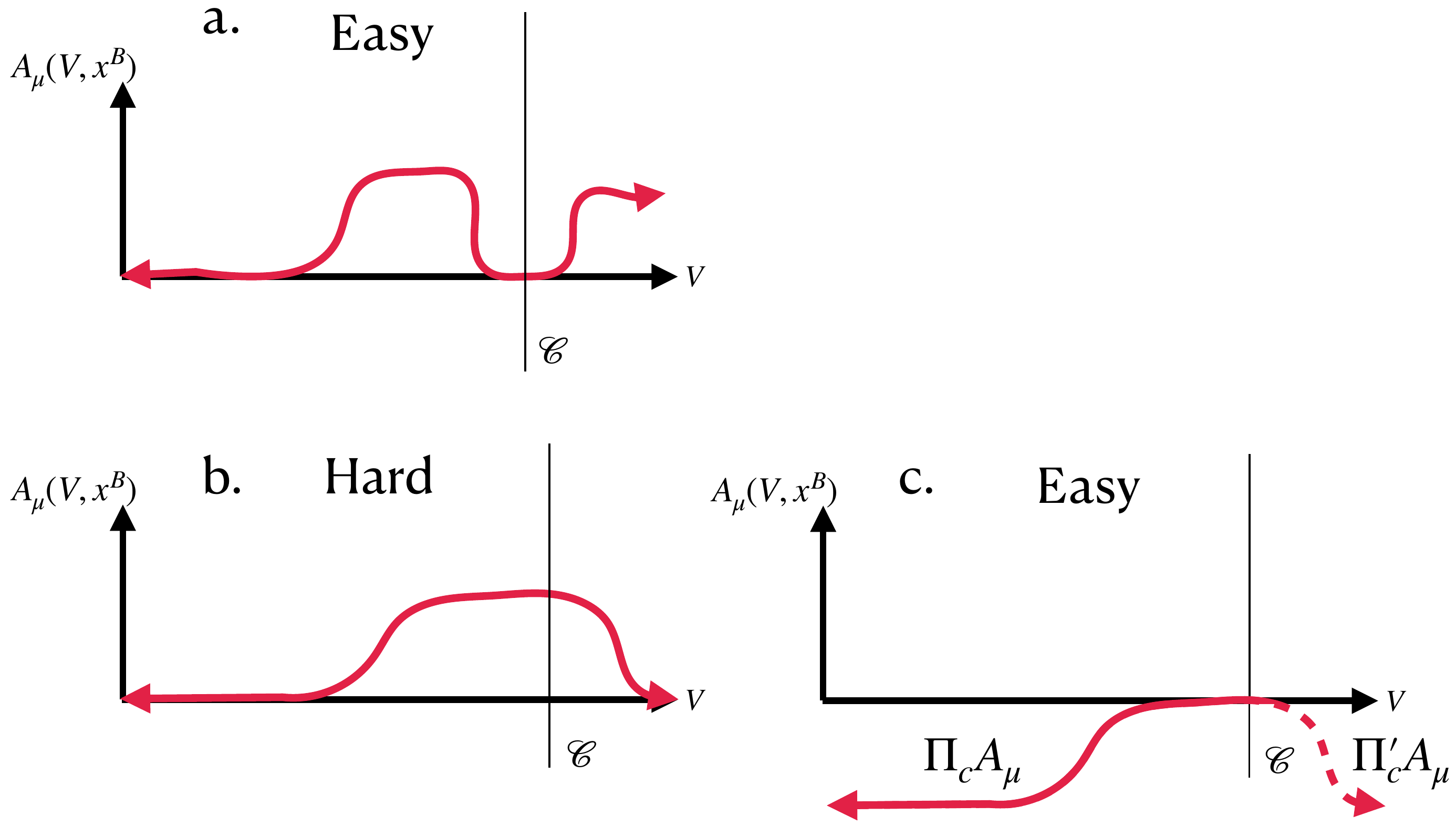}
         \caption{A spatial component $A_\mu$, tangent to the horizon, of the electromagnetic potential is plotted as a function of affine time $V$ parametrizing a horizon generator at angle $x^B$. (a) A field configuration in the ``easy case'' of Casini \textit{et al.} (b) A field configuration in the ``hard case.'' (c) The result of applying a $V$-independent shift to (b), resulting in a field configuration in the easy case. In particular, the final field configuration can be written as a linear combination of two field configurations: one $\Pi_c A$ supported only to the past of the horizon cut $\mathscr{C}$, and another $\Pi'_c A$ supported only to the future. However, neither of the field configurations $\Pi_c A$ and $\Pi'_c A$ decay asymptotically in time---they exhibit electromagnetic memory \cite{Bieri_2013}, and thus do not lie in the standard Hilbert space of scattering theory due to their soft radiation content.}
         \label{fig:EasyHardCases}
     \end{figure}

 In this section, for simplicity, we will work with the  electromagnetic field on the horizon rather than the graviton field. Directly analogous considerations can be used to quantize the graviton field on the horizon, such as is done in \cite{Kudler-Flam_2025}. We begin this section by giving a brief summary of the arguments we will use, to give an intuitive understanding of the methods to follow.
 
 On the horizon, the distinguishability to the past of a cut $\mathscr{C}$ is a property of the corresponding algebra  $\mathfrak{A}_c$ to the past of that cut. As we will explain, we can get a simple description of the decoherence---as well as any other distinguishability measure that we consider---if the operator $\op{U}(A)$ that generates Alice's radiation is an element of $\mathfrak{A}_c$. This has been dubbed the easy case by Casini \textit{et al.} \cite{Casini_2019}, and such a radiation field is depicted in Fig. \ref{fig:EasyHardCases}(a) However, at finite times during Alice's experiment, $\op{U}(A)$ is {\em not} an element of $\mathfrak{A}_c$ or its commutant, because the radiation ``straddles'' the cut $\mathscr{C}$ and the operator cannot be split into separate operators in the respective algebras. Casini \textit{et al.} dub this the hard case, which is depicted in Fig. \ref{fig:EasyHardCases}(b).

 However, the radiation field on the horizon can be shifted by a smooth $V$-independent one-form, whose exact part is a familiar large-gauge {transformation} \cite{Pasterski_2015,Strominger_2}
 \begin{equation}
 \label{eq:LGT}
A_B \to A_B+h_B(x^A).
 \end{equation}
Using this shift, the radiation field $A_{A}$ can be made to vanish at the cut $\mathscr{C}$. In this way we will be able to factorize the operator generating Alice's radiation
\begin{equation}\label{eq:factorizationSketch}
U(A) = \op{U}(\Pi_{c}A)\op{U}(\Pi_{c}^{\prime}A)\op{U}(A|_c)e^{-i \Delta/2},
\end{equation}
where $\Pi_{c}A$ and $\Pi_{c}^{\prime}A$ {are} radiation fields with support entirely to the past and future of $\mathscr{C}$, respectively,  $\op{U}(A|_c)$ implements the corresponding shift. Here $\Delta$ is a $c$-number corresponding to the ``memory'' of $A$, the definition of which will be given shortly in Sec. \ref{sec:ordinary}. While this formally factorizes the operator, the radiation fields  $\Pi_{c}A$ and $\Pi_{c}^{\prime}A$ do not decay at asymptotically early and late times and contain {an} infinite number of radiative quanta. The corresponding operators are {\em not} elements of the standard von Neumann algebra of observables \cite{Casini_2019}.

The purpose of this section is to enlarge the algebra to include such radiation fields and thereby achieve the factorization. This will result in a simple description of the distinguishability measures of Alice's radiation, which will be considered in Sec.~\ref{sec:ttTheory}. 

\subsection{Ordinary null quantization on the horizon}\label{sec:ordinary}
The algebra of local, gauge-invariant electromagnetic field observables in the bulk consists of the four-potential $\op{A}_a(x)$ smeared with divergence-free test functions $j^a(x)$ whose support decays to zero asymptotically on the spacetime manifold $\mathscr M$,
\begin{equation}
    \op{A}(j) := \int_\mathscr{M} d^4x \sqrt{-g}\op{A}_a(x) j^a(x).
\end{equation}
Such observables can be rewritten in terms of observables intrinsic to the horizon ${\mathscr{H}^+}$ by using the symplectic structure of the classical theory.

Let $\phi_{1,B}(V,x^A)$  and $\phi_{2,B}(V,x^A)$ be two classical four-potential solutions of the source-free Maxwell equations, restricted to the horizon. Working in a Coulomb-like gauge where $A_V = 0$, the symplectic form for the free, radiative degrees of freedom of Maxwell theory on ${\mathscr{H}^+}$ is
\begin{equation}
\mathscr{W}(\phi_1,\phi_2)=2\int_{{\mathscr{H}^+}}dVd^2x^A {\phi_{[1}}_B(V,x^A)\partial_V\phi_{2]}^B(V,x^A).
\end{equation}
Classically, this symplectic structure is associated with a Poisson algebra of electric field observables $
    E(\phi) := \mathscr{W}(A, \phi)$
on the horizon,\footnote{When $s_B$ decays asymptotically on the horizon, $E(s)$ is nothing more than twice the electric field smeared on the horizon with the test function $s_B$. For a nondecaying test function $\phi_B$ this is no longer true, but we retain the $E(\phi)$ notation for simplicity.} smeared with horizon-restricted four-potential test functions $\phi_B$. Here, $E_B = -\partial_V A_B$. The classical Poisson relations are defined by 
\begin{equation}\label{eq:classicalAlgebra}
    \{E(\phi_1),E(\phi_2)\} = \mathscr{W}(\phi_1,\phi_2)\mathbbm{1}.
\end{equation}

The bulk, quantum observables $\mathbf{A}(j)$ can likewise be expressed in terms of an algebra on the horizon by the relation
\begin{equation}
    \op{A}(j) = \mathscr{W}(\mathcal{E} j,\op{A}),
\end{equation}
where $\mathcal{E}_{ab}$ is the advanced-minus-retarded Green's function and
\begin{equation}
    (\mathcal{E}j)_{a}(x) := \int_\mathscr{M} d^4y \sqrt{-g}\mathcal{E}_{ab}(x,y)j^b(y)
\end{equation}
is the solution corresponding to the source $j^a$. {We} use the symplectic form to rewrite the field algebra on the horizon purely in terms of the electric field $\op{E}_B = -\partial_V\op{A}_B$. To do this, we define the electric field observables,
\begin{equation}
    \op{E}(s):=2\int_{{\mathscr{H}^+}} d^2x^A dV \op{E}_B(V,x^A) s^B(V,x^A),
\end{equation}
which are related to the local algebra of the exterior via
\begin{equation}
    \op{A}(j) = \mathscr{W}(\mathcal{E}j,\op{A}) = -\op{E}(\mathcal{E}j).
\end{equation}
Given two solutions of the four-potential on the horizon $(s_1)_A$ and $(s_2)_A$ that fall off to zero asymptotically in $V$ (and therefore do not describe field configurations with memory), the commutation relations follow from the symplectic form and are
\begin{equation}\label{eq:fieldCommutator}
    [\op{E}(s_1),\op{E}(s_2)]=2i\int_{\mathscr{H}^+}d^2x^A dV {s_{[1}}_B(V,x^A)\partial_V s_{2]}^B(V, x^A)\mathbf{1}.
\end{equation}
This defines the ordinary algebra of field observables on the horizon.

We will quantize with respect to the Poincar\'e invariant vacuum. In the case of a general Killing horizon a nearly identical procedure goes through for the Killing-symmetry invariant state \cite{KW_1991}. The Poincar\'e-invariant state $\omega_\Omega$ is a Gaussian state, fully determined by its vanishing one-point function and its two-point function,
\begin{equation}\begin{split}\label{eq:particleVacuumState}
        &\omega_\Omega(\op{E}(s_1)^*\op{E}(s_2)) \\&= \frac{-1}{\pi}\int_{\mathscr{H}^+}dV_1 dV_2d^2x^A\frac{\bar s_{1,B}(V_1,x^A)s_2^B(V_2,x^A)}{\left(V_1-V_2-i0^+\right)^2}.
\end{split}\end{equation}
By specifying the quasifree state $\omega_\Omega$ via the two-point function (\ref{eq:particleVacuumState}), we can also obtain a
$\ast$-representation of the horizon field algebra through the Gelfand--Naimark--Segal (GNS) construction.\footnote{In Sec. \ref{sec:IDT} we used $\mathscr{A}$ with roman numeral subscripts to denote regions of the spacetime accessible to Alice. From here onward we repurpose $\mathscr{A}$ (possibly with some decorations) to denote a $*$-algebra and hope the reader will forgive this reuse of notation.}
Let $\overline{\mathscr{A}}$  denote the $\ast$-algebra generated by the smeared electric-field operators $\op{E}(s)$
with $s_B(V,x^A)$ smooth and decaying as $V\to\pm\infty$, with commutator (\ref{eq:fieldCommutator}). Let $B(\mathcal{F}_0)$ be the set of bounded operators on $\mathcal{F}_0$.
Then there exists a cyclic vector $|\Omega\rangle$ and a representation $\pi_\Omega:\overline{\mathscr{A}}\to{B}(\mathcal{F}_0)$
such that
\begin{equation}
\omega_\Omega(\hat{a})=\langle\Omega|\pi_\Omega(\hat{a})|\Omega\rangle
\end{equation}
for all $\hat{a}\in\overline{\mathscr{A}}$.
$\mathcal{F}_0$ is nothing more than the usual bosonic Fock space over the associated
one-particle Hilbert space $\mathcal{H}_1$:
\begin{equation}
\mathcal{F}_0
 := \mathbb{C}\oplus \bigoplus_{n=1}^{\infty} \mathcal{H}_{n} .
\end{equation}

Here $\mathcal{H}_{n}$ is the $n$-fold symmetrized tensor product of the $one$-particle Hilbert space $\mathcal{H}_{1}$. The subscript ``$0$'' in $\mathcal{F}_{0}$ denotes the fact that {\em all} states in $\mathcal{F}_{0}$ must decay at early and late affine times \cite{ashtekar1987asymptotic}. As mentioned in the Introduction, the relevant infrared property is captured by the ``horizon memory''\footnote{By Maxwell's equations, the exact part of the memory $\Delta_{A}(x^{A})$ can be related {to} the difference in ``large-gauge charges'' associated to the radial component $E_{r}$ of the electric field between early and late times (see., e.g. \cite{He_2014, Pasterski_2015}). The charges generate the exact shifts $h_B=\mathcal D_B\lambda$. These large-gauge charges can be defined on any Killing horizon \cite{Iyer_1994} as well as at null infinity \cite{Kapec_2017}. } of $\phi_B$,
\begin{equation}\label{eq:memoryOfPhi}
(\Delta_\phi)_{B}(x^{A}) := \lim_{V\to+\infty}\phi_B(V,x^A)-\lim_{V\to-\infty}\phi_B(V,x^A) \neq 0,
\end{equation}
and the decay of states simply is the statement that all states in $\mathcal{F}_{0}$ have $\Delta_{A}=0$. The fact that the standard Fock space comprises only zero-memory configurations can be seen by considering the Fourier transform of the inner product \eqref{eq:particleVacuumState}. This yields 
\begin{equation}\label{eq:one-particle-inner-product}
\omega_\Omega\!\big(\op{E}(s_1)^\ast \op{E}(s_2)\big)
 = 
\int_{\mathbb{R}^2} d^2x_A\int_{0}^{\infty}\frac{d\omega}{\pi} \omega 
\overline{\widetilde{s}_{1,B}}(\omega,x^A) \widetilde{s}_2^{ B}(\omega,x^A),
\end{equation}
where $\widetilde{s}_{A}$ is the Fourier transform with respect to affine time and $\overline{\widetilde{s}_{A}}$ denotes complex conjugation. From \eqref{eq:one-particle-inner-product} it follows that any one-particle state with memory has infinite norm in the standard Fock space. Indeed, the low frequency behavior of any solution with memory contains a soft pole,
\begin{equation}
\widetilde{\phi}_B(\omega,x^A) = \frac{(\Delta_\phi)_B(x^A)}{i\omega} + O(1)\qquad (\omega\to 0).
\end{equation}
The corresponding norm of the vector $\op{E}(s)\ket{\Omega}=\ket{s}$ for any $s_{A}$ with memory logarithmically diverges at low frequencies. Thus, $\ket{s}$ is not a vector in $\mathcal{H}_{1}$. Similarly, the Weyl operator 
\begin{equation}
\label{eq:Weyl1}
\op{W}(s) := e^{-i\op{E}(s)}
\end{equation}
satisfying 
\begin{equation}
\label{eq:Weyl2}
    \op{W}(s_1)\op{W}(s_2)^* = \op{W}(s_1-s_2) e^{i\mathscr{W}(s_1,s_2)/2}
\end{equation}
is a well-defined operator on $\mathcal{F}_{0}$ if {and} only if $s_B$ has zero memory. We can construct a von Neumann algebra of the bounded operator $\op{W}(s)$ on $\mathcal{F}_{0}$ by taking the weak closure of such operators with respect to $\mathcal{F}_{0}$. We denote the corresponding ``zero memory'' von Neumann algebra as $\overline{\mathfrak{A}}$.

\subsection{Infrared extension of the Weyl algebra}\label{sec:irExtension}
As we have argued, the key insight required to solve the hard case of Fig. \ref{fig:EasyHardCases} is that, in a gauge theory\footnote{Similar considerations apply to scalar fields where the relevant transformation is not an asymptotic symmetry but is still a well-defined map on initial data.} such as QED or general relativity, a coherent state that fails to factorize across the cut is related to one that \textit{does} factorize by a $V$-independent shift, whose exact part is a large-gauge transformation, as shown in Fig.~\ref{fig:EasyHardCases}. To realize this, we will include the corresponding memory observables and memory-changing Weyl elements into the horizon algebra.

We seek to extend the algebra of the previous subsection to include objects charged under the memory observables. We face a technical difficulty in doing this, however, because a simple argument reveals that it is impossible to represent these operators on the ordinary Fock space. The argument is as follows.

The classical memory of a field configuration $\phi_B$ can be obtained by computing its Poisson bracket with so-called ``memory observables,'' which are in fact already present in our classical Poisson algebra: it consists of nothing more than the special case of electric field observables smeared with a $V$-independent one-form $h_B(x^A)$. As an observable on phase space,
 \begin{equation}
 \Delta(h) := -\mathscr{W}(A,h)/2,
 \end{equation}
and its Poisson relations are

\begin{align}\label{eq:Ephimem}
        \{E(\phi),\Delta(h)\}&=\frac{1}{2}\int_{\mathscr{H}^+} d^2x^A dV h_B(x^A)\partial_V \phi^B(x^A,V)  \notag\\&=\frac{1}{2}\Delta_\phi(h)\mathbbm{1}.
\end{align}

Similar to Eq. \eqref{eq:memoryOfPhi}, $\Delta_\phi(h)$ denotes the \textit{value}, now smeared with $h_B$, of the memory of a classical solution $\phi_B$.
Note that $E(\phi)$ is a good classical observable even if $\phi$ does not decay to zero as $V\to \pm\infty$ on the horizon, because even in that case it has well-defined Poisson brackets with any other observable. 

However, in the quantum theory, the ordinary field algebra, whose construction we reviewed in Sec.~\ref{sec:ordinary}, excludes any test function that fails to decay to zero as $V\to \pm\infty$. Therefore, the corresponding quantum horizon memory observable $\op{\Delta}(h)$ is defined by
\begin{equation}
    [\op{E}(s),\op{\Delta}(h)] := i\{E(s),\Delta(h)\}\mathbf{1} = \frac{i}{2}\Delta_s(h)\mathbf{1}=0
\end{equation}
as well as the relation that $\op{\Delta}_B$ commutes with itself. Thus, if we extend the ordinary field algebra to $\overline{\mathscr{A}}_{\Delta}$ which, in addition to $\op{E}(s)$ also includes memory observables $\op{\Delta}(h)$ for all $h_B$, then these memory observables are
central elements of $\overline{\mathscr{A}}_{\Delta}$. 

For concreteness, we will refer to GNS Hilbert space $\mathcal{F}_{0}$ of the ordinary field algebra above the  Poincar\'e-invariant vacuum as ``the ordinary Fock space,'' though the same arguments can be applied to the Hilbert space of the Hartle-Hawking or Unruh vacua of a black hole, or the Bunch-Davies vacuum of de Sitter spacetime, or any other state defined by invariance under the symmetries of a particular Killing horizon. The GNS representation of the field algebra above the Poincar\'e vacuum is irreducible, so we learn that $\op{\Delta}(h)$ is represented as a multiple of the identity operator in any irreducible representation. In particular, on $\mathcal{F}_{0}$, the memory operator is simply the zero operator---i.e., $\op{\Delta}_B=0$ on  $\mathcal{F}_{0}$ \cite{ashtekar1987asymptotic,PSW_2022}. One can straightforwardly obtain memory Fock representations $\mathcal{F}_\Delta$ where all states have memory $\Delta_B$ (see, e.g., \cite{ashtekar1987asymptotic,PSW_2022} for details on this construction). In other words, on these representations $\op{\Delta}(h) = \Delta(h)\op{1}$ on $\mathcal{F}_\Delta$. However, in any definite memory representation, the Fock space $\mathcal{F}_{\Delta}$ cannot carry a representation of an operator  ``charged'' under the memory observables. Such operators cannot commute with the memory and therefore cannot be represented on any definite-memory Fock space.

This might seem to be an insurmountable obstacle to our desired factorization (\ref{eq:factorizationSketch}), because the unitaries involved correspond to radiation fields with memory, and are therefore memory-changing. Such operators do not preserve the zero memory character of the standard Fock space. Nevertheless, we seek a representation in which the Poincar\'e-invariant vacuum state exists as a proper state; we seek to extend the vacuum onto this broader class of operators. In fact, several authors have confronted this difficulty, and a variety of different methods have been considered to accommodate a description of electromagnetic (or gravitational) memory in Hilbert spaces. For example, a direct sum over memories might be considered \cite{Herdegen_1997, PSW_2022}. If it is desirable that an infinitesimal generator of the unitary group relating memory sectors be represented as an observable, and if the existence of a Poincar\'e-invariant vacuum state were not required, then even a direct integral over memories might be appropriate \cite{Prabhu_2024, Prabhu_2024b}. Related challenges have arisen in other contexts where unitarily inequivalent sectors of a theory coexist, such as have been addressed by other means in the context of $1+1$-dimensional conformal field theory by Hollands and Longo \cite{Hollands_2017}. In this work we take an algebraic approach, in which we include precisely the algebra elements needed to answer the physical question at hand, and then see what Hilbert space representation naturally emerges. In Appendix \ref{app:memoryHilbert} we elaborate on the different Hilbert space representations one might possibly consider.

Important to us, 
we are only concerned with Weyl unitaries that change the memory.
We now show that the inclusion of these memory-changing coherent operators (i.e., those that produce a coherent state when acting on the vacuum) proceeds without issue.

To include memory-changing operators, we need to include operators which do not commute with the memory.  Furthermore, on this enlarged algebra, we require that the vacuum be a well-defined state. Equation~\eqref{eq:Ephimem} suggests that we might try quantizing the classical {observable} $E(\phi)$ where $\phi$ has nonzero memory. Such observables do not classically commute with memory and would be a good candidate for our desired observables in the quantum theory. However, such observables are not well-defined operators on the vacuum. Indeed, by \eqref{eq:one-particle-inner-product}, the state ``$\op{E}(\phi)\ket{\Omega}$'' would have infinite norm.\footnote{One might object that ``$\op{E}(\phi)$,'' corresponding to an unbounded operator, need only be defined on a dense domain, and that perhaps this domain does not include the vacuum when $\phi$ has memory. In the full algebra, however, the memory-changing unitaries do not form a continuous group in the GNS norm ---a fact that we will revisit when constructing unitaries charged under the memory. This failure of continuity rules out the possibility of any such {infinitesimal} generator if a proper Poincar\'e-invariant vacuum state is to be retained.}  While such operators may be of interest in other circumstances \cite{flanagan_2023, Kudler-Flam_2025, Prabhu_2024b, Prabhu_2024}, this is not a candidate observable to address {these issues} that we are concerned with.

However, the Weyl algebra $\op{W}(s)$ given by~\eqref{eq:Weyl1} and~\eqref{eq:Weyl2} can be straightforwardly extended to include solutions with memory and, as we explain in the following section, these operators are well-defined on the vacuum.  
{More precisely}, we extend the Weyl algebra to include test functions $\phi_B(V,x^A)$ such that  $\partial_V\phi_B(V,x^A)$  vanishes at asymptotically late and early affine times, but whose value in fact approaches some nontrivial function on the horizon; i.e., $\lim_{V\to\pm\infty} \phi_B(V,x^A)$ is a nonzero function of $x^A$ along the asymptotic horizon cross section. The commutation relations follow immediately from the ordinary Weyl relations, so that
\begin{equation}\label{eq:extendedAlgebra}
    \op{U}(\phi_1)\op{U}(\phi_2)^* = \op{U}(\phi_1-\phi_2) e^{i\mathscr{W}(\phi_1,\phi_2)/2}.
\end{equation}

Before we can obtain the Tomita operator it is useful to further extend the algebra to accommodate the decomposition of classical solutions into ``components'' that factorize across the cut $\mathscr{C}$. This will allow us to factorize coherent operators that do \textit{not} factorize across a given cut $\mathscr{C}$ of the horizon in the original, ordinary Weyl algebra.

Now we will make use of the projectors $\Pi_c$ and $\Pi_c'$ introduced earlier in the section, beginning with their proper definition. Given a smooth function $f_B(V,x^A)$ (valued in vectors on the horizon cross section), we now define a projector (i.e., an idempotent map on test functions) $\Pi_c$ that shifts the function so that it evaluates to zero at $V_c$ and forces the function to zero thereafter. We also define a ``complementary'' projector $\Pi_c'$ whose action is reflected about the cut,
\begin{equation}\begin{split}
        \Pi_cf_B(V,x^A):=\ &\Theta(V-V_c)f_B(V_c,x^A)\\&+ \Theta(V_c-V)f_B(V,x^A)-f_B(V_c,x^A),\\
        \Pi'_c f_B(V,x^A):=\ &\Theta(V_c-V)f_B(V_c,x^A)\\&+\Theta(V-V_c)f_B(V,x^A)-f_B(V_c,x^A),
\end{split}\end{equation}
where $\Theta$ is the Heaviside step.

These projectors are chosen so as to set the function to zero at the cut and to the future (or past), respectively, as required to achieve the factorization of Eq.~(\ref{eq:factorizationSketch}) and Fig.~\ref{fig:EasyHardCases}. It should be noted that even if $f_B$ is smooth, then the projection $\Pi_c f_B$ is in general not smooth. One might be concerned that this would obstruct the desired extension to the Weyl algebra \eqref{eq:extendedAlgebra}, which depends on the symplectic product between the solutions. The singularity of the projected functions does not spoil the classical Poisson algebra, however, because $\Pi_c \phi_B$ and $\Pi'_c \phi_B$ are both once-differentiable (in the Sobolev sense)
\begin{equation}\begin{split}
        &\partial_V(\Pi_{c} f_B(V)) = (\partial_Vf_B(V))\Theta(V_c-V),
        \\&\partial_V (\Pi'_cf_B(V)) = (\partial_Vf_B(V))\Theta(V-V_c),
\end{split}\end{equation}
due to the fact that all intermediate $\delta$-functions cancel identically. This has the result of guaranteeing that the symplectic product of any two projected functions ${\mathscr{W}}(\Pi_c f_B,\Pi_c g_B)$ is well-defined, as is the symplectic product of a projected function with a smooth one.

With these tools in hand, a straightforward application of the extended Weyl relations of Eq.~(\ref{eq:extendedAlgebra}) gives an identity that realizes, at the level of the algebra, the factorization property we sought in Eq.~(\ref{eq:factorizationSketch}). This identity {is} the key technical advance of this section, and constitutes the mathematical realization of the observation illustrated in Fig.~\ref{fig:EasyHardCases}.

Letting $(\phi_c)_B(V,x^A):=\phi_B(V_c,x^A)$ we obtain a factorization identity for general coherent algebra elements
\begin{equation}
    \begin{split}
        \op{U}(\phi)=\op{U}(\Pi_c \phi)\op{U}(\Pi'_c \phi)e^{i2\op{\Delta}(\phi_c)}{e^{-i\Delta_\phi(\phi_c)/2}}.
    \end{split}
    \label{eq:factorizeCoherent}
\end{equation}
Note that $\Delta_{\phi}({\phi_c}) = \mathscr{W}(\phi_c,\phi)$ where we trivially extend $(\Delta_\phi)_B$ to act on functions on ${\mathscr{H}^+}$ that are constant in~$V$.
As promised at the start of this section, we have in Eq.~(\ref{eq:factorizeCoherent}) rewritten a unitary\footnote{Note that this is \textit{not} a unitary operator on the ordinary Fock space. Here we use the word ``unitary'' to refer to either a unitary element of a  $*$-algebra, or its representative as a unitary operator in a GNS representation of that algebra.} that does not, strictly speaking, factorize across the cut in terms of a product of unitaries belonging to the Weyl algebras on either side of the cut, at the expense of an additional unitary corresponding to a $V$-independent shift (generated by the memory $\op{\Delta}_B$). In doing so we have, in effect, {transformed} the hard case of Casini \textit{et al.} into the easy case as illustrated in Sec.~\ref{sec:horizonAlgebra}.

We also note that
\begin{equation}\label{eq:projectedCommutator}
    [\op{U}(\Pi_c f),\op{U}(\Pi'_c g)]=0
\end{equation}
for any test functions $f$ and $g$, justifying our claim that they belong to the commuting algebras on either side of the cut $\mathscr{C}$, respectively.

In addition to the ordinary $*$-algebra of fields $\op{E}(s)$ smeared into smooth, asymptotically decaying functions, and the memory observable, the global horizon algebra includes all $\op{U}(\phi)$ for $\phi_B$ in the extended class of test functions. In addition to smooth test functions $\phi_{B}$ with asymptotically-decaying first $V$-derivatives this includes all projections $\Pi_c\phi_B$ or $\Pi^{\prime}_c\phi_B$, for each cut of the horizon $\mathscr{C}$ and its associated projectors $\Pi_{c}$ and $\Pi'_{c}$.

\subsection{Vacuum state and Hilbert space}
\label{sec:vacuumState}
An abstract algebra is of little use without a physical state within which to compute expectation values and  moments of probability distributions for the observables of that algebra. 
To make use of our algebra, and to achieve a Hilbert space representation of it via the GNS construction, we extend the Killing-symmetry invariant state of \cite{Kay:1988mu}. In the present context of a Rindler horizon, as well as at null infinity, this constitutes the Poincar\'e-invariant vacuum state.

On the original Weyl algebra,
\begin{equation}
    \omega_\Omega(\op{W}(s_1)^*\op{W}(s_2))=e^{-\frac{1}{2}\lVert s_1-s_2\rVert^2}e^{i\mathscr{W}(s_1,s_2)/2}.
\end{equation}
Here, the norm is the Klein-Gordon-like norm on the positive frequency part of the test function
\begin{equation}
    \lVert s\rVert^2:= \omega_\Omega(\op{E}(s)^2).
    \label{eq:kgNorm}
\end{equation}
The vacuum state on the extended Weyl algebra is
\begin{equation}\begin{split}
        \label{eq:state}
        \omega_\Omega(&\op{U}(\phi_1)^*\op{U}(\phi_2))
        \\&= \begin{cases}
            e^{\frac{i}{2}\Delta_{\phi_1}(\phi_1-\phi_2)} & \partial_V(\phi_{2,B}-\phi_{1,B})=0 \\
            e^{-\frac{1}{2}\lVert \phi_1-\phi_2\rVert^2}e^{\frac{i}{2}\mathscr{W}(\phi_1,\phi_2)} & \Delta_{\phi_1}= \Delta_{\phi_2} \\
            0 & \Delta_{\phi_1}\ne \Delta_{\phi_2},
        \end{cases}
\end{split}\end{equation}
where in the first case, the condition $\partial_{V}(\phi_{2,B}-\phi_{1,B})=0$ implies that $\phi_{1,B}$ and $\phi_{2,B}$ differ by a $V$-independent shift.  
This state extends onto the memory observable $\op{\Delta}(h)$ by the relation
\begin{equation}
    [\op{U}(\phi),\op{\Delta}(h)]=\op{U}(\phi)\Delta_\phi(h)/2
\end{equation}
and the fact that $\omega_\Omega$ is an eigenstate of $\op{\Delta}(h)$ with eigenvalue zero. This implies that because $\Delta_\phi(h)$ is no longer required to vanish, $\op{\Delta}(h)$ is no longer central in the algebra.

In fact, the Weyl element corresponding to the memory observable was already implicit in the extended Weyl algebra. To see this, consider any two solutions $(f_1)_B(V,x^A)$ and $(f_2)_B(V,x^A)$ such that $\partial_V ((f_1)_B - (f_2)_B) = 0$, in which case $h_B:=(f_1)_B-(f_2)_B$ only depends on the spatial cross sections of the horizon, and $\Delta_{f_1} = \Delta_{f_2}$. Then we find that \begin{equation}
    \op{U}(f_1)\op{U}(f_2)^* = e^{i\Delta_{f_1}(h)/2}e^{i2\op{\Delta}(h)},
\end{equation}
so the (exponentiated) memory observable $\op{\Delta}(h)$ was, in fact, already present in the extended operator algebra.

At this point one may ask whether the same strategy might be employed to extend the electric field algebra itself ``$\op{E}(\phi)$'' to include classical solutions $\phi_B$ with memory. Indeed, there is no obstruction defining an \textit{abstract} algebra corresponding to the classical field algebra with memory, in correspondence with the Poisson relations of Eq.~(\ref{eq:classicalAlgebra}), to obtain an abstract field algebra that includes test functions $\phi$ that do not decay asymptotically on the horizon. However, the vacuum state $\omega_\Omega$ does not extend onto this algebra. This can be seen from the fact that Eq.~(\ref{eq:particleVacuumState}) suffers from an infrared divergence when $s_{1,B}$ and $s_{2,B}$ are both replaced by the same function $\phi_B$ with memory, i.e., a function such that $\lim_{V\to\infty}\phi_B(V,x^A)\ne \lim_{V\to -\infty}\phi_B(V,x^A)$. Thus, although the vacuum $\omega_\Omega$ has a natural extension to the Weyl algebra given in Eq.~(\ref{eq:state}), it does not extend onto an enlarged field algebra ``$\op{E}(\phi)$'' of classical solutions with memory. Therefore, along with the extended Weyl algebra via Eq.~\eqref{eq:state}, the resulting GNS Hilbert space will carry a representation of the \textit{ordinary}, zero-memory field algebra via Eq.~(\ref{eq:particleVacuumState}).

Let $\mathcal{H}$ be the GNS Hilbert space of the fully extended Horizon algebra $\mathscr{A}$ with respect to $\omega_{\Omega}$ where the one-particle inner product is given by
\begin{equation}
    \braket{s_{1}|s_{2}}:=\omega_{\Omega}(\op{E}(s_{1})^{\ast}\op{E}(s_{2}))
\end{equation}
for any $s_{1}$ and $s_{2}$ that decay asymptotically, and the inner product on coherent states is
\begin{equation}\label{eq:coherentStateInner}
    \langle \op{U}(\phi_1)\Omega|\op{U}(\phi_2)\Omega\rangle := \omega_\Omega(\op{U}(\phi_1)^*\op{U}(\phi_2)).
\end{equation}

Note that $\op{\Delta}(h)$ is no longer central in the algebra of observables, so it is no longer represented by a multiple of the identity. Regarding those elements that do commute with the memory, as elements of the von Neumann algebra $\mathfrak{A}$ of bounded operators on $\mathcal{H}$ one has, for \textit{asymptotically decaying} test functions $s$,
\begin{equation}
    e^{-i\op{E}(s)} = \op{W}(s) = \op{U}(s).
\end{equation}
No such relation holds, however, for the elements of the Weyl algebra of operators $\op{U}(\phi)$ when $\phi_B$ has memory. This is because, for any $\phi_B$ with memory, the one-parameter unitary group  $t\mapsto\op{U}(t\phi)$ is neither strongly nor weakly continuous in the GNS representation, as can be seen from Eq. (\ref{eq:state}). This is consistent with the fact that the algebra of observables does not contain a quantum analog of the classical observable $E(\phi)$ for $\phi_B$ with memory, which, if included in the algebra, would constitute the generator of that unitary group. This topological obstruction can be understood as the reason why the ordinary Fock space of scattering theory fails to incorporate physical states with memory.

The fact that the formal ``exponential of a field observable with memory'' should be better behaved {than} the field observable itself has been anticipated in the literature \cite{himwich_2020}. The question of the existence of the {infinitesimal} generator of such unitaries is intimately tied to the existence (or lack thereof) of a Poincar\'e-invariant vacuum state for the resulting algebra. In our context, a Poincar\'e-invariant vacuum is physically reasonable, and indeed the GNS constructions of Secs. \ref{sec:ordinary} and \ref{sec:vacuumState} guarantee the existence of this vacuum by construction. Other physical contexts, however, may favor other representations \cite{flanagan_2023, Kudler-Flam_2025, Prabhu_2024b, Prabhu_2024}.

In fact, the GNS Hilbert space we obtain in Sec. \ref{sec:vacuumState} is the direct sum Hilbert space over memory representations \cite{Kibble_1968, Prabhu_2024}, and its properties are detailed in Appendix \ref{app:memoryHilbert}.\footnote{ While there exist uncountably infinite memory ``sectors'' labeled by memory, unlike in superselection theory, one cannot restrict to any chosen sector because there exist physical memory-changing operators. This is discussed further in Sec.~\ref{sec:scattering}.} 
  The cautious reader may be concerned that this Hilbert space is not separable, and it is sometimes argued that nonseparable Hilbert spaces are not useful. However, as we have already seen, this Hilbert space is the GNS representation of a physically relevant algebra. Furthermore, it will prove very useful insofar as it provides a setting for Tomita-Takesaki theory, which remains well-defined regardless of separability \cite{Takesaki_1970}.

\section{Tomita-Takesaki Theory for General Coherent States and Soft Modes}\label{sec:ttTheory}

\subsection{Standard Tomita-Takesaki theory}
We first review standard Tomita-{Takesaki} theory of the free, radiative fields on the horizon. We recall that the standard von Neumann algebra $\bar{\mathfrak{A}}_{c}$ consists of the (weak closure of) the Weyl operators $\op{W}(s)$ on the zero memory Fock space $\mathcal{F}_{0}$ where the operators have support to the past of a cut $\mathscr{C}$. On $\bar{\mathfrak{A}}_c$ the vacuum state $\ket{\Omega}$ is a cyclic and separating state.\footnote{\label{fn:cyclic} 
We say $|\Omega\rangle$ is \emph{cyclic} for $\mathfrak{A}$ if a dense subset of vectors in $\mathcal H$ are produced by acting all elements of $\mathfrak{A}$ on $|\Omega\rangle$.
We say $|\Omega\rangle$ is \emph{separating} for $\mathfrak{A}$ if $\hat{a}|\Omega\rangle=0$ implies $\hat{a}=0$ for all $\hat{a}\in \mathfrak{A}$.} Given two {cyclic and separating} states $\ket{\Psi},\ket{\Phi}\in \mathscr{F}_0$, one can (densely) define a relative Tomita operator $\hat{S}_{\Psi|\Phi}$ which satisfies 
\begin{equation}
\hat{S}_{\Psi\to\Phi}\hat{a}\ket{\Psi}=\hat{a}^{\dagger}\ket{\Phi}. 
\end{equation}
Given the relative Tomita operator, one can define a ``relative modular operator'' $\hat{\Delta}_{\Psi\to\Phi}$ as 
\begin{equation}
\hat{\Delta}_{\Psi\to\Phi} = \hat{S}^{\dagger}_{\Psi\to \Phi}\hat{S}_{\Psi\to\Phi}.
\end{equation}
The {\em modular operator} is defined as the case where both entries of the relative modular operator refer to the same state---i.e., for $\ket{\Psi}$ the modular operator is $\hat{\Delta}_{\Psi} = \hat{\Delta}_{\Psi\to\Psi}$. This operator generates a strongly continuous, one-parameter group of automorphisms of any von Neumann algebra $\mathfrak{A}$
\begin{equation}
\label{eq:modflow}
 \hat{a}(t)=\hat{\Delta}_{\Psi}^{-it}\hat{a} \hat{\Delta}_{\Psi}^{it} \in {\mathfrak{A}} \quad \quad \forall \hat{a}\in {\mathfrak{A}}, t\in \mathbb{R},
\end{equation}
where $\hat{a}(t)$ is the one-parameter family of algebra elements which constitute the ``modular flow'' of $\hat{a}$. The modular operator satisfies the direct analog of the Kubo--Martin--Schwinger (KMS) conditions now with respect to modular flow: 
\begin{equation}
\hat{\Delta}_{\Psi}\ket{\Psi}=\ket{\Psi}, \quad \quad \braket{\Psi|\hat{a}(t) \hat{b}|\Psi}=\braket{\Psi|{\hat{b}\hat{a}(t+i)}|\Psi} 
\end{equation}
for any $\hat{a},\hat{b}\in {\mathfrak{A}}$. The right-hand side is well-defined because the correlation function $\braket{\Psi|{\hat{b}\hat{a}(t)}|\Psi}$ is analytic on the strip ${0 < \Im t < 1}$ and is bounded and continuous on the boundary of the strip. As we will review in the following subsections, the modular operator plays a significant role in the definition {of} many quantum-information-theoretic quantities. However, an explicit construction of the modular operator for a general state is rarely available in practice. 

A notable exception is the vacuum state $\ket{\Omega}$ where on the algebra $\mathfrak{A}_{c}$, the corresponding operator $\hat{\Delta}^{it}_\Omega$ yields a boost about $V_c$ with rapidity $\xi=2\pi t$.
We will henceforth refer to $t$ as the ``boost parameter.'' The action of such a boost about $V_c$ on the horizon is
\begin{equation}
	\Lambda^{2\pi t}(V,x^A) = (V_c + e^{-2\pi t}(V-V_c), x^A).
\end{equation}
The action of the vacuum modular operator on the one-particle Hilbert space is determined by its action on the corresponding classical solutions.
The classical field $E_A(V,x^B)=-\partial_V A_A(V,x^B)$ transforms under the (pullback $\Lambda^\xi_*$ of a) boost as
\begin{equation}\begin{split}
	\Lambda^\xi_*[E_A(V,x^B)] =  e^{\xi} E_A(V_c+e^{-\xi}(V-V_c),x^B).
\end{split}\end{equation}

Another case in which the modular operator is known exactly is due to Casini \textit{et al.} \cite{Casini_2019} and Longo \cite{Longo_2019}, in the easy case of Sec. \ref{sec:horizonAlgebra}. In that case, the relative modular operator between two coherent states that factorize across the cut as 
\begin{equation}
    |\Psi\rangle = \op{W}(s_\Psi)\op{W}(s_\Psi')|\Omega\rangle,\qquad |\Phi\rangle = \op{W}(s_\Phi)\op{W}(s_\Phi')|\Omega\rangle
\end{equation}
is given to be
\begin{equation}
    \hat{\Delta}_{\Psi\to\Phi} = {\op{W}(s'_\Psi)\op{W}(s_\Phi)}\hat{\Delta}_\Omega \op{W}(s_\Phi)^\dagger\op{W}(s'_\Psi)^\dagger.
\end{equation}
In the hard case, however, no such formula has appeared in the literature, nor has such a formula been derived in the presence of field configurations with memory. In what follows, we provide a formula for the relative Tomita operator between general coherent states.
\subsection{Relative Tomita operator for general coherent states}
From here on we will work within the GNS representation of the full algebra\footnote{The construction also applies to free massive and massless scalar fields, even though the {large-gauge interpretation does not apply}.} defined in Secs.~\ref{sec:ordinary} and~\ref{sec:irExtension}. This horizon algebra includes the ordinary zero-memory field algebra $\op{E}(s)$ and the corresponding Weyl algebra $\op{W}(s)$, as well as the Weyl elements $\op{U}(\phi)$ corresponding to coherent radiation with memory, and the memory observable itself $\op{\Delta}(h)$. Using this algebra, we will provide the relative Tomita operator for the algebra to the past of the cut $\mathfrak{A}_c$. The full GNS Hilbert space of $\mathfrak{A}$ has several possibly unfamiliar properties relative to the standard Tomita-Takesaki theory we just presented. 
First, the Hilbert space $\mathcal{H}$ is nonseparable (i.e., it does not admit a countable basis). This is no obstruction, however, to the formulation of Tomita-Takesaki theory~\cite{Takesaki_1970}. 

To apply the Tomita-Takesaki theory, all that is typically required is the existence of a cyclic and separating state, the definitions of which are given in footnote \ref{fn:cyclic}. The algebra $\mathfrak{A}_c$ to the past of the cut will include the standard Weyl algebra $\op{W}(s)$ as well as elements of the extended $\op{U}(\phi)$ that are supported strictly to the past of the cut. Notably, this does not include $\exp({i\op{\Delta}(h)})$. With the inclusion of the enlarged Weyl algebra including operators with memory $\op{U}(\phi)$, it is straightforward to show that $\Omega$ is again cyclic and separating on the algebra $\mathfrak{A}_c$ to the past of the cut.\footnote{
Acting with $\op{U}(\phi_\Delta)\op{W}(s)\in\mathfrak A_c$, with $s$ ranging densely over the zero-memory one-particle space and $\Delta_B$ over all memory profiles, produces a dense set in every $\mathcal F_\Delta$ and hence in $\mathcal H$. {The analogous future algebra is cyclic on $\Omega$, hence $\Omega$ is separating for $\mathfrak A_c$.} 
} 
{Since the state~\eqref{eq:state} is boost-KMS,} the modular operator for the Poincar\'e-invariant vacuum is unchanged from its definition on the Poincar\'e-invariant vacuum of the standard Fock space. Thus, despite the unusual setting, we are equipped with the necessary tools to proceed with the Tomita-Takesaki theory.
\begin{widetext}
We now construct the relative Tomita operator for two general coherent states $|\Psi\rangle = \op{U}(\phi^\Psi)|\Omega\rangle$ and $|\Phi\rangle = \op{U}(\phi^\Phi)|\Omega\rangle$, where we do not assume that $\phi_B^\Psi$ and $\phi_B^\Phi$ vanish at the cut, or decay asymptotically. Then, restricting these states to the past of $\mathscr{C}$, we will show that the relative Tomita operator of the state $\Psi|_c$ relative to $\Phi|_c$ is

\begin{equation}\label{eq:tomita}
    \begin{split}
        \hat{S}_{\Psi\to\Phi}=e^{-i\Delta_{\phi^\Phi}(\phi^\Phi_c)/2}\op{U}(\Pi_c \phi^\Psi)\op{U}(\Pi_c'\phi^\Phi)\hat{S}_\Omega \op{U}(\Pi_c\phi^\Phi)^\dagger\op{U}(\Pi_c'\phi^\Psi)^\dagger e^{i\Delta_{\phi^\Psi}(\phi^\Psi_c)/2},
    \end{split}
\end{equation}
where $\op{S}_\Omega$ is the Tomita operator for the vacuum. Here $\Delta_\phi(h):=\int d^2x^A h^B(\Delta_\phi)_B$.

To justify this, let us verify that $\hat{S}_{\Psi\to\Phi}$ satisfies the defining property of the relative Tomita operator: $\hat{S}_{\Psi\to\Phi}\hat{a}|\Psi\rangle = \hat{a}^\dagger|\Phi\rangle$ for any $\hat{a}\in\mathfrak{A}_c$ (including the infrared sector) to the past of the horizon cut $\mathscr{C}$. For any element $\hat a$ of $\mathfrak{A}_c$, we have
    \begin{equation}\begin{split}
            &\hat{S}_{\Psi\to\Phi}\hat{a}|\Psi\rangle
            \\ &=e^{-i\Delta_{\phi^\Phi}(\phi_c^\Phi)/2}\op{U}(\Pi_c\phi^\Psi)
            \op{U}(\Pi_c'\phi^\Phi)\hat S_\Omega
            \op{U}(\Pi_c\phi^\Phi)^\dagger\op{U}(\Pi_c'\phi^\Psi)^\dagger e^{i\Delta_{\phi^\Psi}(\phi_c^\Psi)/2}\hat{a}|\Psi\rangle
            \\ &=e^{-i\Delta_{\phi^\Phi}(\phi_c^\Phi)/2}\op{U}(\Pi_c\phi^\Psi)
            \op{U}(\Pi_c'\phi^\Phi)\hat S_\Omega
            \op{U}(\Pi_c\phi^\Phi)^\dagger\op{U}(\Pi_c'\phi^\Psi)^\dagger e^{i\Delta_{\phi^\Psi}(\phi_c^\Psi)/2}\hat{a}\op{U}(\phi^\Psi)|\Omega\rangle.
    \end{split}\end{equation}
    Next we apply the observation of Sec.~\ref{sec:horizonAlgebra} that the easy case differs from the hard case (the case at hand) by a $V$-independent shift, as illustrated in Fig.~\ref{fig:EasyHardCases}. We factor the final unitary using the resulting factorization identity (\ref{eq:factorizeCoherent}),
    \begin{equation}\begin{split}
            &=e^{-i\Delta_{\phi^\Phi}(\phi_c^\Phi)/2}\op{U}(\Pi_c\phi^\Psi)
            \op{U}(\Pi_c'\phi^\Phi)\hat S_\Omega
            \op{U}(\Pi_c\phi^\Phi)^\dagger\op{U}(\Pi_c'\phi^\Psi)^\dagger e^{i\Delta_{\phi^\Psi}(\phi_c^\Psi)/2}\hat{a}\op{U}(\Pi_c \phi^\Psi)\op{U}(\Pi'_c \phi^\Psi)
            e^{i2\op{\Delta}(\phi^\Psi_c)}{e^{-i\Delta_{\phi^\Psi}(\phi^\Psi_c)/2}}|\Omega\rangle.
    \end{split}\end{equation}
    Then because $\Omega$ belongs to the null space of $\op{\Delta}$,
    \begin{equation}\begin{split}
            &=e^{-i\Delta_{\phi^\Phi}(\phi_c^\Phi)/2}\op{U}(\Pi_c\phi^\Psi)
            \op{U}(\Pi_c'\phi^\Phi)\hat S_\Omega
            \op{U}(\Pi_c\phi^\Phi)^\dagger\op{U}(\Pi_c'\phi^\Psi)^\dagger\hat{a}\op{U}(\Pi_c \phi^\Psi)\op{U}(\Pi'_c \phi^\Psi)
            |\Omega\rangle.
    \end{split}\end{equation}
    Using the fact that coherent operators projected onto opposite sides of the cut commute (\ref{eq:projectedCommutator}),
    \begin{equation}\begin{split}
            &=e^{-i\Delta_{\phi^\Phi}(\phi_c^\Phi)/2}\op{U}(\Pi_c\phi^\Psi)
            \op{U}(\Pi_c'\phi^\Phi)\hat S_\Omega
            \op{U}(\Pi_c\phi^\Phi)^\dagger\hat{a}\op{U}(\Pi_c \phi^\Psi)|\Omega\rangle,
    \end{split}\end{equation}
    and the defining property of $\hat S_\Omega$ gives
    \begin{equation}\begin{split}
            &=e^{-i\Delta_{\phi^\Phi}(\phi_c^\Phi)/2}\op{U}(\Pi_c\phi^\Psi)
            \op{U}(\Pi_c'\phi^\Phi) \op{U}(\Pi_c \phi^\Psi)^\dagger\hat{a}^\dagger\op{U}(\Pi_c\phi^\Phi)|\Omega\rangle
            \\ &=e^{-i\Delta_{\phi^\Phi}(\phi_c^\Phi)/2}
            \op{U}(\Pi_c'\phi^\Phi)\hat{a}^\dagger\op{U}(\Pi_c\phi^\Phi)|\Omega\rangle.
    \end{split}\end{equation}
    We again use the fact that $\Omega$ is in the null space of $\op{\Delta}$ to write
    \begin{equation}\begin{split}
            &=e^{-i\Delta_{\phi^\Phi}(\phi_c^\Phi)/2}
            \op{U}(\Pi_c'\phi^\Phi)\hat{a}^\dagger\op{U}(\Pi_c\phi^\Phi)e^{i2\op{\Delta}(\phi_c^\Phi)}|\Omega\rangle.
    \end{split}\end{equation}

    Finally, we apply the fact that $\op{U}(\Pi_c'\phi^\Phi)$ is in the commutant of $\mathfrak{A}_c$ and again use the factorization identity (\ref{eq:factorizeCoherent}) to obtain
    \begin{equation}\begin{split}
            &=\hat{a}^\dagger \op{U}(\phi^\Phi)|\Omega\rangle
            \\&=\hat{a}^\dagger|\Phi\rangle,
    \end{split}\end{equation}
thereby satisfying the defining property of the relative Tomita operator. Thus, we see that Eq.~(\ref{eq:tomita}) does, indeed, specify the relative Tomita operator.
\end{widetext}

 With this, we can compute various quantum information theoretic quantities, such as the Holevo fidelity and relative entropy between general coherent states, in both the ``easy'' and ``hard'' cases of Casini \textit{et al.}, and also for coherent states of soft radiation.

The so-called relative modular operator is the modulus of the relative Tomita operator  $\hat{\Delta}_{\Psi\to\Phi}=\hat{S}^\dagger_{\Psi\to\Phi}\hat{S}_{\Psi\to\Phi}$, and it is useful in the computation of many information-theoretic quantities. In our case it evaluates to
\begin{equation}
    \begin{split}
        \hat{\Delta}_{\Psi\to\Phi}&=\op{U}(\Pi_c'\phi^\Psi+\Pi_c\phi^\Phi)\hat{\Delta}_\Omega\op{U}(\Pi_c\phi^\Phi+\Pi_c'\phi^\Psi)^*.
    \end{split}
\end{equation}
In the case that $\Phi$ is the vacuum $\Omega|_c$ and $\Psi$ has the vanishing one-point function at $\mathscr{C}$, the projection (including the $V$-independent shift) implemented by $\Pi'_c$ acts trivially, and we recover Casini \textit{et al.}'s expression for the relative modular operator in the easy case. We are, however, interested in the fully general case.

Formerly, in the hard case of Casini \textit{et al.}, the lack of a general formulation of the corresponding Tomita-Takesaki theory meant that each new quantum-information theoretic quantity of interest required a new method of computation. Thus, each quantity of interest (e.g., R\'enyi entropies, etc.) would need to be treated on a case-by-case basis. Now, having suitably extended the algebra and the framework of Tomita-Takesaki theory, a variety of information-theoretic applications can be developed with relative ease. We first return to the Holevo fidelity, which was the original motivation for our foray into Tomita-Takesaki theory, in Sec.~\ref{sec:IDT}.

With the relative modular operator in hand, we can explicitly evaluate the distinguishability of the states on Bob's side of the horizon sourced by Alice's superposition. We will express this in terms of the Holevo fidelity, the Uhlmann fidelity, the relative entropy, and the R\'enyi entropies.

\subsection{The Holevo fidelity between coherent states}\label{sec:holevoCoherent}
The (square root of the) Holevo fidelity between two coherent states $\Psi$ and $\Phi$  restricted to the past of $V_c$ is given by
\begin{equation}
	\sqrt{f_\mathrm{H}}(\Psi,\Phi) = \langle\Psi|\hat{\Delta}^{1/2}_{\Psi\to\Phi}
	|\Psi\rangle,
\end{equation}
where the square root of the relative modular operator is
\begin{equation}
	\hat{\Delta}_{\Psi\to\Phi}^{1/2}=\op{U}(\Pi_c'\phi^\Psi+\Pi_c\phi^\Phi)\hat{\Delta}_\Omega^{1/2}\op{U}(\Pi_c\phi^\Phi+\Pi_c'\phi^\Psi)^*.
\end{equation}
We will ultimately be interested in the case where $\Phi = \Omega$, so we restrict our attention to that case for simplicity. The calculation follows by exact analogy for a general coherent state $\Phi$.

Because the expectation value $\langle \Psi| \Delta^{it}|\Psi\rangle$ has an analytic continuation to strip $-1 < \Im{t} < 0$, the conjugation by $\hat{\Delta}_\Omega^{1/2}$ corresponds to a boost with the boost parameter analytically continued to $\xi/(2\pi)=t= -i/2 $, which gives the element $\Lambda_{RT}(V,x^A)=(2V_c-V,x^A)$ of the Lorentz group.

The fidelity is symmetric in its arguments, so for convenience we calculate
\begin{equation}\begin{split}
\sqrt{f_\mathrm{H}}({\Omega},\Psi)&=\langle\op{U}(\Pi_c\phi^\Psi)^*\hat{\Delta}_\Omega^{1/2}\op{U}( \Pi_c\phi^\Psi)\rangle_\Omega \\&=\langle\op{U}(\Pi_c\phi^\Psi)^* \hat{\Delta}_\Omega^{1/2}\op{U}(\Pi_c\phi^\Psi)\hat{\Delta}_\Omega^{-1/2}\rangle_\Omega
\\&= \langle \op{U}(\Pi_c\phi^\Psi)^*\op{U}(-\Lambda_*^{RT} \Pi_c\phi^\Psi)\rangle_\Omega,
\label{eq:fidelityContinuation}
\end{split}\end{equation}
where $\Lambda_*^{RT}$ is the action of $\Lambda_{-i\pi} = \Lambda_{RT}$ and reflects a test function $h^A$  on the horizon about $V_c$:
\begin{equation}
	\Lambda_*^{RT} h^A(V,x^{B}) = h^A(2V_c-V,x^{B}).
\end{equation}
In the second line of Eq.~(\ref{eq:fidelityContinuation}) we have used the fact that ${\Omega}$ is an eigenstate of $\hat\Delta_\Omega$ with eigenvalue $1$.

Thus, we obtain a simple expression for the fidelity between two coherent states restricted to the past of a cut $\mathscr{C}$,
\begin{equation}\label{eq:fidelityFinalAnswer}
	\sqrt{{f}_\mathrm{H}}(\Psi,{\Omega}) = e^{-\frac{1}{2}\lVert \Pi_c\phi^\Psi + \Lambda_*^{RT}\Pi_c\phi^\Psi \rVert^2}.
\end{equation}
As promised, this formula applies regardless of whether the function $\phi^\Psi$ is supported at $V_c$ or not. That is, this gives the Holevo fidelity in both the hard and easy cases of Casini \textit{et al.}, and therefore determines the distinguishability of interior states even at intermediate times, while Alice's superposition is held open as discussed in Sec.~\ref{sec:IDT}.

We previously analyzed an intimately related question \cite{DKSW_2025}, in which we determined the optimal protocol for Alice to follow so that she minimizes the decoherence she finds in the exterior of a black hole. There, it was shown that to minimize her decoherence, Alice should not merely close her superposition, but also subsequently reopen it and reclose it according to a specific pattern. By doing this she, in effect, produces a radiation field on the horizon that more closely resembles a vacuum fluctuation and thus reduces the probability of Bob achieving certainty about the state of her superposition. In that paper, it was shown that a reasonably good procedure (though not the optimal one) would be for the reopening and reclosing process to proceed as the $CRT$ conjugate of her original process, giving rise to what was there referred to as the ``$CRT$ purification'' on the horizon.

The operation $CRT$ on the Hilbert space of a quantum field theory corresponds to the element $\Lambda_{RT}$ of the Lorentz group about $\mathscr{C}$. The squared magnitude of the inner product of the $CRT$ purification with the vacuum is precisely the Holevo fidelity, which we have computed here by much simpler means. As shown in Appendix \ref{app:holevoFidelity} the Holevo fidelity approximates the Uhlmann fidelity by
\begin{equation}\label{eq:uhlmannHolevoBound}
    1-\sqrt{f_\mathrm{H}}(\rho,\sigma)\le \sqrt{1-f_\mathrm{U}(\rho,\sigma)}\le \sqrt{1-f_\mathrm{H}(\rho,\sigma)},
\end{equation}
which bounds Bob's probability of success by Eq. (\ref{eq:uhlmannIntermediateTimes}) and corresponds to the \textit{optimal} purification of \cite{DKSW_2025}, and will be the focus of the next section (Sec.~\ref{sec:uhlmannCoherent}).

It is instructive to consider the special case that the coherent state $\Psi$ corresponds to a classical solution $s_B$ that decays asymptotically (zero memory) and also vanishes at the cut. As was shown in \cite{Frob:2024ijk} it is possible to decompose the exponent of Eq.~(\ref{eq:fidelityFinalAnswer}) as
\begin{equation}\begin{split}
    &{-\frac{1}{2}\lVert \Pi_c s + \Lambda_*^{RT}\Pi_c s \rVert^2}\\&=-\frac{1}{2}\lVert \Pi_c s\rVert^2 - \frac{1}{2}\lVert \Lambda_*^{RT}\Pi_c s\rVert^2 -  {\Re}\omega(\op{E}(\Pi_cs), \op{E}(\Lambda_*^{RT}\Pi_c s))
    \\&=-\lVert \Pi_c s\rVert^2 -  {\Re}\omega(\op{E}(\Pi_c s), \op{E}(\Lambda_*^{RT}\Pi_c s)).
\end{split}\end{equation}
The first term counts the number of particles in the coherent state corresponding to $s_B(V,x^A)$ while the final term is the {symmetric part} of the two-point function, namely \cite{KW_1991},
\begin{equation}\label{eq:KWRealPart}\begin{split}
        \Re\omega(\op{E}(&\Pi_c s), \op{E}(\Lambda_*^{RT}\Pi_c s)) = \frac{-1}{\pi}\int_{\mathbb{R}^2\times \mathbb{R}^2} dV_1dV_2 d^2x^B\\&\times\operatorname{P.V.}\left[\frac{\Pi_c s^A(V_1,x^B) \Lambda_*^{RT}\Pi_cs_{A}(V_2,x^B)}{(V_1-V_2)^2}\right],
\end{split}\end{equation}
where the principal value acts trivially because the test functions have disjoint support.
Following \cite{Frob:2024ijk}, then, in the easy case the fidelity cleanly splits into two factors for finite $V_c$,
\begin{align}\label{eq:mixedStateFidelity}
    \sqrt{f_\text{H}}(\Psi|_c, \Omega|_c) = e^{-\lVert \Pi_c s\rVert^2}e^{-{\Re}\omega(\op{E}(\Pi_c s), \op{E}(\Lambda_*^{RT}\Pi_c s))}.
\end{align}
In the case that the radiation remains zero forever after the cut, e.g., in the case that Alice has finished her experiment, the first factor is the fidelity due to the expected number of particles in the state $\langle N\rangle_{\Psi} = \lVert s\rVert^2$, while the second piece is caused by the uncertainty in Bob's ability to measure these particles, due to vacuum fluctuations. Only the latter factor depends on $V_c$, and
its exponent monotonically decays toward zero as $V_c$ increases as can be seen from evaluating (\ref{eq:KWRealPart}) and its first $V_c$ derivative.

This implies that as Bob is allowed more time in the interior, he can better distinguish the configuration of the interior field from vacuum fluctuations. This is complementary to the derivation of \cite{DKSW_2025} showing Alice can minimize her decoherence in the exterior by performing her experiment so as to ``mimic'' a vacuum fluctuation (although the actual improvement in her coherence is minuscule if her superposition has been open for an appreciable time).

In the case in which no time constraints are placed on Bob, then the states on the Rindler horizon would become pure states, in which case there is guaranteed to be some measurement available to Bob that will saturate the bound (\ref{eq:uhlmannIntermediateTimes}) on his one-shot probability $P_\mathrm{Bob}^\mathrm{certain}(\mathscr{B})$ to unambiguously distinguish the state.

The infinite time case corresponds to the fidelity along the entire horizon,\footnote{
    Perhaps surprisingly, this expression is \textit{not} identical to the $V_c\to\infty$ limit of the finite-time, mixed state fidelity of Eq.~(\ref{eq:mixedStateFidelity}). This discrepancy is due to the fact that $\lim_{V_c\to\infty}\lVert \Lambda_*^{RT}\Pi_c s\rVert \ne \lVert \lim_{V_c\to\infty}\Lambda_*^{RT}\Pi_c s\rVert$.} which is
\begin{equation}
    \sqrt{f_\mathrm{U}}(\Psi,{\Omega}) = e^{-\frac{1}{2}\lVert \phi^\Psi \rVert^2} = e^{-\frac{1}{2}\langle\hat N\rangle_\Psi},
\end{equation}
where $\langle\hat N\rangle_\Psi$ is the expected number of particles in the coherent state $\Psi$. As one would expect, we have recovered the formula of \cite{DSW_2021,DSW_2022,DSW_2023,DSW_2024} for the final decoherence $1-\sqrt{f_\mathrm{U}}$ of Alice in terms of the number of entangling particles $\langle \hat N\rangle_\Psi$ in the limit that we account for the distinguishability of all entangling degrees of freedom on the horizon. The discrepancy between this formula and the distinguishability of the states according to Bob at finite times as given in Eq. (\ref{eq:mixedStateFidelity}) is notable. Evidently, even at very large $V_c$ and thus very small $\Re\omega(\op{E}(\Pi_c s), \op{E}(\Lambda_*^{RT}\Pi_c s)) \simeq 0$, the uncertainty due to vacuum fluctuations poses a significant obstacle to Bob's ability to perform discerning measurements of the field in his region. If, even at very late times, it is possible for Bob to infer with certainty the full future time evolution on the horizon (by, for example, learning by some other means that Alice is finished with her experiment), he gains a significant amount of additional which-path information. This is demonstrated by the exponent of the fidelity passing from $\langle\hat N\rangle_\Psi$ to $\langle \hat N\rangle_\Psi / 2$ in the pure-state case.

\subsection{The Uhlmann fidelity between general coherent states}\label{sec:uhlmannCoherent}
Recall the definition (\ref{eq:uhlmannDefinition1}) of the Uhlmann fidelity $f_\mathrm{U}(\rho,\sigma)$,
\begin{equation}\label{eq:uhlmannDefinition2}
    f_\mathrm{U}(\rho,\sigma) := \sup_{\pi}|\langle \rho|\sigma\rangle_\pi|^2.
\end{equation}
As before, the supremum runs over all Hilbert space representations $\pi$ of the $*$-algebra such that $|\rho\rangle$ and $|\sigma\rangle$ are vectors in the corresponding Hilbert space and are purifications of the states $\rho$ and $\sigma$, respectively. This supremum over all purifications makes the Uhlmann fidelity very difficult to compute in general. For practical purposes, therefore, the Holevo fidelity is often more useful, because the latter bounds the former from above and below by
\begin{equation}\label{eq:uhlmannHolevoBoundBody}
    1-\sqrt{f_\mathrm{H}}(\rho,\sigma)\le \sqrt{1-f_\mathrm{U}(\rho,\sigma)}\le \sqrt{1-f_\mathrm{H}(\rho,\sigma)}
\end{equation}
as shown in Appendix \ref{app:holevoFidelity}.

If one restricts to the space of coherent states, then one can achieve a better estimate than the Holevo fidelity. In a related article \cite{DKSW_2025} we, together with J. Kudler-Flam and R.\ M. Wald,  obtained a formula for the optimal purification of the states in $\mathcal{G}$. For simplicity we consider the case of a single coherent state $\Psi$ along with the vacuum $\Omega$. Let $s_B$ be the classical solution corresponding to $\Psi$.  Restricted to a subalgebra, Ref.~\cite{DKSW_2025} provides an explicit formula for the optimal purification of $\Psi$ into the space of pure \text{coherent} states. That is to say, it was shown that when the purification $|\Psi\rangle$ of $\Psi$ extremizes
\begin{equation}
    \max_{|\Psi\rangle\in\mathcal{G}}|\langle \Psi|\Omega\rangle|
\end{equation}
in the subspace $\mathcal{G}$ of \textit{coherent} states within the Hilbert space, then the optimal coherent state is
\begin{equation}
|\Psi\rangle_\mathrm{max}=\op{U}(\Pi_c s + R_c\Pi_c s  + s_c)|\Omega\rangle,
\end{equation} where the optimal recovery $R_c$ acting on some solution $f_B(V,x^A)$ is
\begin{equation}
    R_c f_B(V,x^A)  :=\frac{1}{\pi}\int_{V_c}^{2V_c} dV'
    \frac{f_B(2V_c-V',x^A)}{V-2V_c+V'}\sqrt{\frac{V-V_c}{V'-V_c}}.
\end{equation}

It is natural to expect that $\max_{|\Psi\rangle\in\mathcal{G}}|\langle \Psi|\Omega\rangle| = \sup_{\pi}|\langle \Psi|\Omega\rangle_\pi|$, i.e., this is the optimal purification in absolute terms, not just in the space of coherent states above the vacuum. In that case the Uhlmann fidelity becomes
\begin{equation}
    f_\text{U}(\Psi,\Omega)=e^{-\left\lVert \Pi_c s +  R_c\Pi_c s\right\rVert^2}.
\end{equation}
Even if it should turn out that an alternative recovery protocol (leveraging purifications outside coherent states) can be found that achieves a larger overlap between the states than the coherent-state protocol of \cite{DKSW_2025}, this nevertheless dramatically outperforms the $CRT$ purification of the Holevo fidelity for many states. Therefore, this gives an improved bound on the ultimate coherence of Alice's superposition even if it turns out not to be strictly equal to the Uhlmann fidelity.

In the case of field configurations without memory that vanish at the cut, a calculation directly analogous to that of the preceding section again finds that the distinguishability according to Bob at late finite times is controlled by $\sqrt{f_\mathrm{U}}\to \exp{(-\langle\hat N\rangle_\Psi)}$, whereas in the case that he can infer the entire future state on the horizon we have $\sqrt{f_\mathrm{U}}=\exp{(-\frac{1}{2}\langle\hat N\rangle_\Psi)}$. 

\subsection{Relative entropy of general coherent states}
\label{sec:relEntropy}
The calculation of the relative entropy between a coherent state and the vacuum forms the primary result of Casini \textit{et al.} In the easy case in which the coherent state factorizes across the cut of the initial data surface, their calculation mirrors ours exactly. In the hard case, however, their calculation requires much greater effort. In the following calculation, the two cases are treated at once, with the hard case being reduced to the level of simple calculation that could previously be employed in the easy case. This highlights the usefulness of the extended Tomita-Takesaki theory for analyzing local quantum information theoretic questions such as the relative entropy to the past of a horizon cut $\mathscr{C}$.

Let $|\Omega\rangle$ and $|\Psi\rangle$ be $CRT$ (i.e., ``natural cone'') purifications of the states $\Omega|_c$ and $\Psi|_c$, related by $\op{U}(\phi)|\Omega\rangle = |\Psi\rangle$. Let $K\phi$ and $\bar K\phi$ be the positive and negative frequency parts of $\phi$, respectively. Then,
% Let $\tau(\xi) = \Theta(\cos\Im\xi)-\Theta(-\cos\Im\xi)$.
\begin{equation}\begin{split}
        {S}_{\Psi|\Omega} =&-\langle \Psi|\ln\Delta_{\Psi\to\Omega}|\Psi\rangle
        \\=&i\frac{d}{dt}\langle\Psi| \op{U}(\Pi'_c\phi)\Delta_\Omega^{it}\op{U}(\Pi'_c\phi)^*|\Psi\rangle\bigg\rvert_{t=0}
        \\=&i\frac{d}{dt}\langle\Omega| \op{U}(\Pi_c\phi)^*\Delta_\Omega^{it}\op{U}(\Pi_c\phi)|\Omega\rangle\bigg\rvert_{t=0}
        \\&=i\frac{d}{dt}\exp\big(-\lVert\Pi_c\phi-e^{2\pi t}\Lambda^{2\pi t}_*\Pi_c\phi\rVert^2/2\\&+i\mathscr{W}(\Pi_c\phi,{e^{2\pi t}}\Lambda^{2\pi t}_*\Pi_c\phi)/2\big)\bigg\rvert_{t=0}
        \\=&\frac{-1}{2}\bigg[i\frac{d}{dt}\lVert\Pi_c\phi-e^{2\pi t}\Lambda^{2\pi t}_*\Pi_c\phi\rVert^2
        \\&+\frac{d}{dt}\mathscr{W}(\Pi_c\phi,e^{2\pi t}\Lambda^{2\pi t}_*\Pi_c\phi)\bigg]_{t=0}
        \\=&\frac{-1}{2}\frac{d}{dt}\bigg[\mathscr{W}(K\Pi_c\phi-e^{2\pi t}K\Lambda^{2\pi t}_*\Pi_c\phi,
        \\&\bar{K}\Pi_c\phi-e^{2\pi t}\bar{K}\Lambda^{2\pi t}_*\Pi_c\phi)
        \\&+\mathscr{W}(\Pi_c\phi,e^{2\pi t}\Lambda^{2\pi t}_*\Pi_c\phi)\bigg]_{t=0},
\end{split}\end{equation}
where in going from the second line to the third we have used the fact that $\op{U}(\phi)|\Omega\rangle = |\Psi\rangle$ and the identity of Eq.~(\ref{eq:factorizeCoherent}). Using the chain rule for the symplectic product
\begin{equation}
    \frac{d}{dt}\mathscr{W}(g,h) = \mathscr{W}\left(\frac{d}{dt}g,h\right)+\mathscr{W}\left(g,\frac{d}{dt}h\right),
\end{equation}
one finds that the contribution from the Klein-Gordon norm vanishes, giving
\begin{equation}\begin{split}
        {S}_{\Psi|\Omega}&=-\frac{1}{2}\frac{d}{dt}\mathscr{W}(\Pi_c\phi,e^{2\pi t}\Lambda^{2\pi t}_*\Pi_c\phi)\bigg\rvert_{t=0}
        \\&=-\frac{1}{2}\mathscr{W}\bigg(\Pi_c\phi,
        2\pi\Lambda^{2\pi t}_*\Pi_c\phi+e^{2\pi t}\frac{d}{dt}\Lambda^{2\pi t}_*\Pi_c\phi\bigg)\bigg\rvert_{t=0}
        \\&=-\frac{1}{2}\mathscr{W}\bigg(\Pi_c\phi,\frac{d}{dt}\Lambda^{2\pi t}_*\Pi_c\phi\bigg)\bigg\rvert_{t=0}
        \\&=2\pi\int_{-\infty}^{V_c} dV d^2x^A (V_c-V)(\partial_V\phi)^2\\&\ + \pi \lim_{V\to-\infty} (V_c-V)
        \\&\times\int_{\mathbb{R}^2} d^2x^A \partial_V \phi_B(V,x^A)(\phi^B(V,x^A)-\phi^B(V_c,x^A)).
\end{split}\end{equation}
At present we are considering a Hilbert space of functions whose first derivatives decay so that the final term is zero.
This leaves\footnote{During the preparation of this manuscript, we became aware that Stefan Hollands and Akihiro Ishibashi have independently derived directly parallel expressions, also allowing for the presence of soft radiation (private communication).}
\begin{equation}\label{eq:calorimetricEntropy}
{S}_{\Psi|\Omega}=2\pi\int_{-\infty}^{V_c} dV d^2x^A (V_c-V)(\partial_V\phi_B)^2.
\end{equation}
This is in agreement with the formula of \cite{Casini_2019}, but its validity is now established even for coherent states corresponding to classical solutions that do not decay at infinity, such as arise in the presence of memory effects.

As Eq. (\ref{eq:calorimetricEntropy}) makes clear, the relative entropy $S_{\Psi|\Omega}$ is directly related to the square of the affine derivative of the potential, $(\partial_V\phi_B)^2$. The relative entropy would therefore appear to be much more closely related to the energy of the potential than it is to any ``soft'' quantity, such as the horizon memory $\Delta_B$. 

At the same time, we have seen that fidelity between states that differ by a substantial number of soft photons or gravitons can be made arbitrarily small. Such ``soft radiation'' therefore carries an enormous amount of which-state information as measured by the fidelity. Evidently the relative entropy (\ref{eq:calorimetricEntropy}) and the fidelity capture different aspects of the quantum information content of the radiation.

The gedanken experiment of \ref{sec:IDT} helps to illustrate the difference between the two, and in particular, the relative sensitivities of each quantity to the soft radiation produced by Alice. To illustrate the point, we consider two scenarios: (I)  Alice opens her superposition, and then holds it open forever, and (II) Alice recombines her superposition shortly after opening it. In both scenarios we analyze the relative entropy and the Uhlmann fidelity of the resulting coherent state restricted to the past of an arbitrary cut $V_c$. 

In scenario (I), suppose that Alice opens her superposition over an affine time interval $V\in(V_0,V_1)$ by keeping one branch of the superposition stationary for all times, and displacing the other branch to a new position where it is then left stationary. Then the relevant, entangling radiation content on the horizon will be determined by the radiation field $\phi_B$ sourced by the displaced trajectory, relative to the vacuum. Because the integrand of Eq. \eqref{eq:calorimetricEntropy} is only supported when the derivative of the field is nonzero, the relative entropy becomes
\begin{equation}
    S_{\Psi|\Omega} = 2\pi \int_{V_0}^{V_1}dVd^2x^A(V_c-V)(\partial_V\phi_B)^2.
\end{equation}
At late cut times $V_c \gg V_1$ this simplifies further. We write the time at the cut in terms of Alice's proper time $t_c$, which is related to $V_c$ by $V_c = e^{\kappa t_c}$, and obtain
\begin{equation}\label{eq:entropyApprox}
    S_{\Psi|\Omega} \sim 2\pi e^{\kappa t_c}\int_{V_0}^{V_1}dVd^2x^A(\partial_V\phi_B)^2.
\end{equation}
From this we learn two things: (i) at late times, the relative entropy grows exponentially in Alice's proper time $t_c$, and (ii) the relative entropy is determined by the hard part of the field configuration, in terms of its affine time derivative. In fact, one could replace Alice's entangling field configuration with a different one, $\phi_\text{hard}$, in which Alice opens and then immediately closes her superposition, and her protocol could be chosen such that \begin{equation}\int_{V_0}^{V_1}dVd^2x^A(\partial_V\phi_{\text{hard},B})^2 = \int_{V_0}^{V_1}dVd^2x^A(\partial_V\phi_B)^2.
\end{equation} In that case, the relative entropies would be essentially identical at late times, despite the fact that $\phi_{\text{hard},B}$ contains no soft radiation whatsoever, because Alice has closed her superposition.  {The} growth of the relative entropy with time illustrates that even an arbitrarily small amount of energy deposited at early times can give rise to a very large relative entropy as measured at sufficiently late times. More to the point, however, this example illustrates that there are situations in which the relative entropy is not directly sensitive to the soft radiation content of field configurations.

We contrast this with the Uhlmann fidelity, in the same setup in which Alice holds her superposition open. At late times the Uhlmann fidelity 
\begin{equation}
    f_\text{U}(\Psi,\Omega)=e^{-\left\lVert \Pi_c s +  R_c\Pi_c s\right\rVert^2}
\end{equation}
decays exponentially \cite{DKSW_2025} in $t_c$ as
\begin{equation}
    \ln f_\text{U}(\Psi,\Omega)\sim -\kappa t_c \quad \quad \textrm{ \textrm{(I)}} 
\end{equation}
for $V_{c}\gg V_{1}$. Thus, we see that, while Alice's superposition is open, the fidelity decays exponentially, due to the linearly increasing number of soft photons in the optimal purification of her superposition \cite{DKSW_2025}. 

In scenario (II), Alice, instead, closes her superposition shortly after opening it and sources $\phi_{\text{hard},B}$ on the horizon. The resulting fidelity would become essentially $t_c$ independent at late times \cite{DKSW_2025}
\begin{equation}
    \ln f_\text{U}(\Psi,\Omega)\sim \textrm{const} \quad \quad \textrm{ \textrm{(II)}} 
\end{equation}
for $V_{c}\gg V_{1}$. This can be understood as the result of the fact that the ``optimal recovery protocol'' for Alice is, in the case that she has already closed her superposition long ago, simply to leave it closed. However, from the discussion above, the relative entropy is still well-approximated by \eqref{eq:entropyApprox} in scenario (II). 

Thus, we learn that, in this example, although the relative entropy is not sensitive to the soft radiation content of the field $\phi_B$, the fidelity is sensitive to its soft photon content (or, more precisely, the soft photon content of its optimal purification). This is a reflection of the fact that the fidelity obtains its operational interpretation in terms of one-shot unambiguous state discrimination, and that whether or not Alice holds her superposition open has an enormous influence on the ease with which Bob may distinguish the resulting change in the field from a vacuum fluctuation.\footnote{The relative entropy, on the other hand, obtains its operational interpretation in a multishot setting \cite{cmp/1104248844}, where vacuum fluctuations average out over many trials.}

For general coherent states $\Psi$ and $\Phi$, the fidelity bounds the relative entropy by \cite{GudderMarchandWyss1979Bures}
\begin{equation}\label{eq:entropyBound}
    e^{S_{\Psi|\Phi}}\ge 1/ f_{\text{U}}(\Psi,\Phi).
\end{equation}
Again letting $\Psi=\Omega|_c$, we see that, in the case under consideration at late times, this is a very weak bound for $\phi_{\text{hard},B}$: the relative entropy grows exponentially in $t_c$, while the fidelity becomes $t_c$-independent. When $\phi_B$ contains soft radiation, however, this bound can become nontrivial, because in that case both quantities grow with $t_c$. The relative entropy grows because of the time-dependent coefficient in front of the purely hard integral of Eq. \eqref{eq:entropyApprox}. The fidelity, on the other hand, is sensitive to the soft photon content of $\phi_B$. 
Since we are considering the case in which Alice never recombines her superposition, in the limit that $t_c\to\infty$, the corresponding states of soft radiation are precisely orthogonal and have zero fidelity. Likewise, their relative entropy is infinite. If, on the other hand, Alice had opened and then immediately closed her superposition, we would replace $\phi_B$ with $\phi_{\text{hard},B}$. The fidelity would stay finite, while the relative entropy would nonetheless diverge. Evidently, it is the fidelity that reflects the quantum information content of soft radiation most faithfully.

\section{Discussion}\label{sec:discussion}

In this paper, we have argued that the framework of approximate quantum error correction is the natural language for understanding the gedanken experiments of \cite{Mari_2016, DSW_2021, DSW_2022, Danielson_thesis, Satishchandran:2025cfk}. From this perspective, the trade-off---indeed, the equivalence of Eq. (\ref{eq:exactChannelIDT})---between decoherence due to entangling radiation and decoherence due to an eavesdropper learning which-path information is clear. We were then able to explicitly compute the information in soft radiation accessible to an observer behind the horizon. We conclude with a few remarks that may deserve further investigation.

\subsection{Connections to scattering theory}
\label{sec:scattering}
Much of the technical content of this article has, for the sake of simplicity, focused on applications to a Rindler horizon. The results, however, apply to horizons more generally, as well as at null infinity of asymptotically flat spacetimes. In fact, much of the related literature until this point has focused \textit{primarily} on soft radiation at null infinity \cite{Bondi_1962, Sachs_1962, Strominger_2}, i.e., in the context of scattering theory. In scattering theory, questions of the coherence of scattering states \cite{Carney_2017, CCNS_2017} and the quantum information content of soft radiation have been complicated by the ubiquitous production of radiation states with memory in scattering processes. This generic production of radiation with memory is signaled by the ubiquitous appearance of ``infrared divergences'' in conventional scattering calculations.

The fact that soft modes, in the infinite time limit, become totally orthogonal is the origin of the observation of Carney \textit{et al.} that, at future infinity, an electron is totally decohered by soft radiation \cite{Carney_2017, CCNS_2017}. This strong degree of entanglement between hard and soft radiation is essential to the proper treatment of quantum information-theoretic questions in scattering theory \cite{Carney_2017, CCNS_2017}. The present article presents a framework with which to analyze questions such as the distinguishability of soft radiation, and the resulting decoherence of the entangled matter, at finite cuts of null infinity, as this entanglement is still accumulating. These techniques can be applied without appealing to the so-called ``inclusive'' quantities typically leveraged to remedy infrared divergences. If one is interested, for example, in the amount of entanglement between hard and soft quanta, an inclusive approach gives an incomplete account, because these techniques discard information about the soft radiation present in a given state. The present work sets the stage for more detailed analysis of the quantum information content of soft radiation at null infinity, and the related phenomena of decoherence in scattering theory, at both finite and asymptotic times. Indeed, the latter topic is the focus of upcoming work~\cite{DSW_2026}.

\subsection{Decoherence by causal horizons }
\label{sec:causal}
While we have made a substantial effort in this paper to generalize our previous work \cite{DSW_2022,DSW_2023,DSW_2024,DKSW_2025}, our results still fall short of fully justifying the conclusions of the gedanken experiment presented in the Introduction. In particular, our analysis here has focused on the decoherence induced by Killing horizons in the vicinity of Alice’s laboratory. Nevertheless, the structure of the argument suggests a more general conclusion: namely, a universal bound on the coherence of Alice’s quantum superposition in the presence of any causal horizon.

To make this precise, let $\Gamma$ be a future- and past-inextendible timelike curve in a spacetime $\mathscr{M}$ representing the worldline of Alice’s laboratory. We assume that the proper time along $\Gamma$ extends to $-\infty$ as $\tau \to -\infty$ ensuring that Alice can, in principle, carry out her experiment for an arbitrarily long duration of proper time $T$. We do not, however, assume that her proper time similarly extends to $+\infty$ since $\Gamma$ may terminate at a singularity.

If $D(\Gamma)$ is the domain of communication of $\Gamma$ (i.e., the causal diamond associated with $\Gamma$) then Alice's worldline possesses a causal horizon if the complement $\mathscr{M}\setminus D(\Gamma)$ is nonempty. The future causal horizon $\mathscr{H}^{+}$ is the future boundary of $D(\Gamma)$. Suppose that the degrees of freedom beyond the horizon constitute the complete complementary output of Alice's channel and carry nonzero information, $\mathcal{I}(\mathcal{N}^c)>0$, about her experiment. The inaccessible region could then, in principle, contain ``Bob(s)'' who access that information. If, moreover, the complementary outputs form a nested cumulative family as $T$ increases, these hypothetical Bob(s) can accumulate a nondecreasing amount of information. By the arguments presented in the Introduction, Alice is commensurately decohered. Under this nesting assumption, her decoherence is therefore nondecreasing with $T$, and its effect is {\em as if} she is becoming entangled with degrees of freedom on the other side of the horizon. 

In all previous work \cite{Gralla_2024, Biggs_2024, Wilson-Gerow_2024, Li_2025, Li_2025b, Li_2025c, Kawamoto_2025} as well as the present paper thus far, $D(\Gamma)$ is the exterior of a stationary black hole, or the de Sitter static patch, and $\mathscr{H}^{+}$ is a Killing horizon. However, from the above arguments, we see that an analogous decoherence effect {can} occur in any spacetime with a causal horizon---e.g., Alice falling toward a black hole, or a big crunch cosmology. Indeed, the arguments and conclusions of Sec.~\ref{sec:IDT} can be straightforwardly generalized to this case. Assuming Bob can obtain any information whatsoever about Alice's state on his side of the horizon, so that $\mathcal{I}(\mathcal{N}^c)>0$, one again obtains an exact IDT for Alice's quantum channel $\mathcal{N}$ given by 
\begin{equation}
D(\mathcal{N}) = \mathcal{I}(\mathcal{N}^{c})>0,
\end{equation}
where the positivity follows in any situation in which Bob, on the other side of the horizon, can gain any information whatsoever about Alice's state. Additionally, under the same nesting assumption, Bob's accessible information, and hence her decoherence, cannot decrease in time.
However, a nonstationary spacetime need not admit preferred notions of ``photon number'' or global thermality, which played a crucial role in previous analyses of decoherence. Nevertheless, the causal and information-theoretic framework introduced here remains well-defined in this more general setting. We leave a detailed investigation of these issues to future work \cite{Nishkal_2025}.

\acknowledgments{
    We thank Jonah Kudler-Flam for crucial insights and initial collaboration. We also thank Stefan Hollands, Nicholas LaRacuente, Jonathan Sorce, Robert Wald, and Edward Witten for stimulating discussions. D.L.D. acknowledges support as a Black Hole Initiative Fellow, as a  Fannie and John
    Hertz Foundation Fellow holding the Barbara Ann Canavan Fellowship, as an Eckhardt Graduate Scholar
    at the University of
    Chicago, and as a Long-Term TURIS-IQOQI Fellow of the Austrian Academy of Sciences and the University of Vienna. D.L.D. acknowledges the Julian Schwinger Foundation for financial support at the 2025 Peyresq Spacetime Meeting, where useful discussions influenced this work. This work was performed in part at Aspen Center for Physics, which is supported by National Science Foundation (NSF) grant PHY-2210452.
    This research was supported in part by NSF Grant No. PHY-2403584 and John Templeton Foundation Grant No. 62845 to the University of Chicago.
    This publication is funded in part by Gordon and Betty Moore Foundation Grant No. 13526, and John Templeton Foundation Grant No. 63445. 
	}

\section*{DATA AVAILABILITY}
There are no publicly available research data or software supporting this manuscript. Requests for further information or data should be sent to the authors.

\appendix

\section{CHANNEL INFORMATION AND CHANNEL DECOHERENCE}
\label{app:IDT}
The primary purpose of this appendix is to elaborate on the definitions, properties, and utility of the channel decoherence and channel information of Sec. \ref{sec:IDT}. In Sec. \ref{app:IDTproof} of this appendix we establish the result of Eq. \eqref{eq:exactChannelIDT}: the exact information-decoherence trade-off of Sec. \ref{sec:IDT}.  In Sec. \ref{sec:consistency} we establish that the channel decoherence recovers a familiar notion of decoherence when applied to a simple experiment, and is therefore a generalization of the familiar notion of ``decoherence.'' Section \ref{app:remarks} highlights three desirable properties of the definition of channel decoherence: \ref{app:falseDeco}. its elimination of the problem of ``false decoherence'' \cite{Unruh_2000}, \ref{app:intrinsic}. its independence of a ``choice of state,'' which guarantees it reflects the intrinsic decoherence of the channel, and \ref{app:ancilla}. its stability under the introduction of an entangled, auxiliary system, due to extremization over all such systems. The validity and utility of the results of Sec. \ref{sec:IDT}, and the conclusions drawn from there, rest on the content of this appendix. Due to its rather technical nature, however, we have deferred this material until this point so as to preserve the clarity of the primary physics points of the paper.
\subsection{Proof of the equivalence between information capacity and complementary decoherence capacity}\label{app:IDTproof}
Quantum channels are the most general maps describing the physical evolution of a system. Here we assume, for the sake of pedagogy, that one has already obtained a Hilbert space representation. For the corresponding definitions working directly with algebraic states on the $C^*$ algebra of bounded operators, see \cite{stinespring1955positive}. One such map is the channel $\mathcal N$ taking Alice's subsystem between its initial and final states. If Alice's subsystem comprises operators on the Hilbert space $\mathcal{H}_A$, then
\begin{align}
    \begin{aligned}
        \label{eq:channel}
        \mathcal{N}: \mathcal{S}(\mathcal{H}_A) &\rightarrow \mathcal{S}(\mathcal{H}_A)
        \\
        \omega_A &\mapsto \Tr_{{\mathscr{H}^+}}\left[ \hat{U}^{\dagger}( \omega_A \otimes \omega_{QFT})\hat{U}\right].
    \end{aligned}
\end{align}
Here, $\mathcal{S}(\mathcal{H})$ is the space of quantum states on the Hilbert space $\mathcal{H}$, and $\hat U$ is the unitary evolution operator on the combined system of Alice's particle and the quantum field. The state $\omega_{QFT}$ is fixed, independent of $\omega_A$, and may be taken pure without loss of generality by including any purifying degrees of freedom in $\mathcal H_{\mathscr H^+}$; the partial trace is over this environment Hilbert space. Given a suitable Hilbert space representation, all quantum channels can be represented by this form, namely tensoring in an auxiliary state, global unitary evolution, and partial trace~\cite{stinespring1955positive}.

There is a unique (up to isometric equivalence) \textit{complementary channel}, $\mathcal{N}^c$, that performs the same evolution, but instead traces over the quantum state of Alice's system,
\begin{align}
    \begin{aligned}\label{eq:complementary}
        \mathcal{N}^c:\mathcal{S}(\mathcal{H}_A) &\rightarrow \mathcal{S}(\mathcal{H}_\mathscr{H^+})
        \\
        \omega_A &\mapsto \Tr_{A}\left[ \hat{U}^{\dagger}( \omega_A \otimes \omega_{QFT})\hat{U}\right].
    \end{aligned}
\end{align}
This is the quantum state available to Bob. {Unitary evolution constrains the information jointly available through $\mathcal N$ and $\mathcal N^c$.}

We recall the definition, introduced in Sec.~\ref{sec:IDT}, of  the intrinsic quantum decoherence of a quantum channel $\mathcal{N}$.
\begin{definition*}[Channel Decoherence]\label{eq:channelDecoherence2}

        {
        \begin{equation}
            D(\mathcal{N}) :=
            1- \frac{\left[\sup_{\mathcal{R}}F(\mathcal{R}\circ\mathcal{N},\operatorname{Id})-1/d\right]_+}{1-1/d}.
        \end{equation}
        }
    \end{definition*}

Here we have used the channel fidelity $F$ between quantum channels, which is defined in terms of the  fidelity $f_\text{U}$ as
\begin{align}
    F(\mathcal N,\mathcal M):=\inf_k\inf_\omega f_\text{U}\left((\mathrm{Id}_k\otimes\mathcal N)(\omega),(\mathrm{Id}_k\otimes\mathcal M)(\omega)\right),
    \label{eq:entFidelity2}
\end{align}
where $k$ is the dimension of an ancillary Hilbert space into which the state $\omega$ is purified. {Throughout this appendix, the channel fidelity $F$ is defined using the ``squared'' convention for the Uhlmann fidelity $f_\mathrm{U}$ of Eq.~(\ref{eq:uhlmannDefinition1}).}

The infimum over $k$ enforces that a perfect recovery channel must not only restore the initial reduced state, but must also restore its correlations with any external system, as implied by Eq.~(\ref{eq:noDecoherence}).

As a preliminary sign of the usefulness of the channel decoherence, note that when $D(\mathcal{N})$ is very small (say $\epsilon \ll 1$), this can be understood as an indicator of quantum coherence, because one may take $\omega$ to be \textit{any} initial state for Alice and the quantum channel $\mathcal{N}$ will be almost perfectly reversible via the decoding channel, the errors being of size $\epsilon$.

In Appendix Sec.~\ref{sec:consistency} we show that the definition of channel decoherence is consistent with the ordinary notion of decoherence of Eq. (\ref{eq:dAlice}). In Appendix Sec.~\ref{app:remarks} we remark on various desirable properties of the definition (\ref{eq:channelDecoherence}) such as its ability to distinguish false decoherence~\cite{Unruh_2000} from physical decoherence.

We recall the definition, also introduced in Sec.~\ref{sec:IDT}, of the information capacity of a channel $\mathcal N$.
\begin{definition*}[Channel Information]\label{def:chInfo2}
    {
    \begin{equation}
        \mathcal I(\mathcal N) :=
        1- \frac{\left[\mathscr L(\mathcal N)-1/d\right]_+}{1-1/d},
    \end{equation}
    }
\end{definition*}
where $\mathscr L(\mathcal N)$ is defined as
{
\begin{align}
    \mathscr{L}(\mathcal{N}):=\sup_{\chi}F(\mathcal{N},\mathcal{C}_{\chi}),
    \qquad \mathcal{C}_{\chi}(\omega):=\Tr(\omega)\chi .
\end{align}
}

We will show that this measure of information capacity is tightly related (in fact, equal to) to the complementary channel decoherence. First, we must establish an intermediate result, that the channel loss $\mathscr{L}(\mathcal{N})$ is equivalent to the channel fidelity between $\mathcal{N}$ and a ``discard-and-recover'' channel:
\begin{equation}\label{lem:infoEntFidelity}
    \mathscr{L}(\mathcal{N}) =\sup_\mathcal{R'}F(\mathcal N,\mathcal R'\circ\operatorname{Id}^c).
\end{equation}
To see this, note that an example of a channel complementary to $\operatorname{Id}$ is $\Tr$, whose target is one-dimensional. In this case the domain of $\mathcal R'$ will be one-dimensional, so it must map all inputs onto a fixed state, $\chi$. Thus,
{
\begin{equation}
    \sup_\mathcal{R'}F(\mathcal N,\mathcal R'\circ\operatorname{Id}^c)
    =\sup_{\chi}F(\mathcal N,\mathcal C_\chi)
    =\mathscr L(\mathcal N).
\end{equation}
}
\paragraph{Exact information-disturbance trade-off.}
Having given sensible measures of the decoherence capacity $D(\mathcal N)$ of a quantum channel as well as its information content $\mathcal{I}(\mathcal N)$ we apply B\'eny and Oreshkov's Theorem 1 \cite{Beny:2010byk}, which is a version of what is known as an information-disturbance trade-off between the information {conveyed} by the channel $\mathcal N^c$ to the environment (in our case, the horizon) and the disturbance the original channel $\mathcal N$ exerts on the system under consideration (in our case, Alice's particle). In general, we can also allow for another channel $\mathcal M$ to act on the system; since squaring is order preserving on $[0,1]$, the theorem remains valid for the squared fidelity used here. The \textit{information-disturbance trade-off} is
\begin{equation}
    \sup_{\mathcal R'}F(\mathcal N^c,\mathcal R'\circ\mathcal M^c)=	\sup_\mathcal RF(\mathcal R\circ \mathcal N, \mathcal M).
\end{equation}
Specializing to the case where $\mathcal M = \operatorname{Id}$, our desired result follows immediately from Eq.~(\ref{lem:infoEntFidelity}): \begin{equation}\label{thm:decoherenceInformation}
    D(\mathcal N) = \mathcal{I}(\mathcal{N}^c).
\end{equation}
Therefore, the channel decoherence is \textit{equal} to the channel information, as claimed in Eq. (\ref{eq:exactChannelIDT}).

\subsection{The channel decoherence generalizes an ordinary notion of decoherence}
\label{sec:consistency}
In the specific case of Alice's experiment, the generalized definition of decoherence $D(\mathcal N)$ {recovers} the familiar notion of decoherence, as we will see here.

{By joint concavity of the square root Uhlmann fidelity, the infimum in Eq.~(\ref{eq:entFidelity2}) may be taken over pure states on $\mathcal H_A\otimes\mathcal H_R$.}

For simplicity, let us assume that Alice's system consists of a spin. Since any state can be purified using an ancilla of at most the {same dimension}, the ancillary Hilbert space dimension $k$ need not be any larger than that of Alice's spin $\mathcal{H}_A$, which is the domain of $\mathcal N$ in this example. If $\mathcal H_R$ denotes the ancillary, or reservoir, Hilbert space, then a general pure state in $\mathcal H_A\otimes\mathcal H_R$ can be expressed in the $z$-spin basis as
\begin{align}\label{eq:optimalState}
    |\rho_{AR}\rangle=\sum_{ij} c_{ij}\ket{i}_A\ket{j}_R \quad \text{with } \sum_{ij}|c_{ij}|^2 = 1,
\end{align}
where $\ket{i} \in \{\ket{\psi_1,\downarrow}, \ket{\psi_2,\uparrow}\}$.
If $\mathcal N$ implements the effects of decoherence on Alice's superposition at asymptotically late times, then
\begin{align}
    \begin{aligned}
        &{(\operatorname{Id}_R\otimes\mathcal N)(|\rho_{AR}\rangle\langle\rho_{AR}|)}\\
        &\quad=\sum_{ijkl}c_{ij}c_{kl}^*\ket{i}\bra{k}_A
        \otimes \ket{j}\bra{l}_R \bra{\Psi_k}\Psi_i\rangle_{{\mathscr{H}^+}}.
    \end{aligned}
\end{align}
The state of Alice's spin and ancilla immediately after Alice completes her experiment is well-approximated by this asymptotic state.

If $\hat{K}_s$ are {Kraus} operators implementing the recovery channel $\mathcal{R}$, then
\begin{align}
    \begin{aligned}
        &{(\operatorname{Id}_R\otimes\mathcal R\circ\mathcal N)(|\rho_{AR}\rangle\langle\rho_{AR}|)}\\
        &=\sum_{ijkl}\sum_s c_{ij}c_{kl}^*\hat K_s\ket{i}\bra{k}_A\hat K_s^\dagger\\
        &\quad\otimes\ket{j}\bra{l}_R\langle\Psi_k|\Psi_i\rangle_{{\mathscr H}^+}.
    \end{aligned}
\end{align}

{For an input Schmidt diagonal in the $z$-spin basis,} one obtains
\begin{align}
    \begin{aligned}
        &{(\operatorname{Id}_R\otimes\mathcal R\circ\mathcal N)(|\rho_{AR}\rangle\langle\rho_{AR}|)}\\
        &=\sum_{ik}\sum_s c_{ii}c_{kk}\hat K_s\ket{i}\bra{k}_A\hat K_s^\dagger\\
        &\quad\otimes\ket{i}\bra{k}_R\langle\Psi_k|\Psi_i\rangle_{{\mathscr H}^+}.
    \end{aligned}
\end{align}
Here $c_{ii}\in\mathbb R$.

$\mathcal{R}$ maps Alice's Hilbert space to itself, so the Kraus matrices $\hat{K}_s$ are \textit{square} matrices.
    The corresponding Uhlmann fidelity is
    \begin{align}
        \begin{aligned}
            &{f_\mathrm U\!\left((\operatorname{Id}_R\otimes\mathcal R\circ\mathcal N)(|\rho_{AR}\rangle\langle\rho_{AR}|),|\rho_{AR}\rangle\langle\rho_{AR}|\right)}
            \\&=\langle \rho_{AR}|\left(\operatorname{Id}_R\otimes\mathcal{R}\circ \mathcal{N}\right)(|\rho_{AR}\rangle\langle\rho_{AR}|) |\rho_{AR}\rangle
            \\
            &={\sum_{i,k,s}c_{ii}^2c_{kk}^2 \langle i|\hat K_s|i\rangle\langle k|\hat K_s^\dagger|k\rangle_A\langle\Psi_k|\Psi_i\rangle_{{\mathscr H}^+}}.
        \end{aligned}
    \end{align}
    {
    Let $p:=|\mathscr P|$ and absorb the phase of $\mathscr P$ into a unitary recovery. Let $\op{X},\op{Y},\op{Z}$ denote the Pauli operators in the ordered basis
$\{|\psi_1,\downarrow\rangle,|\psi_2,\uparrow\rangle\}$. The induced channel is the qubit dephasing channel
    \begin{equation}
        \mathcal N_p(\rho)=q\rho+(1-q)\op{Z}\rho \op{Z},
        \qquad q:=\frac{1+p}{2}.
    \end{equation}
    For the identity recovery and any purification $|\rho\rangle$ of an input state $\rho$,
    \begin{equation}
        f_\mathrm{U}\!\left((\operatorname{Id}_R\otimes\mathcal N_p)(|\rho\rangle\langle\rho|),|\rho\rangle\langle\rho|\right)
        =q+(1-q)|\Tr(\rho \op{Z})|^2\ge q,
    \end{equation}
    with equality at $\rho=\mathbf 1/2$. Conversely, if $\{\hat{R}_s\}$ are Kraus operators for an arbitrary recovery and $\hat{R}_s=a_s\mathbf 1+b_s\op{Z}+c_s\op{X}+d_s\op{Y}$, then the maximally entangled input $\Phi$ gives
\begin{equation}\begin{split}f_\mathrm{U}\!\big((&\operatorname{Id}_R\otimes\mathcal R\circ\mathcal N_p)(\Phi),\Phi\big)
        \\&=q\sum_s|a_s|^2+(1-q)\sum_s|b_s|^2\le q,
    \end{split}\end{equation}
    where trace preservation implies $\sum_s(|a_s|^2+|b_s|^2+|c_s|^2+|d_s|^2)=1$. Since the channel fidelity minimizes over inputs, the identity recovery is optimal and
    \begin{align}
        \max_{\mathcal R}F(\mathcal R\circ\mathcal N,\operatorname{Id})
        &=\frac{1+|\mathscr P|}{2}, \notag\\
        D(\mathcal N)
        &=1-2\left[\frac{1+|\mathscr P|}{2}-\frac12\right]_+
        =1-|\mathscr P|,
    \end{align}
    }
    where $\mathscr{P}:={\langle\Psi_1|\Psi_2\rangle_{{\mathscr H}^+}}$.
    This agrees with the ordinary notion of decoherence $D_\text{Alice}|_c = 1-|{\langle\Psi_1|\Psi_2\rangle_{{\mathscr{H}^+}}}|$ and of Eq.~(\ref{eq:dAlice}) that is applied in, e.g., \cite{DSW_2021, DSW_2022}. Thus, we see that $D(\mathcal{N})$ is a generalization of the ordinary notion of decoherence to general quantum channels. This also agrees with the long-time limit of $\mathcal{I}_\text{Bob} = 1-\sqrt{f_\text{U}}(\Psi_1,\Psi_2) = 1-|\langle\Psi_1|\Psi_2\rangle|$, as anticipated by Eq.~(\ref{thm:decoherenceInformation}).

\subsection{Remarks on channel decoherence}\label{app:remarks}
In Sec.~\ref{sec:IDT} we treat Alice’s exterior experiment (and its horizon generalization) as a quantum channel
$\mathcal{N}$ on her laboratory degrees of freedom, with the radiative data on $\mathscr{H}^+\union\scri^+$ playing the
role of an environment. The channel decoherence $D(\mathcal{N})$, defined in
Eqs.~(\ref{eq:entFidelity2})--(\ref{eq:channelDecoherence2}), is the operational measure of \emph{irreversible} loss of
coherence that enters the exact information-decoherence trade-off \eqref{eq:exactChannelIDT} and, through it, the horizon
monotonicity statements of Sec.~\ref{sec:secondLawforHorizons}. Since this notion is deliberately more refined than the
na\"ive “trace over the field and read off off-diagonals'' criterion---especially at finite times, where one encounters
Unruh’s false decoherence---we record here a few remarks clarifying the roles of the decoding optimization, the
extremization over input states, and the ancillary reference system that stabilizes the definition.\paragraph{Good decoding channels eliminate false decoherence.}
\label{app:falseDeco}
The definition of channel decoherence quantifies the irreversible decoherence incurred by a quantum channel. Any disturbance of the quantum state which can be reversed purely by manipulating the output of the quantum channel (e.g., by operations performable entirely in Alice's own laboratory) is automatically corrected by the application of the recovery channel $\mathcal{R}$. In the preceding example, the decoding channel had the effect of correcting the phases of Alice's superposition.

The inclusion of a decoding channel in the definition of channel decoherence has the effect of eliminating the problem of false decoherence, as discussed in \cite{Unruh_2000}. Suppose, for instance, Alice were to create a superposition and hold it open for some time, but \textit{not} recombine its spatial branches. Formally, the joint state of her particle and the sourced field are
\begin{equation}
	\frac{1}{\sqrt{2}}\left(|\psi_1,\downarrow\rangle\otimes|\Psi_1\rangle+|\psi_2,\uparrow\rangle\otimes|\Psi_2\rangle\right),
\end{equation}
where $\psi_1\ne\psi_2$. Due to the distinct Coulomb fields of the left and right branches of the superposition, this state is totally decohered.\footnote{As given, the state is merely a formal expression, because the ``Coulomb subtraction'' procedure of \cite{DSW_2023} cannot be defined for a charge in superposition. However, following \cite{Unruh_2000}, the field can be given a small mass to make the analysis precise.} This is, however, false decoherence, because it can be trivially reversed by Alice. The optimization over decoding channels in Definition~\ref{eq:channelDecoherence2} eliminates such false decoherence: in this case, the optimal decoding channel would involve Alice adiabatically recombining the branches of her spatial superposition, thus recombining the branches of the Coulomb field.

\paragraph{Infimum over initial states.}\label{app:intrinsic} A key change in perspective from \cite{DSW_2022,
DSW_2023} is that we are characterizing decoherence as a property of the quantum channel, independent of the actual state of Alice. Since we will quantify coherence in terms of the recoverability of Alice's initial state, we must be careful to avoid any definition that trivializes the recovery process. The following example illustrates the type of ``trivial case'' any sensible definition must discount.

If Alice's objective is merely to recover Alice's initial state $\omega_A$ after performing her experiment, then regardless of the experiment that she does, there will always exist a decoding channel that achieves this:
\begin{align}
	\begin{aligned}
		\mathcal{R}: ~&\mathcal{S}(\mathcal{H}_A ) \rightarrow \mathcal{S}(\mathcal{H}_A )
		\\
		&\omega'_{A} \mapsto \Tr\left[ \omega'_{A} \right] \omega_A.
	\end{aligned}
\end{align}
It is straightforward to check that this linear map is both completely positive and trace preserving. The result is that $f_\mathrm{U}(\mathcal{R} \circ \mathcal{N}(\omega_A),\omega_A) = 1$. Of course, we should not conclude that $\mathcal{N}$ cannot decohere Alice's experiment, because this decoding channel requires Alice to know in advance the state of the system that she is to recover.
Had Alice begun with a different state, $\chi_A\ne\omega_A$, then the above decoding channel would be catastrophic. It is for this reason that the channel fidelity of Eq. (\ref{eq:entFidelity2}) takes an infimum over initial states: to ensure that it defines a fidelity intrinsic to the channel, and not merely a characterization of its performance on one highly specialized state. Thus, our decoherence measure provides a measure of unavoidable decoherence due to the channel itself, and not merely due to a poor choice of initial state.

\paragraph{The need for the ancillary reference system.}\label{app:ancilla}
What motivates the inclusion of an ancillary identity channel in Eq.~(\ref{eq:entFidelity2})? This serves to stabilize the norm, as the usual one-norm generally may increase even when acting trivially on the $k$ ancillary degrees of freedom---a peculiar consequence of entanglement. To properly characterize the channel decoherence when acting on mixed states, we must consider how the channel acts on the purifications of these states. To illustrate this point, we consider a simple gedanken experiment involving an Einstein--Podolsky--Rosen (EPR) pair.

Suppose that Alice begins with one half of an EPR pair, the other half being held by Bob in his respective laboratory. EPR pairs are valuable resources, allowing for useful protocols such as quantum teleportation, so Alice wants to maintain the coherence of the joint state of her particle and Bob's so that she retains this valuable resource at later times. Alice performs her Stern-Gerlach experiment while Bob does nothing at all. If the quantum channel irreversibly ruins the state of the EPR pair, even when acting trivially on Bob, we must consider this to be a form of decoherence. Only when an ancillary reference system is included, can the channel decoherence properly account for this form of decoherence, in which entanglement is degraded even while the reduced density matrix of Alice's subsystem is preserved.

\section{THE HOLEVO FIDELITY}\label{app:holevoFidelity}
The purpose of this appendix is to establish bounds, from above and below, between the Uhlmann fidelity and the Holevo fidelity between two quantum states. Since both fidelities are used in the article, it is important to know how they relate to one another. The most important result of this appendix is to show that the Holevo fidelity can be used as an approximant for the Uhlmann fidelity. This is useful in part because the Uhlmann fidelity has a clear operational meaning in terms of unambiguous state discrimination, as reviewed in Sec. \ref{sec:IDT}, while the Holevo fidelity is easily computed, using the methods of Tomita-Takesaki theory.  The Uhlmann fidelity, on the other hand, is generally quite a difficult quantity to evaluate.

The Holevo fidelity is defined in Tomita-Takesaki theory as
\begin{equation}
    \sqrt{f_\mathrm{H}}(\rho,\sigma):=\langle \rho|\hat\Delta^{1/2}_{\rho\to\sigma}|\rho\rangle,
\end{equation}
where $|\rho\rangle$ is the $CRT$ (i.e., natural cone) purification of the algebraic state $\rho$. When $\rho$ and $\sigma$ can be represented as density matrices this reduces to
\begin{align}
    \sqrt{f_\mathrm{H}}(\rho, \sigma) = \Tr \left[ \sqrt{\rho}\sqrt{\sigma}\right].
\end{align}

The Holevo fidelity is bounded by the Uhlmann fidelity,
\begin{align}
    f_\mathrm{H}(\rho,\sigma) \leq f_\mathrm{U}(\rho,\sigma),
    \label{eq:holevoBound}
\end{align}
{where equality holds, in particular, when the two states commute.}

Both the Uhlmann and Holevo fidelities satisfy respective Fuchs--Van deGraaf--type inequalities \cite{fuchsVanDegraf, kholevo1972quasiequivalence},
\begin{equation}
    1-\sqrt{f}(\rho,\sigma)\le\frac{1}{2}\lVert\rho-\sigma\rVert_1\le \sqrt{1-f(\rho,\sigma)}.
    \label{eq:fuchsVanDeGraaf}
\end{equation}
Taken together, Eqs. (\ref{eq:holevoBound}) and (\ref{eq:fuchsVanDeGraaf}) imply
\begin{align}
    \begin{aligned}
        1-\sqrt{f_\mathrm{U}}(\rho,\sigma)\le 1-\sqrt{f_\mathrm{H}}(\rho,\sigma)\le \frac{1}{2}\lVert\rho-\sigma\rVert_1\\\le \sqrt{1-f_\mathrm{U}(\rho,\sigma)}\le \sqrt{1-f_\mathrm{H}(\rho,\sigma)},
    \end{aligned}
\end{align}
which includes a bound between the Uhlmann and Holevo fidelities,
\begin{equation}\label{eq:uhlmannHolevoBoundAppendix}
    1-\sqrt{f_\mathrm{H}}(\rho,\sigma)\le \sqrt{1-f_\mathrm{U}(\rho,\sigma)}\le \sqrt{1-f_\mathrm{H}(\rho,\sigma)}.
\end{equation}

\section{APPROXIMATION OF THE CHANNEL DECOHERENCE}
\label{sec:approximateCapacity}
	Here we compute bounds on the channel decoherence $D(\mathcal{N})$, or equivalently the channel information, which are more readily computable in general situations than the channel decoherence itself. In particular, given any two states in the range of the complementary channel, such as might arise in Alice's experiment from Bob's point of view, we give a relatively simple lower bound on the decoherence.

    The computation relies on a number of bounds and technical arguments, which are not essential to the main arguments of the paper, so we have deferred their presentation until now. Nonetheless, the final bound is important for establishing the computational utility of the channel decoherence, beyond its utility as an operationally useful quantity and its satisfaction of the exact information-decoherence trade-off.  

{Choosing an idempotent complement $\operatorname{Id}^c$ of $\operatorname{Id}$,} Bény and Oreshkov prove \cite{Beny:2010byk} that
\begin{equation}\begin{split}
 \frac{1}{2}\sqrt{1-\sqrt{{F}(\mathcal N^c,\mathcal N^c\circ\operatorname{Id}^c)}}&\le
 \sqrt{1-\sqrt{\sup_\mathcal R{F}(\mathcal R\circ\mathcal N,\operatorname{Id})}}\\
 &\le\sqrt{1-\sqrt{{F}(\mathcal N^c,\mathcal N^c\circ\operatorname{Id}^c)}}.
\end{split}\end{equation}

Because $F(\mathcal N^c,\mathcal N^c\circ\operatorname{Id}^c)\le \sup_{\mathcal R}F(\mathcal R\circ\mathcal N,\operatorname{Id})$ and ${F}\le\sqrt{{F}}\le1$, this gives rise to a weaker form of information-disturbance trade-off (IDT) reminiscent of the diamond norm IDT of \cite{kretschmann_information-disturbance_2006},\footnote{Explicitly, the diamond-norm IDT is $\frac{1}{4}\inf_{\mathcal{R}}\parallel \mathcal{R} \circ \mathcal{N} - \text{Id}\parallel_{\diamond}^2 \leq \parallel \mathcal{N}^c -\Phi\parallel_{\diamond}
    \leq 2\inf_{\mathcal{R}}\parallel \mathcal{R}\circ \mathcal{N} -\text{Id}\parallel_{\diamond}^{1/2}
$, where $\lVert \,\cdot\,\rVert_\diamond$ denotes the diamond norm of \cite{kretschmann_information-disturbance_2006} and the channel $\Phi$ takes any input state and replaces it with some fixed state.}
\begin{equation}\begin{split}
 \frac{1}{2}\sqrt{1-\sqrt{{F}(\mathcal N^c,\mathcal N^c\circ\operatorname{Id}^c)}}&\le
 \sqrt{1-\sup_\mathcal R{F}(\mathcal R\circ\mathcal N,\operatorname{Id})}\\
 &\le\sqrt{1-{F}(\mathcal N^c,\mathcal N^c\circ\operatorname{Id}^c)}.
\end{split}\end{equation}
This can be reexpressed in terms of the channel decoherence,
{
\begin{equation}\begin{split}\label{eq:approximationIDT}
 \frac{1}{4}\left[1-\sqrt{{F}(\mathcal N^c,\mathcal N^c\circ\operatorname{Id}^c)}\right]&\le (1-1/d)D(\mathcal N)\\
 \le\min\!\bigl\{1-1/d,&\,1-{F}\bigl(\mathcal N^c,\mathcal N^c\circ\operatorname{Id}^c\bigr)\bigr\},
\end{split}\end{equation}
}
where $d$ is the dimension of the {input Hilbert space} on which $\mathcal{N}$ acts (with $1/d:=0$ for an infinite-dimensional input space). Thus, we see that the channel decoherence can be approximated without solving the supremum-infimum optimization problem required to evaluate $D(\mathcal{N})$ directly.

{
The preceding inequalities are extremely general in that they hold for arbitrary quantum channels. We now obtain a direct lower bound whenever two states in the range of the complementary channel are known. Let $\omega_i=\mathcal N^c(\rho_i)$, $i=1,2$, for two input states $\rho_1$ and $\rho_2$. The Bures angle between two states is
\begin{equation}
\theta_\mathrm{B}(\rho,\sigma):=\arccos\sqrt{f_\mathrm{U}(\rho,\sigma)}.
\end{equation}
It obeys the triangle inequality. Thus, for any state $\chi$,
\begin{equation}\begin{split}
\theta_\mathrm{B}(\omega_1,\omega_2)
&\le \theta_\mathrm{B}(\omega_1,\chi)
+\theta_\mathrm{B}(\chi,\omega_2),\\
\max_{i=1,2}\theta_\mathrm{B}(\omega_i,\chi)
&\ge \frac12\theta_\mathrm{B}(\omega_1,\omega_2).
\end{split}\end{equation}
Since $\cos\theta$ is decreasing for $0\le\theta\le\pi/2$,
\begin{equation}\begin{split}
\min_{i=1,2}f_\mathrm{U}(\omega_i,\chi)
&\le \cos^2\!\left[\frac12\theta_\mathrm{B}(\omega_1,\omega_2)\right]\\
&=\frac{1+\sqrt{f_\mathrm{U}(\omega_1,\omega_2)}}{2}.
\end{split}\end{equation}
The defining infimum in the channel fidelity may be evaluated on either input state $\rho_i$, so
\begin{equation}\begin{split}
\mathscr L(\mathcal N^c)
&\le \sup_\chi\min_{i=1,2}f_\mathrm{U}(\omega_i,\chi)\\
&\le \frac{1+\sqrt{f_\mathrm{U}(\omega_1,\omega_2)}}{2}.
\end{split}\end{equation}
Using Eq.~(\ref{eq:exactChannelIDT}) and the definition of the channel decoherence, for $d>1$ we obtain
\begin{equation}\begin{split}
(1-1/d)D(\mathcal N)
&=1-1/d-\left[\mathscr L(\mathcal N^c)-1/d\right]_+\\
&\ge 1-1/d-\left[\frac{1+\sqrt{f_\mathrm{U}(\omega_1,\omega_2)}}{2}-1/d\right]_+\\
&=\frac12\left[1-\sqrt{f_\mathrm{U}(\omega_1,\omega_2)}\right].
\end{split}\end{equation}
Applying this to the particular case of Sec.~\ref{sec:exactChannelIDT} in which the channel $\mathcal{N}^{\mathscr{H}^+}_\text{Alice}(t)$ models the influence of the horizon on Alice's quantum particle, taking $\omega_1=\Psi_1|_c$ and $\omega_2=\Psi_2|_c$ gives
\begin{equation}\begin{split}
D(\mathcal N)
&\ge 1-\sqrt{f_\mathrm{U}(\Psi_1|_c,\Psi_2|_c)}\\
&=D_\mathrm{Alice}|_c.
\end{split}\end{equation}
Because the Holevo fidelity is often easier to compute explicitly than the Uhlmann fidelity, it is useful to apply Raggio's inequality \cite{Raggio_1982}, which in the squared conventions used here is
\begin{equation}
f_\mathrm{H}(\rho,\sigma)\le f_\mathrm{U}(\rho,\sigma)\le \sqrt{f_\mathrm{H}(\rho,\sigma)}.
\end{equation}
Therefore, for Alice's experiment we obtain the less tight but more readily computable bound
\begin{equation}
D(\mathcal N)\ge 1-f_\mathrm{H}({\Psi_1}|_c,{\Psi_2}|_c)^{1/4}.
\end{equation}
More generally, for any two states $\omega_1,\omega_2$ in the range of $\mathcal N^c$,
\begin{equation}
\frac12\left[1-f_\mathrm{H}(\omega_1,\omega_2)^{1/4}\right]
\le (1-1/d)D(\mathcal N).
\end{equation}
}

\section{R\'ENYI ENTROPIES FOR GENERAL COHERENT STATES}\label{app:renyi}
In Sec. \ref{sec:relEntropy} we apply the {infrared} extension of Tomita-Takesaki theory to compute the relative entropy of general coherent states. We make no assumption about whether the field configurations vanish at the cut demarcating the restriction of the state, nor do we assume that the field configurations are zero-memory configurations that decay asymptotically. We include this for completeness, but defer it until this point because it does not have direct bearing on the main physics content of the article, which pertains more to the fidelities and the relative entropy.

We proceed by considering any two coherent states, {expressed} as states on the horizon $\mathcal{H}^+$, $|\Psi\rangle = \op{U}(\phi^\Psi)|\Omega\rangle$ and $|\Phi\rangle = \op{U}(\phi^\Phi)|\Omega\rangle$, and their restrictions to the past of the cut $\mathscr{C}$, namely $\Psi|_c$ and $\Phi|_c$.
Define
\begin{equation}
    \phi^{\Psi\to\Phi}_B := \Pi_c\phi^\Psi_B-\Pi_c\phi^\Phi_B.
\end{equation}
Then, {for $0<\alpha<1$,} the order $\alpha$ {Petz} R\'enyi entropy of $\Psi$ relative to $\Phi$ is
{
\begin{equation}\begin{split}
        {S}^\alpha_{\Psi|\Phi} = \frac{1}{\alpha-1}&\ln\!\left\langle \Psi\left|\hat\Delta_{\Psi\to\Phi}^{1-\alpha}\right|\Psi\right\rangle
        \\=\frac{1}{\alpha-1}&\ln\!\left\langle\op{U}(\phi^{\Psi\to\Phi})^*\hat\Delta_\Omega^{1-\alpha}
        \op{U}(\phi^{\Psi\to\Phi})\right\rangle_\Omega\\
        =\frac{1}{\alpha-1}&\bigg[
        \frac{-1}{2}\lVert \phi^{\Psi\to\Phi}-e^{i2\pi\alpha}\Lambda_*^{i2\pi\alpha}\phi^{\Psi\to\Phi}\rVert^2
        \\&+\frac{i}{2}e^{i2\pi\alpha}{\mathscr W}(\phi^{\Psi\to\Phi},\Lambda_*^{i2\pi\alpha}\phi^{\Psi\to\Phi})\bigg].
\end{split}\end{equation}
}
As was the case for the relative entropy, this formula applies regardless of whether the coherent states lie in the easy or hard cases of Casini \textit{et al.} This formula is consistent with that of \cite{Frob:2024ijk}, with its validity now established in the case that the coherent states contain soft radiation, so the classical solutions do not decay asymptotically.

\section{HILBERT SPACE REPRESENTATION}
\label{app:memoryHilbert}
The discussion of Sec.~\ref{sec:horizonAlgebra} focused on an extension of the algebra and an accompanying extension of the Poincar\'e-invariant vacuum state, as an algebraic state. That gave a Hilbert space by the GNS construction, on which our Tomita-Takesaki theory acts. We arrived at this construction in the process of answering some practical questions in quantum field theory, but an equivalent Hilbert space has, in fact, arisen under various guises in the prior literature \cite{PSW_2022, Kibble_1968}. Here we discuss the features of this Hilbert space representation. The body of the article took an algebraic viewpoint, with the Hilbert space emerging as an immediate by-product (via GNS) after the Poincar\'e-invariant vacuum extends onto the full horizon algebra. Many of the results in the article can also be understood by understanding the properties of the resulting Hilbert space, so we discuss the definition and properties of that Hilbert space here, to provide a complementary perspective on the construction of Sec. \ref{sec:horizonAlgebra}.

The Hilbert space $\mathcal{H}$ is nothing more than the direct sum over definite-memory Fock spaces. Given a Fock space $\mathcal{F}_{0}$ of radiation on the horizon, one obtains a unitarily inequivalent Fock space $\mathcal{F}_\Delta$ by the pullback of a state in $\mathcal{F}_0$ along the algebra automorphism \cite{ashtekar1987asymptotic} $\op{E}(\phi)\mapsto \op{E}(\phi')+\phi_\Delta(\phi')\mathbf{1} $ (and the corresponding GNS construction of a new Hilbert space), where $(\phi_\Delta)_A(V,x^A)$ is a classical solution with memory, and $\phi'_A$ is a test function on the horizon.

One can then consider the direct sum over all such so-called ``memory representations,''
\begin{equation}
    \mathcal{H}:=\bigoplus_\Delta \mathcal{F}_{\Delta},
\end{equation}
where the sum runs over all allowed memory profiles. Our enlarged algebra is naturally represented on this Hilbert space, and the state $\omega_\Omega$ corresponds to the Poincar\'e invariant vacuum state in the $\Delta=0$ term in the direct sum.

Clearly this Hilbert space is not separable: for each memory profile $\Delta_B$ there is a normalized vacuum vector $\Omega_\Delta \in \mathcal{F}_\Delta$, and vectors from different memory sectors are mutually orthogonal. Since the set of allowed profiles $\Delta_B(x^A)$ (e.g., smooth vector fields on $\mathbb{S}^2$) is uncountable, this gives an uncountable orthonormal family in $\mathcal{H}$, which is impossible in a separable Hilbert space. Separability aside, as we have seen, this space is very useful for information-theoretic computations. Due to the direct sum structure, the group action that permutes the memory representations is not continuous in the Hilbert space topology. Thus, the Hilbert space does not carry a representation of the infinitesimal generators of the memory-changing unitaries. This is another way to understand the final discussion of Sec.~\ref{sec:vacuumState}. In some physically realistic scenarios it may be useful to include such infinitesimal generators, and in that case, one may wish to consider a direct integral representation such as those constructed in \cite{Prabhu_2024, Prabhu_2024b, Herdegen_1997, Kudler-Flam_2025}. In Sec.~\ref{sec:vacuumState}, the difficulty in representing such generators arose due to the inability to define a suitable action of the Poincar\'e-invariant vacuum on such operators. In a direct integral representation, one sacrifices the existence of a normalizable Poincar\'e-invariant vacuum, in exchange for the representation of such operators.

\bibliography{main}

@article{Sion_1958,
  author  = {Sion, Maurice},
  title   = {On general minimax theorems},
  journal = {Pacific Journal of Mathematics},
  volume  = {8},
  number  = {1},
  pages   = {171--176},
  year    = {1958},
  doi     = {10.2140/pjm.1958.8.171}
}

@book{Takesaki_1970,
    author = "Takesaki, M.",
    title = "{Tomita's Theory of Modular Hilbert Algebras and its Applications}",
    doi = "10.1007/bfb0065832",
    publisher = "Springer-Verlag",
    series = "Lecture Notes in Mathematics",
    year = "1970",
    address = {Berlin},
}

@article{Raggio_1982,
  author = {Raggio, Guido A.},
  title = {Comparison of Uhlmann's transition probability with the one induced by the natural positive cone of von Neumann algebras in standard form},
  journal = {Letters in Mathematical Physics},
  volume = {6},
  number = {3},
  pages = {233--236},
  year = {1982},
  doi = {10.1007/BF00403277}
}

@article{flanagan_2023,
  author         = "Flanagan, {\'E}anna {\'E}. and Shehzad, I. Zayn",
  title          = "{The classical dynamics of gauge theories in the deep infrared}",
  journal        = {J. High Energy Phys.},
  year           = "2023",
  volume         = "05",
  pages          = "185",
  doi            = "10.1007/JHEP05(2023)185",
    number = {2023},
}

@article{himwich_2020,
  author         = "Himwich, Elizabeth and Narayanan, Sruthi A. and
                    Pate, Monica and Paul, Nisarga and Strominger, Andrew",
  title          = "{The Soft S-Matrix in Gravity}",
  journal        = {J. High Energy Phys.},
  volume         = "09",
  year           = "2020",
  pages          = "129",
    number = {2020},
    doi = {10.1007/JHEP09(2020)129},
}

@phdthesis{Danielson_thesis,
	author = {Daine Lee Danielson},
	month = {August},
	school = {The University of Chicago},
	title = {Gravitationally Mediated Entanglement and Decoherence},
	year = {2025},
    url = {http://knowledge.uchicago.edu/record/15737},
    doi = {10.6082/uchicago.15737}
}

@article{Hollands_2023,
    author = "Hollands, Stefan and Ranallo, Alessio",
    title = "{Channel Divergences and Complexity in Algebraic QFT}",
    doi = "10.1007/s00220-023-04855-x",
    journal = "Commun. Math. Phys.",
    volume = "404",
    number = "2",
    pages = "927--962",
    year = "2023"
}

@article{Eldar_2003,
	author = {Eldar, Y.C.},
	doi = {10.1109/TIT.2002.807291},
	journal = {IEEE Transactions on Information Theory},
	number = {2},
	pages = {446-456},
	title = {A semidefinite programming approach to optimal unambiguous discrimination of quantum states},
	volume = {49},
	year = {2003}}

@article{KW_1991,
	author = {Kay, Bernard S. and Wald, Robert M.},
	doi = {10.1016/0370-1573(91)90015-E},
	journal = {Phys. Rep.},
	pages = {49--136},
	reportnumber = {PRINT-88-0840 (CHICAGO)},
	title = {{Theorems on the Uniqueness and Thermal Properties of Stationary, Nonsingular, Quasifree States on Space-Times with a Bifurcate Killing Horizon}},
	volume = {207},
	year = {1991}}

@article{H_2019,
    author = "Hollands, Stefan and Wald, Robert M. and Zahn, Jochen",
    title = {{Quantum instability of the Cauchy horizon in Reissner{\textendash}Nordstr{\"o}m{\textendash}deSitter spacetime}},
    doi = "10.1088/1361-6382/ab8052",
    journal = {Classical Quantum Gravity},
    volume = "37",
    number = "11",
    pages = "115009",
    year = "2020"
}

@article{KP_2025,
    author = "Kudler-Flam, Jonah and Penington, Geoff",
    title = "{It costs nothing to teleport information into a black hole}",
    year = "2025",
    journal = {Int. J. Mod. Phys. D},
    volume = {34},
    pages = {2543002},
    doi = {10.1142/S0218271825430023},
}

@article{Casini_2019,
	adsurl = {https://ui.adsabs.harvard.edu/abs/2019PhRvD..99l5020C},
	author = {{Casini}, Horacio and {Grillo}, Sergio and {Pontello}, Diego},
	doi = {10.1103/PhysRevD.99.125020},
	eid = {125020},
	journal = {\prd},
	month = jun,
	number = {12},
	pages = {125020},
	title = {{Relative entropy for coherent states from Araki formula}},
	volume = {99},
	year = 2019}

@book{ashtekar1987asymptotic,
	author = {Ashtekar, Abhay},
	publisher = {Bibliopolis},
	title = {Asymptotic quantization: based on 1984 Naples lectures},
	volume = {2},
	year = {1987},
    address = {Naples},
}

@unpublished{Nishkal_2025,
       author = {{Rao}, Nishtal and  {Satishchandran},{{Gautam}}},
 title = "{Causal Horizons Decohere Quantum Superpositions}",
note = {to be published}
}

@article{Satishchandran:2025cfk,
    author = "Satishchandran, Gautam",
    title = "{Black Holes, Entanglement and Decoherence}",
    year = {2026},
    journal = {J. Phys. Conf. Ser.},
    volume = {3177},
    pages = {012002},
    doi = {10.1088/1742-6596/3177/1/012002},
}

@article{Kapec_2017,
  author        = {Kapec, Daniel and Lysov, Vyacheslav and Strominger, Andrew},
  title         = {Asymptotic Symmetries of Massless QED in Even Dimensions},
  journal       = {Adv. Theor. Math. Phys.},
  volume        = {21},
  number        = {7},
  pages         = {1747--1767},
  year          = {2017},
  doi           = {10.4310/ATMP.2017.v21.n7.a6},
}

@article{Longo_2019,
	adsurl = {https://ui.adsabs.harvard.edu/abs/2019LMaPh.109.2587L},
	author = {{Longo}, Roberto},
	doi = {10.1007/s11005-019-01196-6},
	journal = {Lett. Math. Phys.},
	month = dec,
	number = {12},
	pages = {2587-2600},
	title = {{Entropy of coherent excitations}},
	volume = {109},
	year = 2019}

@article{Kawamoto_2025,
  author        = {Kawamoto, Shoichi and Lee, Da-Shin and Yeh, Chen-Pin},
  title         = {Decoherence by black holes via holography},
  year          = {2026},
  doi           = {10.1007/JHEP01(2026)154},
    journal = {J. High Energy Phys.},
    volume = {01},
    number = {2026},
    pages = {154},
}

@article{Li_2025,
  author        = {Li, Ran},
  title         = {Decoherence of quantum superpositions by Reissner--Nordstr{\"o}m black holes},
  journal       = {Phys. Rev. D},
  volume        = {111},
  pages         = {024040},
  year          = {2025},
  doi           = {10.1103/PhysRevD.111.024040},
}

@article{Li_2025c,
  author        = {Li, Ran and Man, Zhong-Xiao and Wang, Jin},
  title         = {Decoherence of quantum superpositions in near-extremal Reissner--Nordstr{\"o}m black holes with quantum gravity corrections},
  journal       = {J. High Energy Phys.},
  volume        = {08},
  pages         = {079},
  year          = {2025},
  doi           = {10.1007/JHEP08(2025)079},
    number = {2025},
}

@article{Bieri_2013,
  author        = {Bieri, Lydia and Garfinkle, David},
  title         = {An electromagnetic analog of gravitational wave memory},
  journal       = {Classical Quantum Gravity},
  volume        = {30},
  pages         = {195009},
  year          = {2013},
  doi           = {10.1088/0264-9381/30/19/195009},
}

@article{Li_2025b,
  author        = {Li, Ran},
  title         = {Note on the local calculation of decoherence of quantum superpositions in de Sitter spacetime},
  journal       = {Phys. Rev. D},
  volume        = {111},
  pages         = {044022},
  year          = {2025},
  doi           = {10.1103/PhysRevD.111.044022},
}

@misc{Biggs_2024,
  author        = {Biggs, Anna and Maldacena, Juan},
  title         = {Comparing the decoherence effects due to black holes versus ordinary matter},
  eprint        = {2405.02227},
  archivePrefix = {arXiv},
  primaryClass  = {hep-th},
  doi           = {10.48550/arXiv.2405.02227}
}

@article{Wilson-Gerow_2024,
  author        = {{Wilson-Gerow}, Jordan and Dugad, Annika and Chen, Yanbei},
  title         = {Decoherence by warm horizons},
  journal       = {Phys. Rev. D},
  volume        = {110},
  pages         = {045002},
  year          = {2024},
  doi           = {10.1103/PhysRevD.110.045002},
}

@article{Gralla_2024,
  author        = {Gralla, Samuel E. and Wei, Hongji},
  title         = {Decoherence from Horizons: General Formulation and Rotating Black Holes},
  journal       = {Phys. Rev. D},
  volume        = {109},
  number        = {6},
  pages         = {065031},
  year          = {2024},
  doi           = {10.1103/PhysRevD.109.065031},
}

@article{helstrom1969quantum,
	author = {Helstrom, Carl W},
	journal = {Journal of Statistical Physics},
	pages = {231--252},
	publisher = {Springer},
	title = {Quantum detection and estimation theory},
	volume = {1},
	year = {1969},
    doi = {10.1007/BF01007479},
}

@misc{1996quant.ph.11010F,
	adsurl = {https://ui.adsabs.harvard.edu/abs/1996quant.ph.11010F},
	archiveprefix = {arXiv},
	author = {{Fuchs}, Christopher A.},
	doi = {10.48550/arXiv.quant-ph/9611010},
	eprint = {quant-ph/9611010},
	primaryclass = {quant-ph},
	title = {{Information Gain vs. State Disturbance in Quantum Theory}}}

@article{Uhlmann_1976,
	author = {Uhlmann, Armin},
	journal = {Reports on Mathematical Physics},
	number = {2},
	pages = {273--279},
	publisher = {Elsevier},
	title = {The ``transition probability'' in the state space of a *-algebra},
	volume = {9},
	year = {1976},
    doi = {10.1016/0034-4877(76)90060-4},
}

@article{fuchsVanDegraf,
    author = "Fuchs, Christopher A. and van de Graaf, Jeroen",
    title = "{Cryptographic distinguishability measures for quantum-mechanical states}",
    doi = "10.1109/18.761271",
    journal = "IEEE Trans. Info. Theor.",
    volume = "45",
    number = "4",
    pages = "1216--1227",
    year = "1999"
}

@article{kholevo1972quasiequivalence,
	author = {Kholevo, AS},
	journal = {Theoretical and Mathematical Physics},
	number = {2},
	pages = {1071--1082},
	publisher = {Springer},
	title = {On quasiequivalence of locally normal states},
	volume = {13},
	year = {1972},
    doi = {10.1007/BF01035528},
}

@article{kretschmann_information-disturbance_2006,
    author = "Kretschmann, Dennis and Schlingemann, Dirk and Werner, Reinhard F.",
    title = "{The Information-Disturbance Tradeoff and the Continuity of Stinespring's Representation}",
    doi = "10.1109/TIT.2008.917696",
    journal = "IEEE Trans. Info. Theor.",
    volume = "54",
    number = "4",
    pages = "1708--1717",
    year = "2008",
}

@misc{ksw_channelFidelity_2007,
    author = "Kretschmann, Dennis and Schlingemann, Dirk and Werner, Reinhard F.",
    title = "{A Continuity Theorem for Stinespring's Dilation}",
    eprint = "0710.2495",
    archivePrefix = "arXiv",
    primaryClass = "quant-ph",
}

@inproceedings{Unruh_2000,
	address = {Berlin, Heidelberg},
	author = {Unruh, William G.},
	booktitle = {Relativistic Quantum Measurement and Decoherence},
	editor = {Breuer, Heinz-Peter and Petruccione, Francesco},
	isbn = {978-3-540-45369-7},
	pages = {125--140},
	publisher = {Springer},
	title = {False Loss of Coherence},
	year = {2000}}

@article{cmp/1104248844,
	author = {Fumio Hiai and D{\'e}nes Petz},
	doi = {10.1007/BF02100287},
	journal = {Commun. Math. Phys.},
	number = {1},
	pages = {99 -- 114},
	publisher = {Springer},
	title = {{The proper formula for relative entropy and its asymptotics in quantum probability}},
	url = {https://doi.org/},
	volume = {143},
	year = {1991}}

@book{Lebesgue_1904,
  author    = {Lebesgue, Henri},
  title     = {Le\c{c}ons sur l'int\'egration et la recherche des fonctions primitives},
  address   = {Paris},
  publisher = {Gauthier{-}Villars},
  year      = {1904},
  langid    = {french}
}

@article{Fierz_1936,
  author    = {Fierz, Markus},
  title     = {Über die künstliche Umwandlung des Protons in ein Neutron},
  journal   = {Helvetica Physica Acta},
  year      = {1936},
  volume    = {9},
  number    = {IV},
  pages     = {245},
  publisher = {Birkh{\"a}user},
  doi       = {10.5169/seals-110626},
  issn      = {0018-0238},
  url       = {https://www.e-periodica.ch},
}

@article{Ashtekar_2018,
    author = "Ashtekar, Abhay and Campiglia, Miguel and Laddha, Alok",
    title = "{Null infinity, the BMS group and infrared issues}",
    doi = "10.1007/s10714-018-2464-3",
    journal = {Gen. Relativ. Gravit.},
    volume = "50",
    number = "11",
    pages = "140--163",
    year = "2018"
}

@article{Faulkner_2013,
    author = "Faulkner, Thomas and Lewkowycz, Aitor and Maldacena, Juan",
    title = "{Quantum corrections to holographic entanglement entropy}",
    doi = "10.1007/JHEP11(2013)074",
    journal = {J. High Energy Phys.},
    volume = "11",
    pages = "074",
    year = "2013",
    number = {2013},
}

@article{Bondi_1962,
  author  = {Bondi, Hermann and van der Burg, M. J. G. and Metzner, A. W. K.},
  title   = {Gravitational Waves in General Relativity. VII. Waves from Axi-Symmetric Isolated Systems},
  journal = {Proceedings of the Royal Society of London. Series A. Mathematical and Physical Sciences},
  volume  = {269},
  number  = {1336},
  pages   = {21--52},
  year    = {1962},
  doi     = {10.1098/rspa.1962.0161}
}

@article{Sachs_1962,
  author  = {Sachs, Rainer K.},
  title   = {Gravitational Waves in General Relativity. VIII. Waves in Asymptotically Flat Space-Time},
  journal = {Proceedings of the Royal Society of London. Series A. Mathematical and Physical Sciences},
  volume  = {270},
  number  = {1340},
  pages   = {103--126},
  year    = {1962},
  doi     = {10.1098/rspa.1962.0206}
}

@article{Casini:2008cr,
  author        = {Casini, Horacio},
  title         = {Relative entropy and the {Bekenstein} bound},
  journal       = {Classical Quantum Gravity},
  volume        = {25},
  number        = {20},
  pages         = {205021},
  year          = {2008},
  doi           = {10.1088/0264-9381/25/20/205021}
}

@article{Mari_2016,
  author        = {Mari, Andrea and De Palma, Giacomo and Giovannetti, Vittorio},
  title         = {Experiments testing macroscopic quantum superpositions must be slow},
  journal       = {Scientific Reports},
  volume        = {6},
  pages         = {22777},
  year          = {2016},
  doi           = {10.1038/srep22777},
  url           = {https://doi.org/10.1038/srep22777},
}

@article{GudderMarchandWyss1979Bures,
  author  = {Gudder, Stanley and Marchand, Jean-Paul and Wyss, Walter},
  title   = {Bures distance and relative entropy},
  journal = {Journal of Mathematical Physics},
  volume  = {20},
  number  = {9},
  pages   = {1963--1966},
  year    = {1979},
  month   = sep,
  doi     = {10.1063/1.524296}
}

@article{Lewkowycz:2013nqa,
  author        = {Lewkowycz, Aitor and Maldacena, Juan},
  title         = {Generalized gravitational entropy},
  journal       = {J. High Energy Phys.},
  volume        = {08},
  pages         = {090},
  year          = {2013},
  doi           = {10.1007/JHEP08(2013)090},
    number = {2013},
}

@article{Almheiri:2019hni,
  author        = {Almheiri, Ahmed and Hartman, Thomas and Maldacena, Juan and Shaghoulian, Edgar and Tajdini, Amirhossein},
  title         = {Replica Wormholes and the Entropy of Hawking Radiation},
  journal       = {J. High Energy Phys.},
  volume        = {05},
  pages         = {013},
  year          = {2020},
  doi           = {10.1007/JHEP05(2020)013},
    number = {2020},
}

@article{Hollands_2017,
   title={Non-equilibrium Thermodynamics and Conformal Field Theory},
   volume={357},
   ISSN={1432-0916},
   url={http://dx.doi.org/10.1007/s00220-017-2938-2},
   DOI={10.1007/s00220-017-2938-2},
   number={1},
   journal={Communications in Mathematical Physics},
   publisher={Springer Science and Business Media LLC},
   author={Hollands, Stefan and Longo, Roberto},
   year={2017},
   month=jul, pages={43–60} }

@misc{Kudler-Flam_2025,
    author = "Kudler-Flam, Jonah and Prabhu, Kartik and Satishchandran, Gautam",
    title = "{Vacua and infrared radiation in de Sitter quantum field theory}",
    eprint = "2503.19957",
    archivePrefix = "arXiv",
    primaryClass = "hep-th",
}

@article{He_2014,
  author        = {He, Temple and Mitra, Prahar and Porfyriadis, Achilleas P. and Strominger, Andrew},
  title         = {New Symmetries of Massless QED},
  journal       = {J. High Energy Phys.},
  volume        = {10},
  pages         = {112},
  year          = {2014},
  doi           = {10.1007/JHEP10(2014)112},
    number = {2014},
}

@article{Almheiri_2019,
    author = "Almheiri, Ahmed and Engelhardt, Netta and Marolf, Donald and Maxfield, Henry",
    title = "{The entropy of bulk quantum fields and the entanglement wedge of an evaporating black hole}",
    doi = "10.1007/JHEP12(2019)063",
    journal = {J. High Energy Phys.},
    volume = "12",
    pages = "063",
    year = "2019",
    number = {2019},
}

@article{Penington_2019,
    author = "Penington, Geoffrey",
    title = "{Entanglement Wedge Reconstruction and the Information Paradox}",
    doi = "10.1007/JHEP09(2020)002",
    journal = {J. High Energy Phys.},
    volume = "09",
    pages = "002",
    year = "2020",
    number = {2020},
}

@article{Puzzuoli_2016,
    author = "Puzzuoli, Daniel and Watrous, John",
    title = "{Ancilla dimension in quantum channel discrimination}",
    doi = "10.1007/s00023-016-0537-y",
    year = {2017},
    journal = {Ann. Henri Poincar\'e},
    volume = {18},
    pages = {1153--1184},
}

@article{Raginsky_2001,
  author  = {Maxim Raginsky},
  title   = {A fidelity measure for quantum channels},
  journal = {Physics Letters A},
  volume  = {290},
  number  = {1-2},
  pages   = {11--18},
  year    = {2001},
  doi     = {10.1016/S0375-9601(01)00640-5},
}

@article{Iyer_1994,
	author = {Iyer, Vivek and Wald, Robert M.},
	doi = {10.1103/PhysRevD.50.846},
	journal = {Phys. Rev. D},
	pages = {846--864},
	title = {{Some properties of Noether charge and a proposal for dynamical black hole entropy}},
	volume = {50},
	year = {1994}}

@article{PSW_2022,
    author = "Prabhu, Kartik and Satishchandran, Gautam and Wald, Robert M.",
    title = "{Infrared finite scattering theory in quantum field theory and quantum gravity}",
    doi = "10.1103/PhysRevD.106.066005",
    journal = "Phys. Rev. D",
    volume = "106",
    number = "6",
    pages = "066005",
    year = "2022"
}

@article{Kibble_1968,
    author = "Kibble, T. W. B.",
    title = "{Coherent Soft-Photon States and Infrared Divergences. I. Classical Currents}",
    doi = "10.1063/1.1664582",
    journal = {J. Math. Phys. (N.Y.)},
    volume = "9",
    number = "2",
    pages = "315--324",
    year = "1968"
}

@article{Carney_2017,
    author = "Carney, Daniel and Chaurette, Laurent and Neuenfeld, Dominik and Semenoff, Gordon Walter",
    title = "{Dressed infrared quantum information}",
    doi = "10.1103/PhysRevD.97.025007",
    journal = "Phys. Rev. D",
    volume = "97",
    number = "2",
    pages = "025007",
    year = "2018"
}

@article{DSW_2022,
	adsurl = {https://ui.adsabs.harvard.edu/abs/2022IJMPD..3141003D},
	author = {{Danielson}, Daine L. and {Satishchandran}, Gautam and {Wald}, Robert M.},
	doi = {10.1142/S0218271822410036},
	eid = {2241003},
	journal = {Int. J. Mod. Phys. D},
	month = oct,
	number = {14},
	pages = {2241003},
	title = {{Black holes decohere quantum superpositions}},
	volume = {31},
	year = 2022}

@article{Kay:1988mu,
	author = {Kay, Bernard S. and Wald, Robert M.},
	doi = {10.1016/0370-1573(91)90015-E},
	journal = {Phys. Rep.},
	pages = {49--136},
	reportnumber = {PRINT-88-0840 (CHICAGO)},
	title = {{Theorems on the Uniqueness and Thermal Properties of Stationary, Nonsingular, Quasifree States on Space-Times with a Bifurcate Killing Horizon}},
	volume = {207},
	year = {1991}}

@misc{DSW_2026,
    author = {Danielson, Daine L. and Satishchandran, Gautam and Wald, Robert M.},
    title = {Bulk Locality and Asymptotic States in Quantum Field Theory and Quantum Gravity},
    note = {in preparation},
}

@article{stinespring1955positive,
	author = {Stinespring, W Forrest},
	journal = {Proceedings of the American Mathematical Society},
	number = {2},
	pages = {211--216},
	title = {Positive functions on C$^*$-algebras},
	volume = {6},
	year = {1955},
    doi = {10.1090/S0002-9939-1955-0069403-4},
}

@article{hawking1975particle,
	author = {Hawking, Stephen W},
	journal = {Communications in mathematical physics},
	number = {3},
	pages = {199--220},
	publisher = {Springer},
	title = {Particle creation by black holes},
	volume = {43},
	year = {1975},
    doi = {10.1007/BF02345020},
}

@misc{Strominger_2,
	adsurl = {https://ui.adsabs.harvard.edu/abs/2017arXiv170305448S},
	archiveprefix = {arXiv},
	author = {{Strominger}, Andrew},
	doi = {10.48550/arXiv.1703.05448},
	eprint = {1703.05448},
	primaryclass = {hep-th},
	title = {{Lectures on the Infrared Structure of Gravity and Gauge Theory}}}

@article{Belin_2022,
    author = "Belin, Alexandre and Myers, Robert C. and Ruan, Shan-Ming and S{\'a}rosi, G{\'a}bor and Speranza, Antony J.",
    title = "{Does Complexity Equal Anything?}",
    reportNumber = "CERN-TH-2021-181, YITP-22-02",
    doi = "10.1103/PhysRevLett.128.081602",
    journal = "Phys. Rev. Lett.",
    volume = "128",
    number = "8",
    pages = "081602",
    year = "2022"
}

@article{Belin_2023,
	author = {Belin, Alexandre and Myers, Robert C. and Ruan, Shan-Ming and S{\'a}rosi, G{\'a}bor and Speranza, Antony J.},
	date = {2023-01-26},
	year = {2023},
	doi = {10.1007/JHEP01(2023)154},
	eprinttype = {arxiv},
	issn = {1029-8479},
	journal = {J. High Energy Phys.},
	langid = {english},
	number = {2023},
	pages = {154},
	shortjournal = {{JHEP}},
	title = {Complexity equals anything {II}},
	url = {https://doi.org/10.1007/JHEP01(2023)154},
	urldate = {2025-04-06},
	volume = {01}}

@article{Beny:2010byk,
	author = {B{\'e}ny, C{\'e}dric and Oreshkov, Ognyan},
	date = {2010-03-23},
	year = {2010},
	doi = {10.1103/PhysRevLett.104.120501},
	eprinttype = {arxiv},
	issn = {0031-9007, 1079-7114},
	journal = {Physical Review Letters},
	number = {12},
	pages = {120501},
	shortjournal = {Phys. Rev. Lett.},
	title = {General conditions for approximate quantum error correction and near-optimal recovery channels},
	url = {http://arxiv.org/abs/0907.5391},
	urldate = {2025-02-14},
	volume = {104}}

@article{Brown_2016,
	author = {Brown, Adam R. and Susskind, Leonard},
	date = {2018-04-26},
	year = {2018},
	doi = {10.1103/PhysRevD.97.086015},
	eprinttype = {arxiv},
	journal = {Phys. Rev. D},
	number = {8},
	pages = {086015},
	shortjournal = {Phys. Rev. D},
	title = {Second law of quantum complexity},
	url = {https://arxiv.org/abs/1701.01107},
	volume = {97}}

@article{Brown_2018,
	author = {Brown, Adam R. and Roberts, Daniel A. and Susskind, Leonard and Swingle, Brian and Zhao, Ying},
	date = {2016-04-18},
	year = {2016},
	doi = {10.1103/PhysRevD.93.086006},
	eprinttype = {arxiv},
	journal = {Phys. Rev. D},
	number = {8},
	pages = {086006},
	shortjournal = {Phys. Rev. D},
	title = {Complexity, action, and black holes},
	url = {https://arxiv.org/abs/1512.04993},
	volume = {93}}

@article{DKSW_2025,
    author = "Danielson, Daine L. and Kudler-Flam, Jonah and Satishchandran, Gautam and Wald, Robert M.",
    title = "{How to minimize the decoherence caused by black holes}",
    doi = "10.1103/67vv-km43",
    journal = "Phys. Rev. D",
    volume = "112",
    number = "2",
    pages = "025012",
    year = "2025"
}

@article{DSW_2021,
    author = "Danielson, Daine L. and Satishchandran, Gautam and Wald, Robert M.",
    title = "{Gravitationally mediated entanglement: Newtonian field versus gravitons}",
    doi = "10.1103/PhysRevD.105.086001",
    journal = "Phys. Rev. D",
    volume = "105",
    number = "8",
    pages = "086001",
    year = "2022"
}

@article{DSW_2023,
    author = "Danielson, Daine L. and Satishchandran, Gautam and Wald, Robert M.",
    title = "{Killing horizons decohere quantum superpositions}",
    doi = "10.1103/PhysRevD.108.025007",
    journal = "Phys. Rev. D",
    volume = "108",
    number = "2",
    pages = "025007",
    year = "2023"
}

@article{DSW_2024,
    author = "Danielson, Daine L. and Satishchandran, Gautam and Wald, Robert M.",
    title = "{Local description of decoherence of quantum superpositions by black holes and other bodies}",
    doi = "10.1103/PhysRevD.111.025014",
    journal = "Phys. Rev. D",
    volume = "111",
    number = "2",
    pages = "025014",
    year = "2025"
}

@article{Frob:2024ijk,
    title={Petz–Rényi relative entropy in QFT from modular theory},
    volume={115},
    ISSN={1573-0530},
    url={http://dx.doi.org/10.1007/s11005-025-01923-2},
    DOI={10.1007/s11005-025-01923-2},
    number={2},
    journal={Lett. Math. Phys.},
    publisher={Springer Science and Business Media LLC},
    author={Fröb, Markus B. and Sangaletti, Leonardo},
    year={2025},
    month=mar,
    pages = {30},
}

@article{CCNS_2017,
	author = {Carney, Daniel and Chaurette, Laurent and Neuenfeld, Dominik and Semenoff, Gordon Walter},
	doi = {10.1103/PhysRevLett.119.180502},
	journal = {Phys. Rev. Lett.},
	number = {18},
	pages = {180502},
	title = {{Infrared quantum information}},
	volume = {119},
	year = {2017}}

@article{rudolph_unambiguous_2003,
	author = {Rudolph, Terry and Spekkens, Robert W. and Turner, Peter S.},
	date = {2003-07-17},
	year = {2003},
	doi = {10.1103/PhysRevA.68.010301},
	issn = {1050-2947, 1094-1622},
	journal = {Physical Review A},
	langid = {english},
	number = {1},
	pages = {010301},
	rights = {http://link.aps.org/licenses/aps-default-license},
	shortjournal = {Phys. Rev. A},
	title = {Unambiguous discrimination of mixed states},
	url = {https://link.aps.org/doi/10.1103/PhysRevA.68.010301},
	urldate = {2025-02-27},
	volume = {68}}

@article{Stanford_2014,
	author = {Stanford, Douglas and Susskind, Leonard},
	date = {2014-12-11},
	year = {2014},
	doi = {10.1103/PhysRevD.90.126007},
	eprinttype = {arxiv},
	journal = {Phys. Rev. D},
	number = {12},
	pages = {126007},
	shortjournal = {Phys. Rev. D},
	title = {Complexity and Shock Wave Geometries},
	url = {https://arxiv.org/abs/1406.2678},
	volume = {90}}

@phdthesis{Fuchs_1995,
    author = "Fuchs, Christopher A.",
    title = "{Distinguishability and accessible information in quantum theory}",
    school = "New Mexico U.",
    year = "1995"
}

@article{Engelhardt_2014,
  author         = "Engelhardt, Netta and Wall, Aron C.",
  title          = "{Quantum Extremal Surfaces: Holographic Entanglement Entropy beyond the Classical Regime}",
  journal        = {J. High Energy Phys.},
  volume         = "01",
  pages          = "073",
  year           = "2015",
  doi            = "10.1007/JHEP01(2015)073",
    number = {2015},
}

@article{Bousso_2015,
  author         = "Bousso, Raphael and Fisher, Zachary and Leichenauer, Stefan and Wall, Aron C.",
  title          = "{A Quantum Focussing Conjecture}",
  journal        = "Phys. Rev. D",
  volume         = "93",
  number         = "6",
  pages          = "064044",
  year           = "2016",
  doi            = "10.1103/PhysRevD.93.064044",
}

@article{Harlow_2014,
    author = "Harlow, Daniel",
    title = "{Jerusalem Lectures on Black Holes and Quantum Information}",
    doi = "10.1103/RevModPhys.88.015002",
    journal = "Rev. Mod. Phys.",
    volume = "88",
    pages = "015002",
    year = "2016"
}

@article{Pasterski_2015,
    author = "Pasterski, Sabrina",
    title = "{Asymptotic Symmetries and Electromagnetic Memory}",
    doi = "10.1007/JHEP09(2017)154",
    journal = {J. High Energy Phys.},
    volume = "09",
    pages = "154",
    year = "2017",
    number = {2017},
}

@article{Prabhu_2024b,
    author = "Prabhu, Kartik and Satishchandran, Gautam",
    title = "{Infrared finite scattering theory: scattering states and representations of the BMS group}",
    doi = "10.1007/JHEP08(2024)055",
    journal = {J. High Energy Phys.},
    volume = "08",
    pages = "055",
    year = "2024",
    number = {2024},
}

@article{Kudler-Flam_2023,
    author = "Kudler-Flam, Jonah and Leutheusser, Samuel and Satishchandran, Gautam",
    title = "{Generalized black hole entropy is von Neumann entropy}",
    doi = "10.1103/PhysRevD.111.025013",
    journal = "Phys. Rev. D",
    volume = "111",
    number = "2",
    pages = "025013",
    year = "2025"
}

@article{Strominger_1996,
    author = "Strominger, Andrew and Vafa, Cumrun",
    title = "{Microscopic origin of the Bekenstein-Hawking entropy}",
    reportNumber = "HUTP-96-A002, RU-96-01",
    doi = "10.1016/0370-2693(96)00345-0",
    journal = "Phys. Lett. B",
    volume = "379",
    pages = "99--104",
    year = "1996"
}

@article{Beckenstein_1973,
    author = "Bekenstein, Jacob D.",
    title = "{Black holes and entropy}",
    doi = "10.1103/PhysRevD.7.2333",
    journal = "Phys. Rev. D",
    volume = "7",
    pages = "2333--2346",
    year = "1973"
}

@article{Herdegen_1997,
    author = "Herdegen, Andrzej",
    title = "{Semidirect product of CCR and CAR algebras and asymptotic states in quantum electrodynamics}",
    doi = "10.1063/1.532264",
    journal = {J. Math. Phys. (N.Y.)},
    volume = "39",
    pages = "1788--1817",
    year = "1998"
}

@article{Prabhu_2024,
    author = "Prabhu, Kartik and Satishchandran, Gautam",
    title = "{Infrared finite scattering theory: Amplitudes and soft theorems}",
    doi = "10.1103/PhysRevD.110.085022",
    journal = "Phys. Rev. D",
    volume = "110",
    number = "8",
    pages = "085022",
    year = "2024"
}

@article{Witten,
	author = {Witten, Edward},
	date = {2022-10-03},
	year = {2022},
	doi = {10.1007/JHEP10(2022)008},
	eprinttype = {arxiv},
	journal = {J. High Energy Phys.},
	pages = {008},
	shortjournal = {{JHEP}},
	title = {Gravity and the crossed product},
	url = {https://arxiv.org/abs/2112.12828},
	volume = {10},
    number = {2022},
}
\end{document}